\documentclass{ws-ijmpd}
\usepackage[super,compress]{cite}
\usepackage{slashed}
\usepackage{color}
\usepackage{hyperref}
\usepackage[normalem]{ulem}

\newcommand{\sub}[2]{#1_{\mathrm{#2}}}
\newcommand{\GeV}{~\mathrm{GeV}}
\newcommand{\TeV}{~\mathrm{TeV}}
\newcommand{\cm}{~\mathrm{cm}}
\newcommand{\s}{\mbox{s}}

\begin{document}

\markboth{M. Garny, A. Ibarra, S. Vogl}
{Signatures of Majorana dark matter with $t$-channel mediators}

%
\catchline{}{}{}{}{}
%

\title{Signatures of Majorana dark matter with $t$-channel mediators}

\author{MATHIAS GARNY}

\address{CERN Theory division,\\
CH-1211 Geneva 23, Switzerland\\
mathias.garny@cern.ch}

\author{ALEJANDRO IBARRA}

\address{Physik-Department T30d, Technische Universit\"at M\"unchen, \\
James-Franck-Stra\ss e, 85748 Garching, Germany\\
ibarra@tum.de}

\author{STEFAN VOGL}

\address{Oskar Klein Centre, Department of Physics, Stockholm University, \\
Stockholm, SE-10961, Sweden\\
stefan.vogl@fysik.su.se}

\maketitle


\begin{abstract}

Three main strategies are being pursued to search for non-gravitational dark matter signals: direct detection, indirect detection and collider searches. Interestingly, experiments have reached sensitivities in these three search strategies which may allow detection in the near future. In order to take full benefit of the wealth of experimental data, and in order to confirm a possible dark matter signal, it is necessary to specify the nature of the dark matter particle and of the mediator to the Standard Model. In this paper, we focus on a simplified model where the dark matter particle is a Majorana fermion that couples to a light Standard Model fermion via a Yukawa coupling with a scalar mediator. We review the observational signatures of this model and we discuss the complementarity among the various search strategies, with emphasis in the well motivated scenario where the dark matter particles are produced in the early Universe via thermal freeze-out.  
\end{abstract}


\ccode{Preprint numbers: CERN-PH-TH-2015-036, TUM-HEP 985/15}


\section{Introduction}

Since Zwicky conjectured the existence of dark matter~\cite{Zwicky:1937zza} (DM) to explain his own observations on the dynamics of galaxies in the Coma Cluster~\cite{Zwicky:1933gu}, a lot of effort has been invested in determining the properties of this new form of matter. On the one hand, current cosmological and astrophysical observations are consistent  with the dark matter being constituted by an electrically neutral and colorless particle, stable on cosmological time scales, non-relativistic at the time of structure formation, and with an abundance today which amounts to $\sim 80\%$ of the total matter density of our Universe. On the other hand, the particle physics properties of the dark matter, such as the mass, the spin, the lifetime,  or the nature and strength of its couplings to the luminous matter, are either unknown or poorly constrained by observations or by theoretical considerations. In fact, viable dark matter models have been constructed with masses ranging between  $\sim 1$ $\mu$eV and  $10^{16}$ GeV, and interaction cross sections ranging between  $10^{-35}$ pb and  $\sim 1$ pb (for a review, see Refs.~\citen{Bertone:2004pz,Bergstrom:2012fi,Jungman:1995df}).

One of the most pressing open questions in Fundamental Physics consists precisely in pinning down the Particle Physics properties of the dark matter particle. To this end, three different approaches are being pursued: direct detection, which aims to the observation of the nuclear recoil induced by the scattering with a dark matter particle, indirect detection, which aims to the observation of the photons, antimatter particles or neutrinos hypothetically produced in dark matter annihilations or decays, and collider searches, which aim to the production at a collider of dark matter particles and their subsequent detection through the observation of large amounts of missing energy in the final state. Given the intrinsic complexity of searching for a particle with unknown properties in an experiment with large and, in some instances, poorly understood backgrounds, it is of paramount importance to theoretically identify the cleanest and most powerful search strategy, as well as the synergies among the various strategies. This program, unfortunately, relies heavily on the details of the model. 

In this review we focus in scenarios where the dark matter particle is a Majorana fermion with mass in the range $\sim 10-10^4$ GeV that couples to a light Standard Model (SM) fermion via a Yukawa coupling with a scalar mediator. This scenario leads under reasonable assumptions to the correct dark matter abundance and offers a rich phenomenology both in indirect detection, direct detection and collider experiments, which moreover leads to characteristic signals in the concrete scenario where the dark matter particle is degenerate in mass with the scalar mediator. In Section \ref{sec:toy-model} we present the model and we identify its free parameters. In Section \ref{sec:Relic} we analyze the thermal production of the dark matter particle, emphasizing the role of coannihilations in the mass degenerate scenario. In Section \ref{sec:ID} we review the signals of the model in gamma-ray, antimatter and neutrino detectors, as well as the constraints on the model parameters that stem from current searches, while in Sections \ref{sec:DD} and \ref{sec:collider} the signals in direct detection and collider experiments, respectively. Finally, in Section \ref{sec:complementarity} we discuss the complementarity among the various search strategies and in Section \ref{sec:conclusions} we present our conclusions. \ref{ap:crosssections} contains a collection of relevant cross sections
and \ref{app:DD} a discussion of effective operators for direct detection.

\section{Description of the model}\label{sec:toy-model}

We consider an extension of the Standard Model by one colorless and electrically neutral Majorana fermion, $\chi$, which we assume to constitute the dominant component of dark matter in the Universe, and one complex scalar particle,  $\eta$, which mediates a Yukawa interaction between the dark matter particle and a Standard Model fermion $f$. The general form of the Lagrangian reads:
\begin{align}
{\cal L}={\cal L}_{\rm SM}+{\cal L}_{\chi}+{\cal L}_\eta+
{\cal L}^{\rm fermion}_{\rm int}+{\cal L}^{\rm scalar}_{\rm int}\;.
\end{align}
Here, ${\cal L}_{\rm SM}$ is the Standard Model Lagrangian which includes a potential for the Higgs doublet $\Phi$, $V=m_1^2 \Phi^\dagger \Phi +\frac{1}{2}\lambda_1 (\Phi^\dagger \Phi)^2$. On the other hand ${\cal L}_\chi$ and ${\cal L}_\eta$ are the parts of the Lagrangian involving just the Majorana fermion $\chi$ and the scalar particle $\eta$, respectively, and which are given by:
\begin{align}
\begin{split}
{\cal L}_\chi&=\frac12 \bar \chi^c i\slashed{\partial} \chi
-\frac{1}{2}m_\chi \bar \chi^c\chi\;, \\
{\cal L}_\eta&=(D_\mu \eta)^\dagger  (D^\mu \eta)-m_2^2 \eta^\dagger\eta-
\frac{1}{2}\lambda_2(\eta^\dagger \eta)^2\;,
\end{split}
\end{align}
where $D_\mu$ denotes the covariant derivative. Lastly, ${\cal L}^{\rm fermion}_{\rm int}$ and ${\cal L}^{\rm scalar}_{\rm int}$ denote the fermionic and scalar interaction terms of the new particles to the Standard Model fermions and to the Higgs doublet. The interaction terms depend crucially on the quantum numbers of the dark matter particle and the scalar $\eta$. We first impose a $Z_2$ discrete symmetry under which $\chi$ and $\eta$ are odd while the Standard Model particles are even, in order to guarantee the stability of the dark matter particle. We also impose for simplicity that the dark matter particle only couples to one generation of fermions, which can be ensured by postulating that the extra scalar particle $\eta$ carries a family global quantum number while the dark matter particle does not.\footnote{Lifting this requirement leads in general to too large flavor violating effects. A detailed discussion of the conditions that must be fulfilled in order to satisfy the constraints from flavor physics can be found in the Appendix of Ref.~\citen{Garny:2014waa}, for coupling to quarks, and in Ref.~\citen{Kopp:2014tsa} for coupling to leptons.} Under these assumptions, the possible interaction terms are restricted by the gauge quantum numbers of the dark matter particle and the Standard Model fermion it couples to. 

We will focus in this review in scenarios where the dark matter particle is a $SU(2)_L$ singlet, and therefore does not couple at tree level with the weak gauge bosons. In these scenarios, the dark matter hypercharge must be zero in order to render an electrically neutral particle. Then, the Yukawa coupling of the dark matter to a right-handed quark (lepton) $f_R$ with hypercharge $Y$, requires the scalar $\eta$ to be a color triplet (singlet), singlet under $SU(2)_L$ and with hypercharge $-Y$. With these quantum numbers the only interaction terms in the Lagrangian are:
\begin{align}
  \begin{split}
    {\cal L}^{\rm fermion}_{\rm int}&= -y \bar \chi f_R \eta+{\rm h.c.}\;,  \\
    {\cal L}^{\rm scalar}_{\rm int}&= -\lambda_3(\Phi^\dagger \Phi)
    (\eta^\dagger \eta)\;.
  \end{split}
\label{eq:singlet-psiR}
\end{align}
On the other hand, if the dark matter only couples to a left-handed quark (lepton) with hypercharge $Y$, the scalar $\eta$ must be a color triplet (singlet), doublet under $SU(2)_L$ and with hypercharge $-Y$.  The corresponding interaction terms are:
\begin{align}
  \begin{split}
    {\cal L}^{\rm fermion}_{\rm int}&= -y \bar \chi(f_L i\sigma_2 \eta)+{\rm h.c.}\;,\\
    {\cal L}^{\rm scalar}_{\rm int}&= -\lambda_3(\Phi^\dagger \Phi)
    (\eta^\dagger \eta)
    -\lambda_4(\Phi^\dagger \eta)(\eta^\dagger \Phi)\;.
  \end{split}
\label{eq:singlet-L}
\end{align}
In the Minimal Supersymmetric Standard Model (MSSM) these possibilities are realized if $\chi$ is the bino and $\eta$ is a sfermion. The Yukawa coupling $y=y_{\rm MSSM}$ is then fixed in terms of the $U(1)_Y$ gauge coupling $g'$ to $y_{\rm MSSM}=4g'/(3\sqrt{2})\sim 0.33$ for up-type right-handed quarks, $y_{\rm MSSM}=2g'/(3\sqrt{2})\sim 0.16$ for down-type right-handed quarks and $y_{\rm MSSM}=g'/(3\sqrt{2})\sim 0.08$ for the left-handed quarks. Besides, $y_{\rm MSSM}=\sqrt{2}g' \sim 0.48$ for right-handed leptons and $y_{\rm MSSM}=g'/ \sqrt{2}\sim 0.24$ for left-handed leptons.  For a general mixed neutralino, the couplings to the left- or right-handed fermions are, respectively, given by\cite{Bringmann:2012vr}
\begin{align}
  y_{\rm MSSM, L} &= -\frac{2q_f\mp 1}{\sqrt{2}}g'N_{11}\mp \frac{g}{\sqrt{2}}N_{12}\,,\\
  y_{\rm MSSM, R} &= \sqrt{2}q_fg'N_{11}\;.
\end{align}
In the following we explore the parameter space of the simplified model treating the coupling $y$ as a free parameter. Nevertheless, we refer to the supersymmetric coupling strength as a benchmark value. Furthermore, for simplicity, we assume that $|\lambda_i|\ll 1$, such that the relevant parameters of the model are the dark matter mass $m_\chi$, the mediator mass $m_\eta$ and the Yukawa coupling $y$. In addition, we mostly focus on a coupling to the right-handed fermions for definiteness, although many results can be generalized in a straightforward manner.

Apart from the collider constraints discussed in detail below, the Yukawa interaction involving leptons gives rise to a contribution to the anomalous magnetic moment\cite{Bringmann:2012vr},
\begin{align}
  \Delta a_\ell = -\frac{y^2}{16\pi^2} m_\ell^2 \frac{2m_\chi^6+3m_\chi^4m_\eta^2-6m_\chi^2m_\eta^4+m_\eta^6+6m_\chi^4m_\eta^2\log(m_\eta^2/m_\chi^2)}{6(m_\eta^2-m_\chi^2)^4}\;.
\end{align}
For both electrons and muons the contribution is subdominant compared to the Standard Model expectation for Yukawa couplings compatible with indirect detection constraints and $m_\chi\gtrsim 10$\,GeV\cite{Bringmann:2012vr,Kopp:2014tsa}.

\section{Relic abundance}\label{sec:Relic}

One of the most attractive frameworks to generate the observed dark matter abundance  in our Universe is the freeze-out mechanism. It assumes that the dark matter has interactions with the Standard Model strong enough to keep at high temperatures the dark matter particles in thermal equilibrium with the Standard Model particles. Besides, the interactions must be weak enough to allow the dark matter particles to drop out of equilibrium at sufficiently early times. After this time, dubbed the freeze-out time, the number density per comoving volume of dark matter particles remains practically constant until today, thus constituting the dark matter population we observe in the Universe. Under these assumptions, the dark matter abundance  $\Omega_{\rm DM} h^2=0.1198\pm 0.0015$ \cite{Planck:2015xua}, as measured by the Planck satellite, can be naturally explained. 

Defining the variable $x_{\rm f.o.}\equiv m_{\chi}/T_{\rm f.o.}$, the temperature $T_{\rm f.o.}$ at which the freeze-out takes place is approximately given by the solution of the following equation~\cite{Griest:1990kh}:
\begin{align}
x_{\rm f.o.}=\log\frac{0.038 c(c+2) g_{\rm int} M_{\rm Pl}m_{\chi} \langle \sigma v\rangle}{g_*(x_{\rm f.o.}) ^{1/2} x_{\rm f.o.}^{1/2}}\;,
\label{eq:freeze-out-temperature}
\end{align}
where $c$ is a numerical factor, $c\approx 0.5$, $M_{\rm Pl}=1.22\times 10^{19}\,{\rm GeV}$ is the Planck mass, $g_{*}(x_{\rm f.o.})$ is the number of relativistic degrees of freedom at the freeze-out temperature and $g_{\rm int}$ is the number of dark matter internal degrees of freedom, in this case $g_{\rm int}=2$. Typically $x_{\rm f.o.}$=20-30. 

The relic density of dark matter particles can be calculated solving the Boltzmann equation for the number density of dark matter particles as a function of the temperature. The result approximately reads~\cite{Griest:1990kh}:
\begin{align}
	\Omega_{\chi} h^{2} \simeq  \frac{1.07\times 10^{9}\,{\rm GeV}^{-1}}{J(x_{\rm f.o.})\, g_{*}(x_{\rm f.o.})^{1/2}\,M_{\rm Pl}}\,,
\label{eq:Omegah2}
\end{align}
where $J(x_{\rm f.o.})  =\int_{x_{\rm f.o.}}^{\infty}\,\langle\sigma  v\rangle x^{-2}\,\text{d}x\,$. Here $\langle\sigma  v\rangle$ is the thermally-averaged annihilation cross section, which can be readily calculated using the formalism presented in Ref.~\citen{Gondolo:1990dk}. When expanding the cross section for low velocities as $\sigma v=a+bv^2$, the thermal average is given by $\langle\sigma v\rangle =a+3b/x_{\rm f.o.}$.

As apparent from the previous equations, the thermally-averaged annihilation cross section is crucial for determining the dark matter relic abundance. For a $SU(2)_L$ singlet dark matter particle, the dominant annihilation channel is $\chi\chi\rightarrow f \bar f $, via the $t$-channel exchange of the mediator $\eta$. The velocity independent part is given by 
 \begin{align}
a = \frac{N_c \, m_f^2 \, y^4 \sqrt{m_{\chi }^2-m_f^2}}
{32\, \pi \,  m_{\chi }
   \left(m_{\eta}^2-m_f^2+m_{\chi }^2\right)^2}\,,
 \end{align}
and vanishes in the limit $m_f\rightarrow 0$. Besides the velocity dependent part approximately reads, 
 \begin{align}
b \simeq \frac{N_c m_{\chi }^2 y^4 \left(m_{\eta}^4+m_{\chi }^4\right)}{48 \pi 
   \left(m_{\eta}^2+m_{\chi }^2\right){}^4}\;,
 \end{align}
in the limit $m_f\to 0$ (see \ref{ap:crosssections} for the general expression).

For light fermions, namely $m_f\ll m_\chi$, and $m_\eta/m_\chi \gtrsim 1.5$ the relic density is approximately given by
\begin{align}\label{eq:omegaDM}
\Omega_{\chi} h^2\simeq \frac{0.12}{N_c} \left(\frac{0.37}{y}\right)^4 \left(\frac{m_{\chi}}{100\GeV}\right)^2 \left[ \sum_i \frac{1+m_{\eta_i}^4/m_{\chi}^4}{(1+m_{\eta_i}^2/m_{\chi}^2)^4}  \right]^{-1}\;,
\end{align}
for a freeze-out temperature $T_{\rm f.o.}=m_{\chi}/20$ and $g_*(T_{\rm f.o.})=80$ \cite{Gondolo:1990dk}.

\begin{figure}[t]
  \begin{center}
    \includegraphics[width=0.9\textwidth]{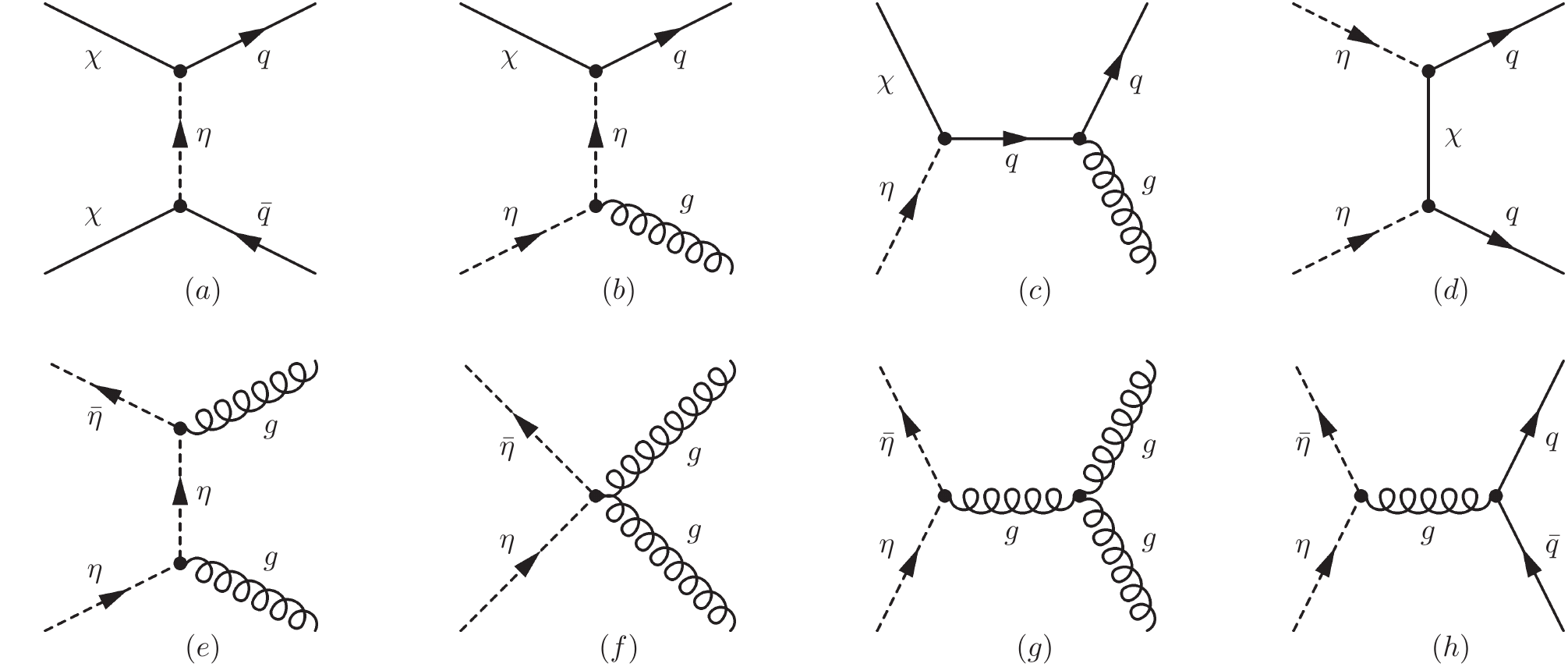}
  \end{center}
  \caption{\label{fig:coannihilation} Feynman diagrams relevant for the computation of the relic density, including coannihilations with a colored mediator.  For an uncolored mediator, similar diagrams involving $SU(2)\times U(1)$ gauge bosons are relevant.}
\end{figure}

This formalism must be modified when the scalar mediator is very degenerate in mass with the dark matter particle. If this is the case, the scalar $\eta$ is still present in the thermal bath and depletes the number of dark matter particles. The effective cross-section is given by
\begin{align}
\sigma _{\rm eff} v= \sum_{ij} \frac{n^{\rm eq}_i n^{\rm eq}_j}{(\sum_k n^{\rm eq}_k)^2} \sigma_{ij}v\,,
\label{eqn:sigma_v_eff}
\end{align}
where the sum extends over all coannihilating particles, including $\chi$. Here, $\sigma_{ij} v$ is the cross section for each of the reactions participating in the freeze-out, shown in Fig.\,\ref{fig:coannihilation}, and $n^{\rm eq}_i=g_i(m_iT/(2\pi))^{3/2}e^{-m_i/T}$ is the number density per comoving volume of particles $i$ with mass $m_i$ in equilibrium at the temperature $T$, with $g_\chi=2$, $g_\eta=g_{\bar\eta}=3 \,(1)$ being the number of internal degrees of freedom of the dark matter and of the colored (uncolored) scalar mediator. 

\begin{figure}[t]
  \begin{center}
    \includegraphics[width=0.49\textwidth]{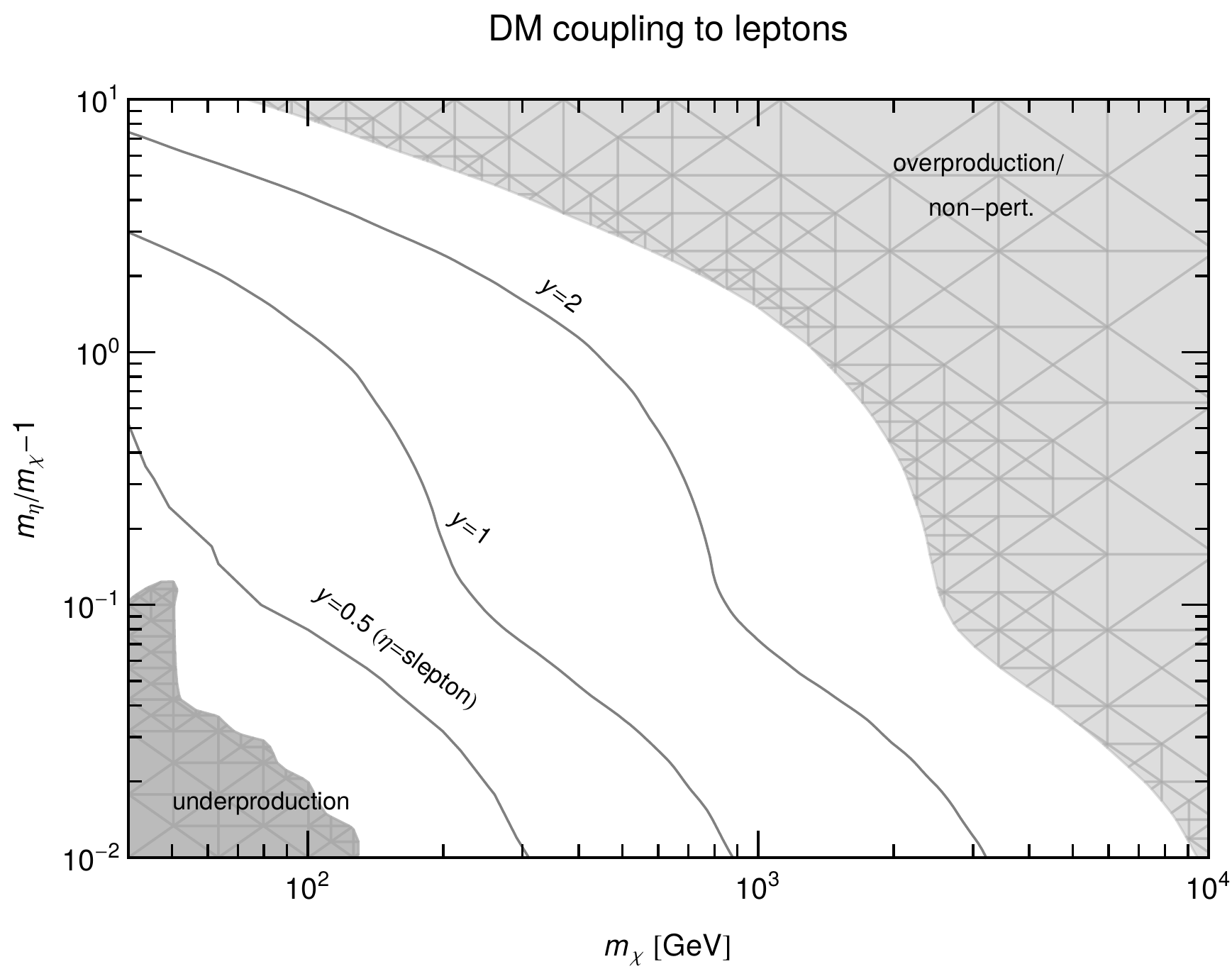}
    \includegraphics[width=0.49\textwidth]{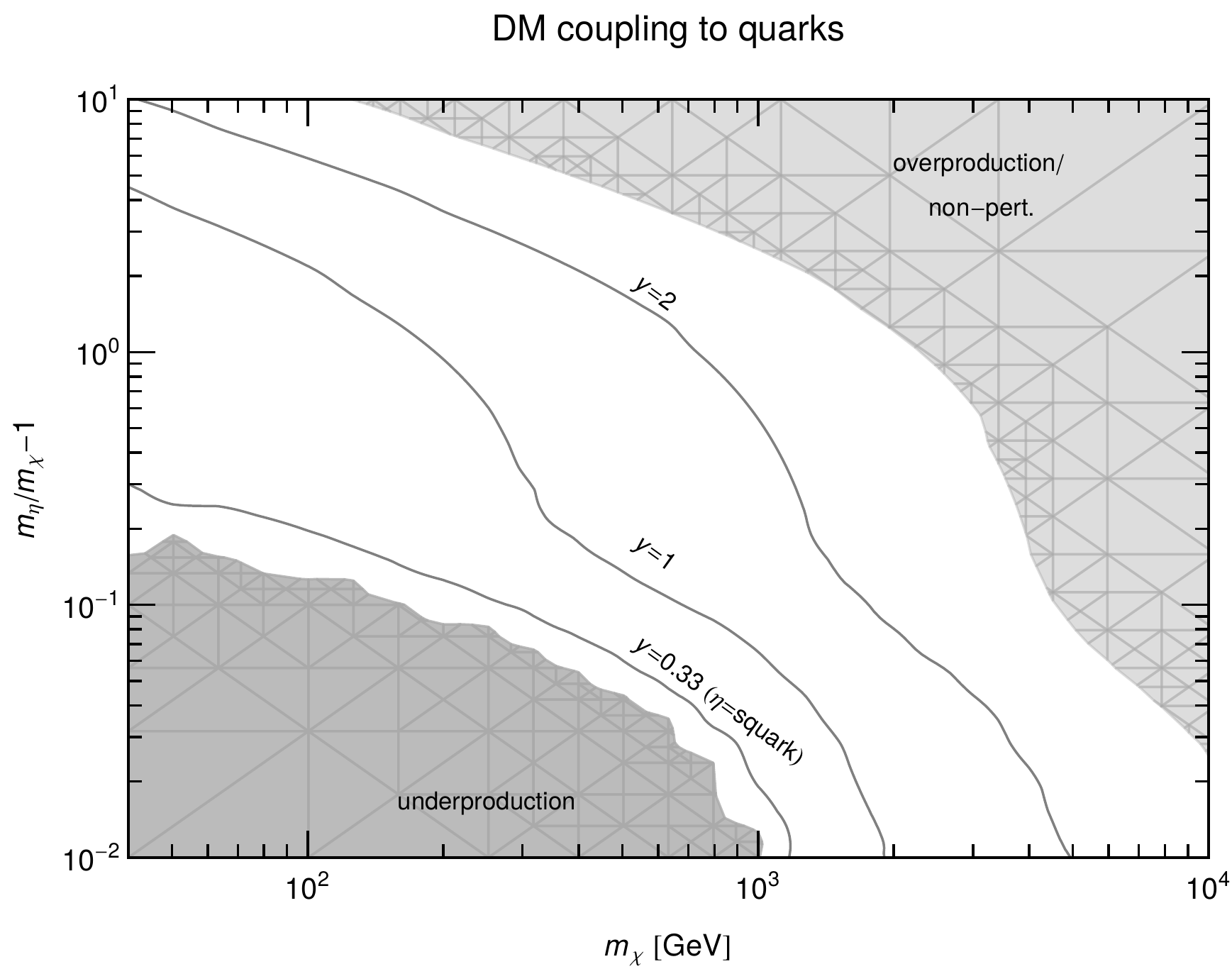}
  \end{center}
  \caption{\label{fig:relic} Parameter space compatible with dark matter production via thermal freeze-out for coupling to leptons (left plot) and to quarks (right plot). The contour lines show the value of the Yukawa coupling $y$ required to reproduce the observed  dark matter density $\Omega_{\rm DM} h^2=0.1198\pm 0.0015$ \cite{Planck:2015xua} via thermal freeze-out. The upper right shaded region leads to $\Omega_\chi>\Omega_{\rm DM}$ unless the coupling becomes non-perturbative, $y>\sqrt{4\pi}$. On the other hand, the lower left shaded region leads to $\Omega_\chi<\Omega_{\rm DM}$ due to strong coannihilations. The plot also shows the lines corresponding to the MSSM values of the Yukawa coupling, namely $y=0.33~(0.5)$ for a right-handed squark (slepton) mediator.
}
\end{figure}

The requirement of reproducing a dark matter density with the value measured by the Planck satellite, $\Omega_{\rm DM} h^2=0.1198\pm 0.0015$ \cite{Planck:2015xua} fixes one of the parameters of the model in terms of the remaining two, {\it e.g.} the Yukawa coupling $y=y_{\rm th}(m_{\chi},m_{\eta})$.  Under this reasonable condition, the parameter space of the model is spanned by only two parameters, that can be chosen to be $m_\chi$, the dark matter mass, and $(m_\eta-m_\chi)/m_\chi$, which measures the mass difference between the scalar mediator and the dark matter particle. Fig.~\ref{fig:relic} shows contour lines with the value of the Yukawa coupling that leads to the observed dark matter abundance via thermal freeze-out, assuming that the dark matter particle couples to leptons (left plot) or to quarks (right plot). As apparent from the plot, the region with large $m_\chi$ and/or large $m_\eta$ leads, for perturbative couplings $y\leq \sqrt{4\pi}$, to a relic abundance which exceeds the observed value. On the other hand, the region with small $m_\chi$ and/or large degeneracy leads to a too small relic abundance, due to the crucial role of coannihilations during freeze-out.
This behavior can be understood from the expression of the effective thermal cross section in the coannihilation regime, Eq.~\ref{eqn:sigma_v_eff}. In this simplified scenario, it is approximately given by:
\begin{align}
\sigma_{\rm eff} v\sim \sigma_{\chi \chi}v + \sigma_{\chi \eta}v\, R + \sigma_{\eta \eta} v\,  R^2\;,
\label{eqn:sigma_v_eff_schem}
\end{align}  
which includes a Boltzmann suppression factor $R=n_\eta^{\rm eq}/n_\chi^{\rm eq}\propto e^{-\frac{m_\eta-m_\chi}{T}}$ for the processes involving the coannihilating particle  $\eta$. The cross sections for $\chi\chi$, $\chi\eta$ and $\eta\eta$ are proportional to $y^4$, $y^2 g^2$ and $g^4$, respectively, $g$ being a gauge coupling constant (see Fig.\,\ref{fig:coannihilation}). Besides, for fixed ratio $m_\eta/m_\chi$, the cross sections must be inversely proportional to $m_\chi^2$. Then, casting the cross sections as $\sigma_{\chi \chi} v = \frac{y^4}{m_{\chi}^2} C_{\chi \chi}$, $\sigma_{\chi \eta} v = \frac{y^2 g^2}{m_{\chi}^2} C_{\chi \eta} R^{-1}$, $\sigma_{\eta\eta} v = \frac{g^4}{m_{\chi}^2} C_{\eta \eta} R^{-2}$, one finds~\cite{Garny:2013ama}
\begin{align}
\Omega_\chi h^2 \propto \frac{1}{\langle \sigma_{\rm eff} v\rangle }\sim \frac{m_{\chi}^2}{y^4 \, \langle C_{\chi \chi}  \rangle + y^2 \,g^2\, \langle C_{\chi \eta}\rangle  + g^4\, \langle C_{\eta \eta} \rangle}\;.
\label{eqn:Omega}
\end{align}
As apparent from this equation, there is always a value of the dark matter mass below which the dark matter abundance lags below the observed value, even for $y=0$, provided $g^4 C_{\eta \eta} \neq 0$, which is the case in the coannihilation regime where the Boltzmann suppression is relatively mild.  In very degenerate scenarios, $m_{\eta}/ m_{\chi} \approx 1.1$, the lower limit on the dark matter mass is $m_{\chi} \gtrsim 200\,\GeV \; (50\,\GeV) $ for dark matter coupling to quarks (leptons).

\section{Indirect detection}\label{sec:ID}

Immediately after the dark matter freeze-out, the annihilation rate of dark matter particles became negligibly small and their number density per comoving volume remained practically constant. In this epoch, the dark matter distribution was roughly homogeneous and isotropic throughout the whole Universe, except for the small density fluctuations generated by the inflaton field. The overdensity regions accreted more and more dark matter particles, thereby generating regions in our Universe with a density of dark matter particles which became orders of magnitude larger than the average density. The baryonic matter followed the gravitational potential wells generated by the dark matter, leading to structures which are identified with the galaxies and clusters of galaxies we observe today. In these regions of enhanced dark matter density, annihilations may still be efficient, producing Standard Model particles which contribute to the total fluxes of gamma-rays, cosmic-rays and (anti-)neutrinos. The search for this exotic component over the expected astrophysical backgrounds then provides information about the dark matter properties.

The annihilation of  dark matter particles with mass $m_{\chi}$ produces Standard Model particles at the position $\vec{r}$, at a rate per unit kinetic energy $T$ and unit volume given by 
\begin{align} 
Q(T,\vec  r)=\frac{1}{2}\frac{\rho^2_{\chi}(\vec{r})}{m^2_{\chi}}\sum_i
  (\sigma v)_i\frac{dN^i}{dT} \;, \label{eqn:source} 
\end{align} 
where $(\sigma v)_i$ is the annihilation cross section times the relative dark matter velocity in the annihilation channel $i$, producing an energy spectrum of Standard Model particles $dN^i/dT$, and $\rho_{\chi}(\vec{r})$ is the dark matter density at the position $\vec{r}$. Unfortunately, in this expression, none of the parameters involved is positively known. 

To assess the prospects to observe a dark matter signal in the sky, it is common to fix the annihilation final state, which then fixes $dN/dT$. Besides, the distribution of dark matter particles in the Milky Way is not precisely known, although it can be inferred from numerical $N$-body simulations. Some popular choices for the dark matter density profile, which illustrate the range of uncertainty in the predictions of the indirect dark matter signatures, are the Navarro-Frenk-White (NFW) profile~\cite{Navarro:1995iw,Navarro:1996gj}:
\begin{align}
  \rho_\text{DM}(r)=\frac{\rho_0}{(r/r_\text{s})
  [1+(r/r_\text{s})]^2}\;,
\end{align}
with scale radius $r_s = 24 ~\rm{kpc}$,~\cite{Cirelli:2010xx} the Einasto profile~\cite{Navarro:2003ew,Graham:2005xx,Navarro:2008kc}:
\begin{align}
  \rho_\text{DM}(r)=\rho_0 \exp\left[-\frac{2}{\alpha}
  \left(\frac{r}{r_s}\right)^\alpha\right]\;,
  \label{eq:Einasto}
\end{align}
with $\alpha=0.17$ and $r_s=20~\rm{kpc}$~\cite{Navarro:2003ew}, and the much shallower isothermal profile~\cite{Bahcall:1980fb}:
\begin{align}
\rho_{\rm DM}(r) = \frac{\rho_0}{1+r^2/r_s^2}\;,
\end{align}
with $r_s=4.4~\rm{kpc}$. In all the cases, the overall normalization factor $\rho_0$ is chosen to reproduce the local dark matter density $\rho_\odot= 0.39~\text{GeV}/\text{cm}^3$~\cite{Catena:2009mf, Weber:2009pt, Salucci:2010qr, Pato:2010yq, Iocco:2011jz} at the distance $r_\odot=8.5$~kpc of the Sun to the Galactic center. 

Under these assumptions for the annihilation final state and the dark matter density distribution, the source term Eq.~(\ref{eqn:source}) depends on two free parameters, the dark matter mass $m_\chi$ and the annihilation cross section in that channel $(\sigma v)$, which span the parameter space of indirect dark matter searches. Various experiments are currently constraining this parameter space using gamma-rays, antimatter or neutrinos. Let us first discuss the possible annihilation channels of the model and then the methods for indirect searches in each of these channels, as well as the present experimental limits.

\subsection{Annihilation channels}

As discussed in the previous section, the cross section for the lowest order annihilation channel $\chi \chi \rightarrow \bar{f} f$ is suppressed by the mass of the final fermion and by the relative velocity of the dark matter particles in the Galactic center, $v \approx 10^{-3}$. Therefore, higher order processes that lift the helicity suppression can give a sizable or even dominant contribution to the total annihilation cross section  for relative velocities relevant to indirect detection. This is the case for the two-to-three annihilation into a fermion-antifermion pair with the associated emission of a gauge boson $\chi\chi\rightarrow f\bar f V$~\cite{Bergstrom:1989jr,Flores:1989ru} and the one loop annihilation into two gauge bosons, $\chi\chi\rightarrow V V'$, which have the largest branching fractions over most of the parameter space, despite the suppression of the cross section by the additional gauge coupling and the three-body phase space for the former process, and by the loop factor for the latter. The two-to-three annihilation $\chi\chi\rightarrow f \bar f h$ \cite{Luo:2013bua}, with $h$ the Higgs boson, also lifts the helicity suppression, however they are subdominant unless the mediator couples strongly to the Higgs, $\lambda_i\sim {\cal O}(1)$ (notice that the $s$-wave amplitude for the one-loop annihilation $\chi \chi\rightarrow hh$, contrary to $\chi\chi \rightarrow V V'$, identically vanishes due to CP conservation). Explicit expressions for the annihilation cross sections for the various channels can be found in \ref{ap:crosssections}.

\begin{figure}[t]
\begin{center}
\includegraphics[width=0.49\textwidth]{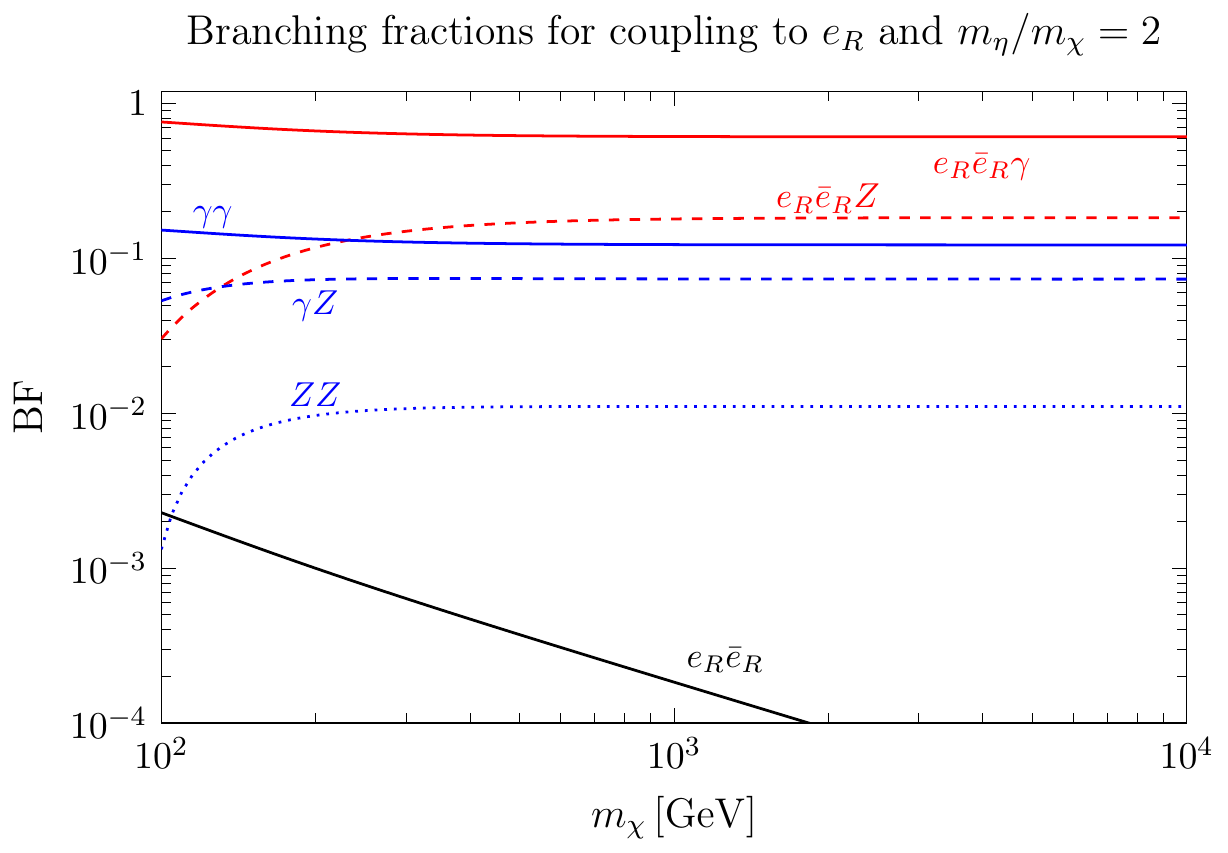}
\includegraphics[width=0.49\textwidth]{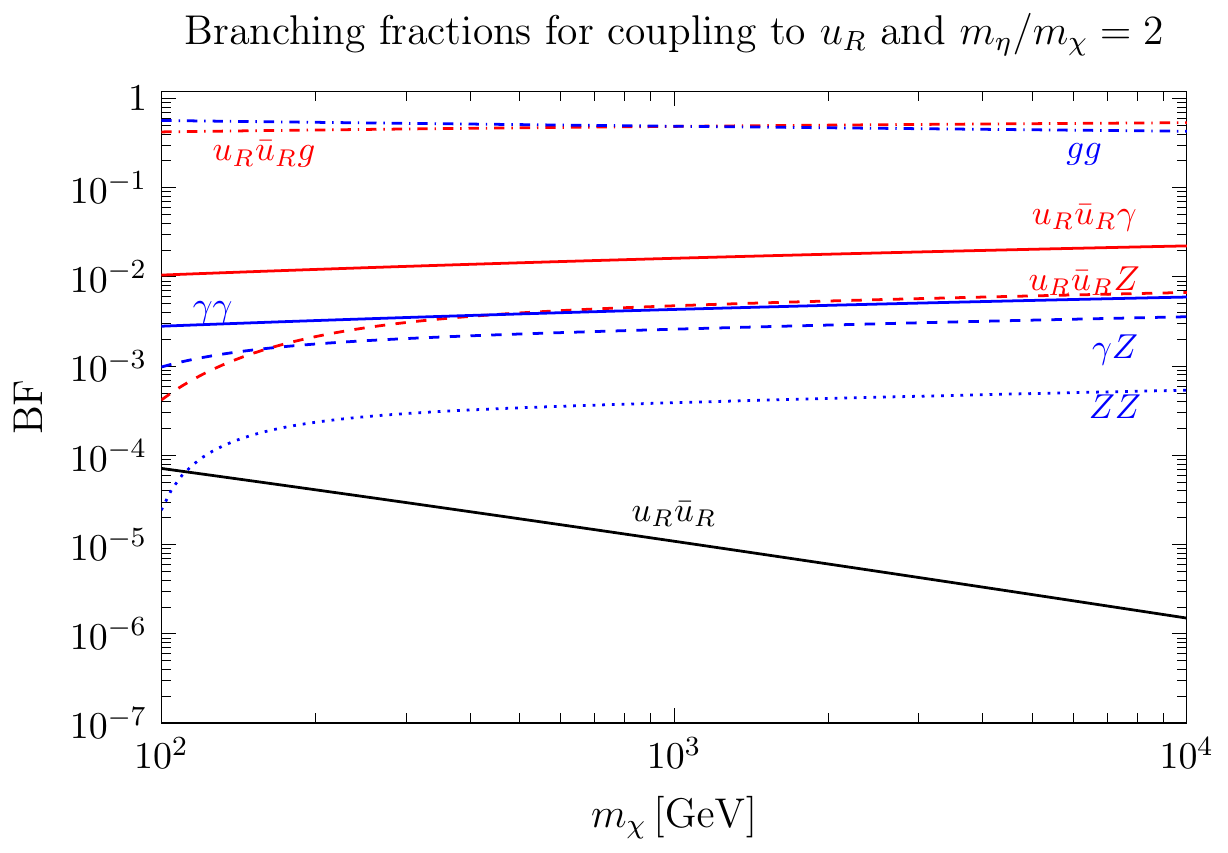} \\
\includegraphics[width=0.49\textwidth]{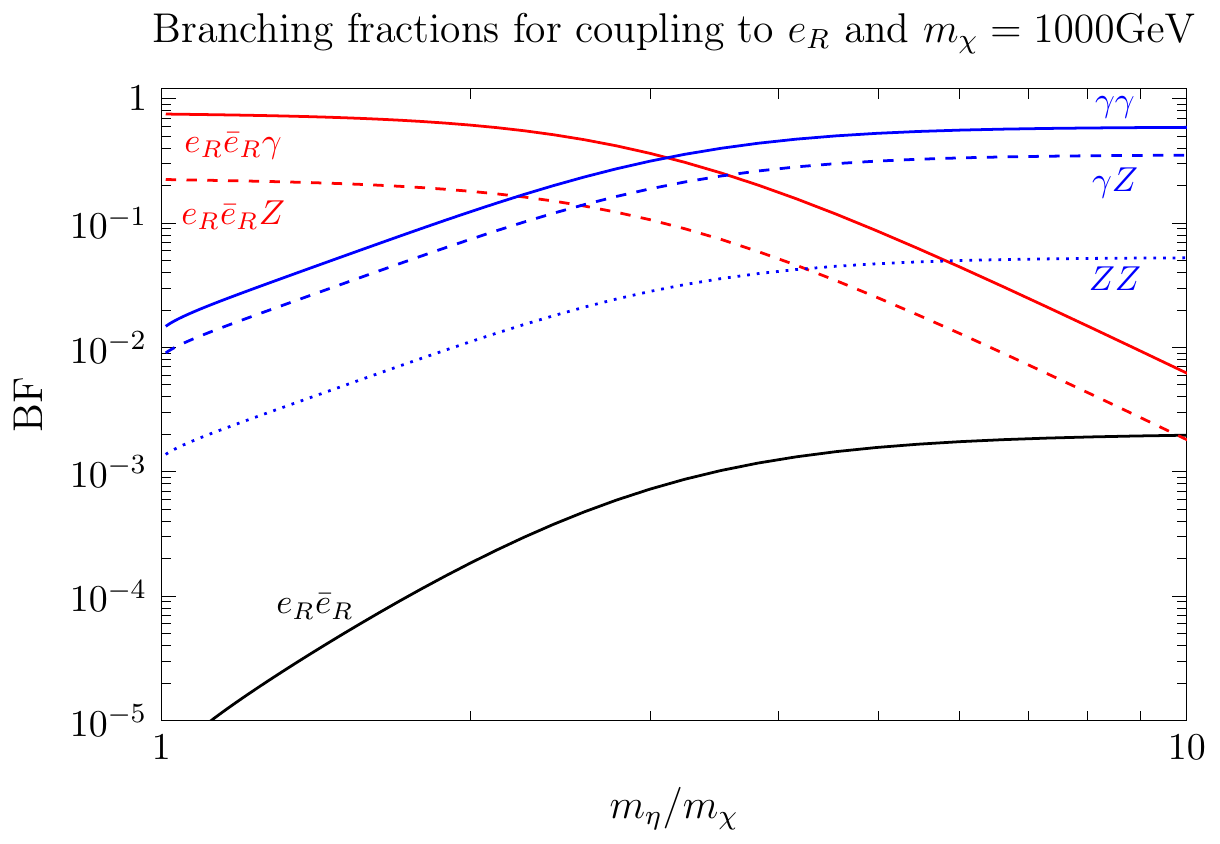}
\includegraphics[width=0.49\textwidth]{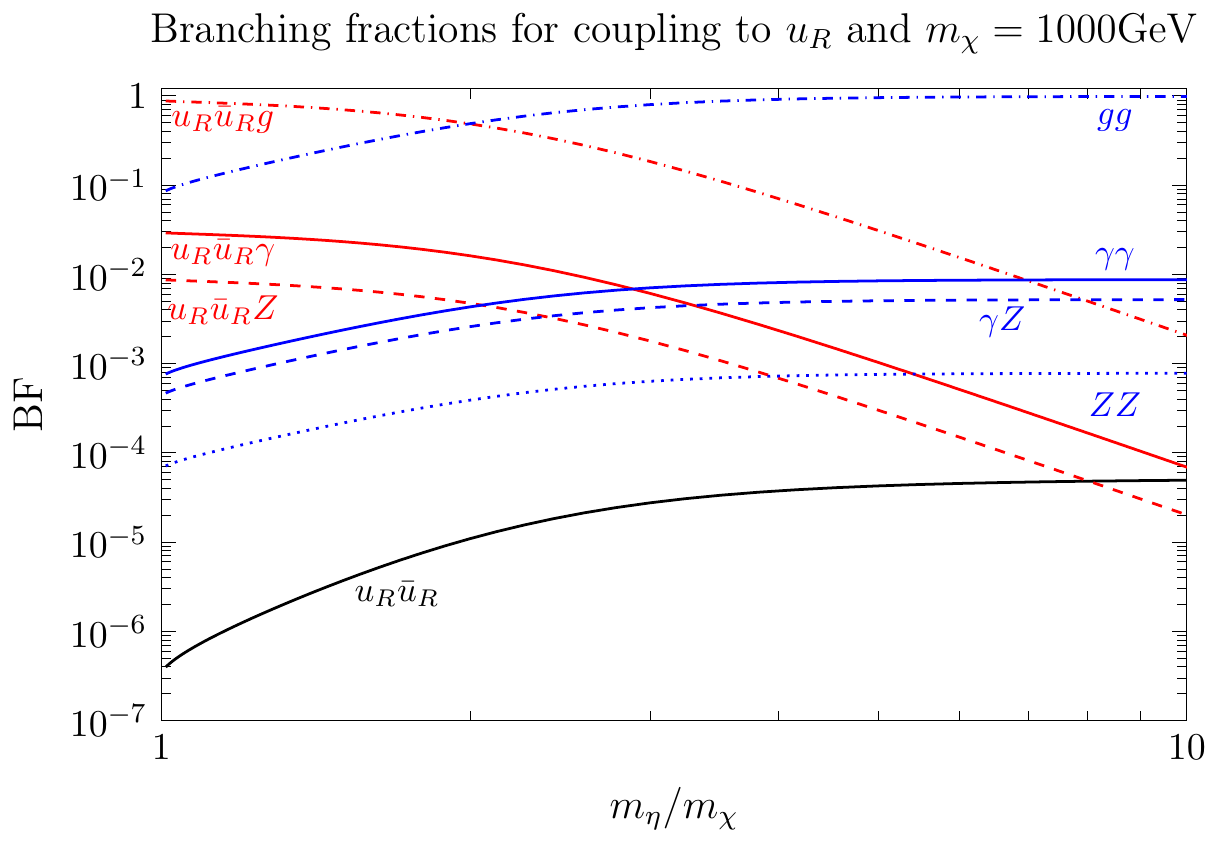} \\
\caption{\small Branching ratios of the various tree-level and one-loop annihilation channels for dark matter coupling to the right-handed electron (left plots) or up-quark (right plots). The top plots show the branching ratios as a function of the dark matter mass $m_\chi$ for fixed $m_\eta/m_\chi=2$, while the bottom plots,  as a function of  $m_\eta/m_\chi$ for fixed $m_\chi=1000\GeV$.}
\label{fig:BRs}
\end{center}
\end{figure}

The relative importance of the different final states mostly depends on the masses of the dark matter particle and the scalar mediator. This is illustrated in Fig.~\ref{fig:BRs}, which shows the dependence of the branching ratios with the dark matter mass $m_{\chi}$ (upper plots) and with the mass degeneracy parameter $m_\eta/m_\chi$ (lower plots), for dark matter coupling to $e_R$ (left plots)  and to $u_R$ (right plots), assuming that all quartic couplings vanish. As apparent from the upper plots, sufficiently away from the production threshold the branching ratios for the two-to-three processes and the one loop processes have constant branching ratios; only the two-to-two tree level annihilation into $f\bar f$ shows a strong dependence on the mass. Moreover, it follows from the lower plots that when the mass difference between the scalar mediator and the dark matter is very large, the two-to-three processes become negligible and the two-to-two dominate the annihilation. Among these, the process with the largest branching fraction is the annihilation into $\gamma\gamma$, for coupling to leptons, and into $gg$, for coupling to quarks. In contrast, as the scalar mediator becomes more and more degenerate in mass with the dark matter particle, the branching ratios for the two-to-three processes become larger and larger, and even become the dominant annihilation channels in the very degenerate limit. The largest cross section in this regime corresponds to the annihilation into $f\bar f \gamma$, for coupling to leptons, and into $f\bar f g$, for coupling to quarks. The ratios of strong and electroweak internal bremsstrahlung cross sections relative to $f\bar f\gamma$, as well as $gg$ relative to $\gamma\gamma$, are shown in Table \ref{tab:ratios}; these values are valid for $m_f\to 0$ and $m_\chi \gg m_Z/2$.

\begin{table}[th]
\tbl{Ratios of annihilation cross sections for a scalar mediator that couples to right-handed up-type quarks, down-type quarks and leptons, respectively, in the limit $m_f\to 0$. The ratios for $\chi\chi\to f\bar fZ$ are valid for $m_\chi\gg m_Z/2$. In the numerical values the strong coupling is evaluated at a scale $\mu=300\,$GeV. 
}
{\begin{tabular}{cccc}
\toprule
 &  $u_R$ & $d_R$ & $\ell_R$\\
\colrule
$\frac{\sigma v_{f\bar fg}}{\sigma v_{f\bar f\gamma }}$  & $\frac{3\alpha_s}{\alpha_{em}}\simeq 38.4$ & $\frac{12\alpha_s}{\alpha_{em}}\simeq 154$  &  - \\
\colrule
$\frac{\sigma v_{f\bar fZ}}{\sigma v_{f\bar f\gamma }}$  & $\tan^2(\theta_w)\simeq 0.30$ & $\tan^2(\theta_w)\simeq 0.30$  & $\tan^2(\theta_w)\simeq 0.30$  \\
\colrule
$\frac{\sigma v_{gg}}{\sigma v_{\gamma\gamma}}$ & $\frac{9\alpha_s^2}{8\alpha_{em}^2}\simeq 184$ &  $\frac{18\alpha_s^2}{\alpha_{em}^2}\simeq 2944$ & - \\
\botrule
\end{tabular}
\label{tab:ratios} }
\end{table}

\subsection{Gamma-rays}

Gamma-rays from dark matter annihilations propagate through our Galaxy in straight lines and practically without losing energy. This allows to devise several search strategies which exploit the spatial distribution expected from the signal events, and which is not possible in antimatter searches. In particular, the most prominent targets for indirect searches for dark matter annihilations with gamma-rays are the Milky Way center and dwarf spheroidal galaxies. Furthermore, the fairly good energy resolution of gamma-ray telescopes allows to search for the characteristic gamma-ray spectrum produced in a given annihilation channel, an approach which is at present not possible with neutrino searches. For these reasons, gamma-rays from self-annihilations are a promising and unique channel for dark matter detection. 

\begin{figure}[t]
  \begin{center}
    \includegraphics[width=0.75\textwidth]{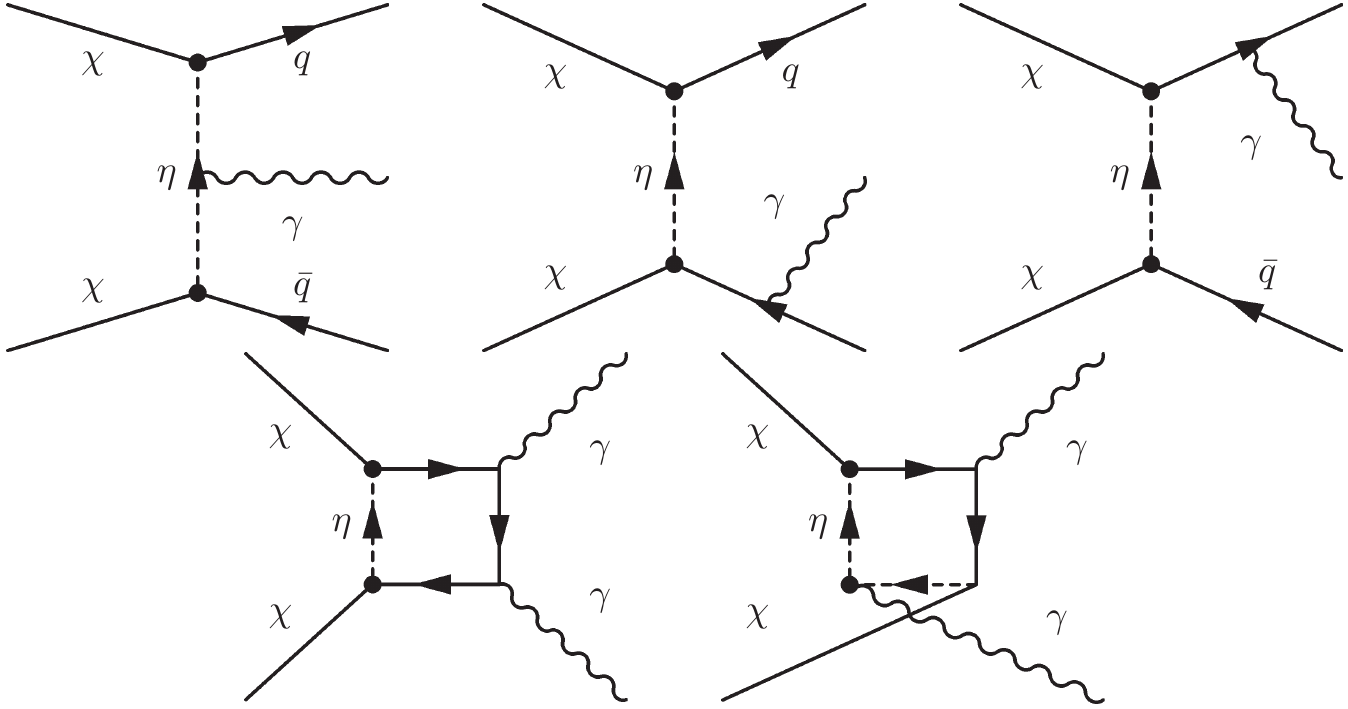}
  \end{center}
  \caption{\label{fig:ID} Feynman diagrams for the annihilation channels which lead to pronounced spectral features close to the end-point $E_\gamma=m_\chi$. The first three contribute to the process of internal bremsstrahlung, while the last two, to the generation of gamma-ray lines.}
\end{figure}

As discussed in the previous section, in the limit of zero relative velocity, the annihilation into a fermion-antifermion pair is helicity- and velocity- suppressed, hence the dominant annihilation channels are the two-to-three process $\chi\chi\rightarrow f\bar f V$ and the one loop process $\chi\chi\rightarrow V V'$, with $V,V'$ gauge bosons (see Fig.~\ref{fig:ID}). Of particular importance for indirect searches are the final states with $V=\gamma$, since they produce a sharp feature in the gamma ray spectrum.~\cite{Srednicki:1985sf,Rudaz:1986db,Bergstrom:1988fp,Bergstrom:1989jr,Flores:1989ru}. More specifically, the two-body annihilation $\chi\chi\rightarrow \gamma\gamma$ produces two monoenergetic photons at $E_\gamma=m_\chi$, hence the energy spectrum is markedly different to the smooth astrophysical background. A related process is the annihilation $\chi\chi\rightarrow \gamma Z$, which produces one monoenergetic photon at $E_\gamma=m_\chi(1-M_Z^2/4m_\chi^2)$. The gamma-ray line generated in the $\gamma Z$ final state is fainter than the line from $\gamma\gamma$, due to the smaller branching ratio of this process and the smaller multiplicity of photons in the final state, and can be neglected in most instances  within the simplified models considered in this review. Finally, when the mass difference between the dark matter particle and the scalar mediator is small, the two-to-three annihilation $\chi\chi\rightarrow f\bar f \gamma$ produces a sharp feature in the spectrum,\cite{Bergstrom:1989jr,Flores:1989ru} due to the enhancement of the scalar $t$-channel propagator  in the non-relativistic limit of the incoming dark matter particles when the momentum of the final fermion is close to zero, and which corresponds to an energy of the photon close to the kinematical end-point of the spectrum of annihilations. Under the above-mentioned conditions, this process generates a signal that resembles a distorted gamma-ray line and cannot be distinguished from the spectrum of monochromatic photons with the energy resolution of current gamma-ray telescopes. Many works have recently investigated the signatures from internal bremsstrahlung, see {\it e.g.}, Refs. \citen{Bergstrom:1989jr,Bergstrom:2004cy,Bergstrom:2005ss,Bringmann:2007nk,Bringmann:2008kj,Bergstrom:2010gh,Bringmann:2011ye,Bringmann:2012vr,Bergstrom:2012vd,Garny:2013ama,Okada:2014zja}. 
In either case, the observation of a sharp feature in gamma-rays would constitute a strong hint for dark matter annihilations as no  astrophysical process can mimic such a signal (a possible exception has been proposed in Ref.~\citen{Aharonian:2012cs}).  

\begin{figure}[t]
  \begin{center}
    \includegraphics[width=0.49\textwidth]{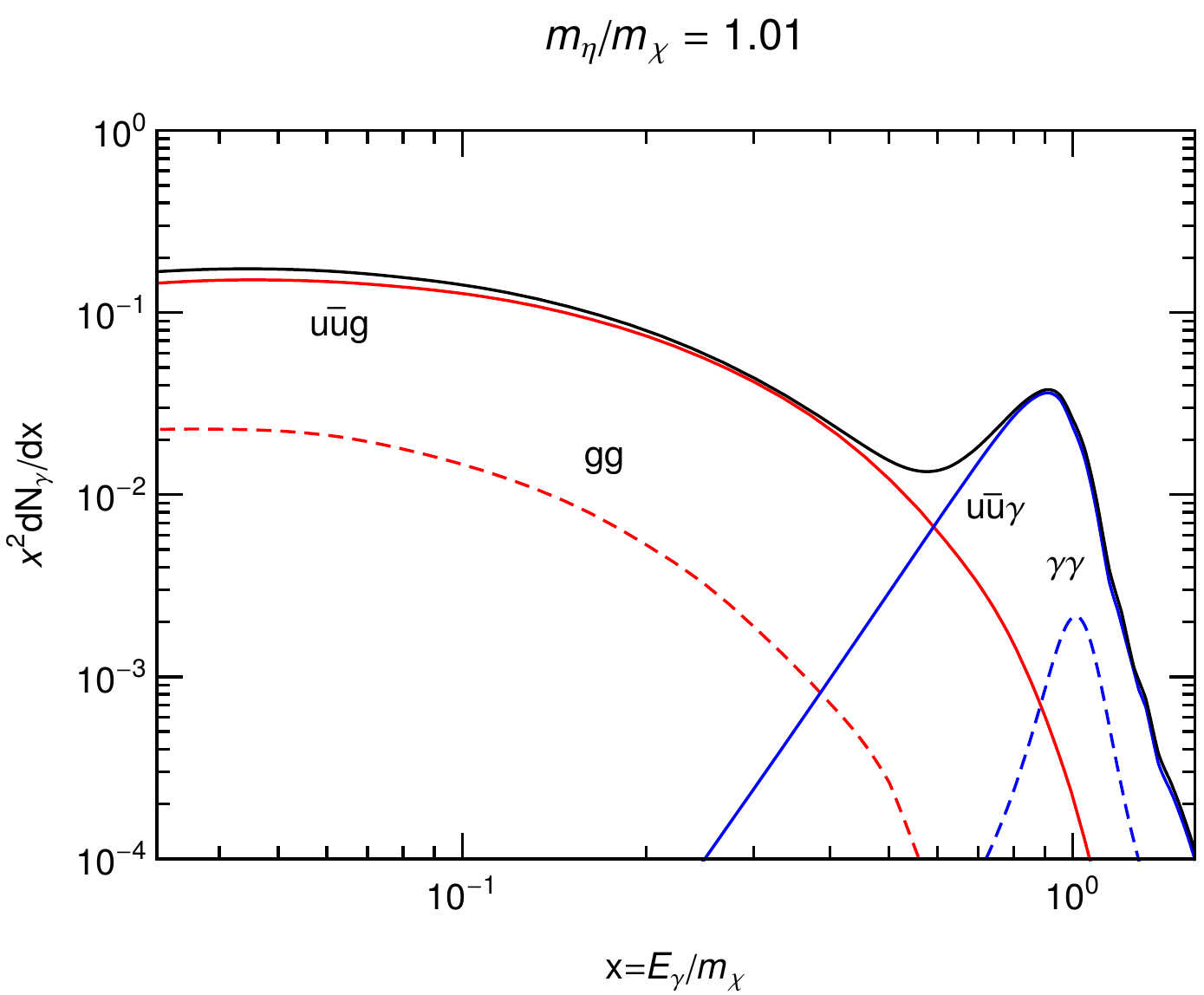}
    \includegraphics[width=0.49\textwidth]{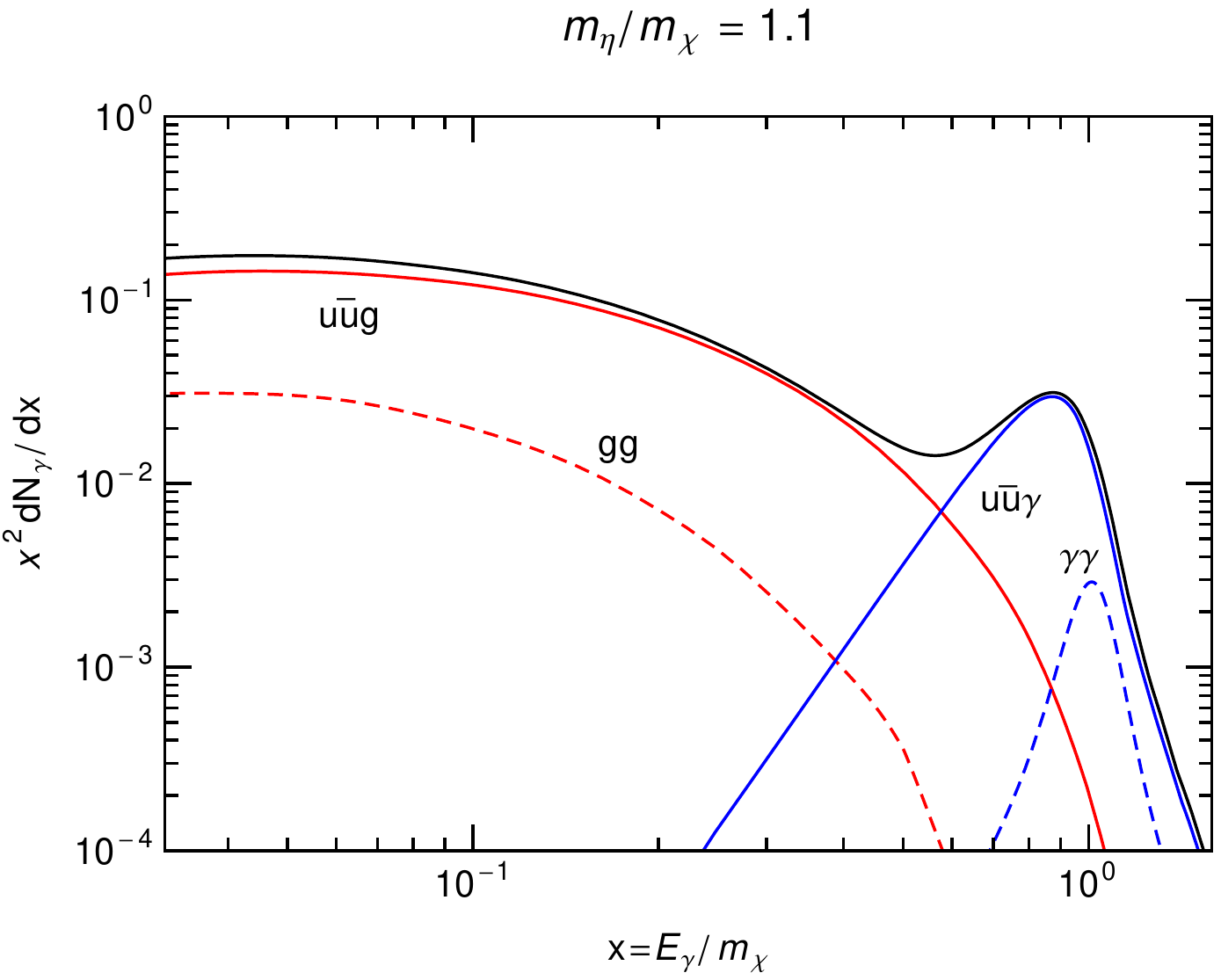} \\
    \includegraphics[width=0.49\textwidth]{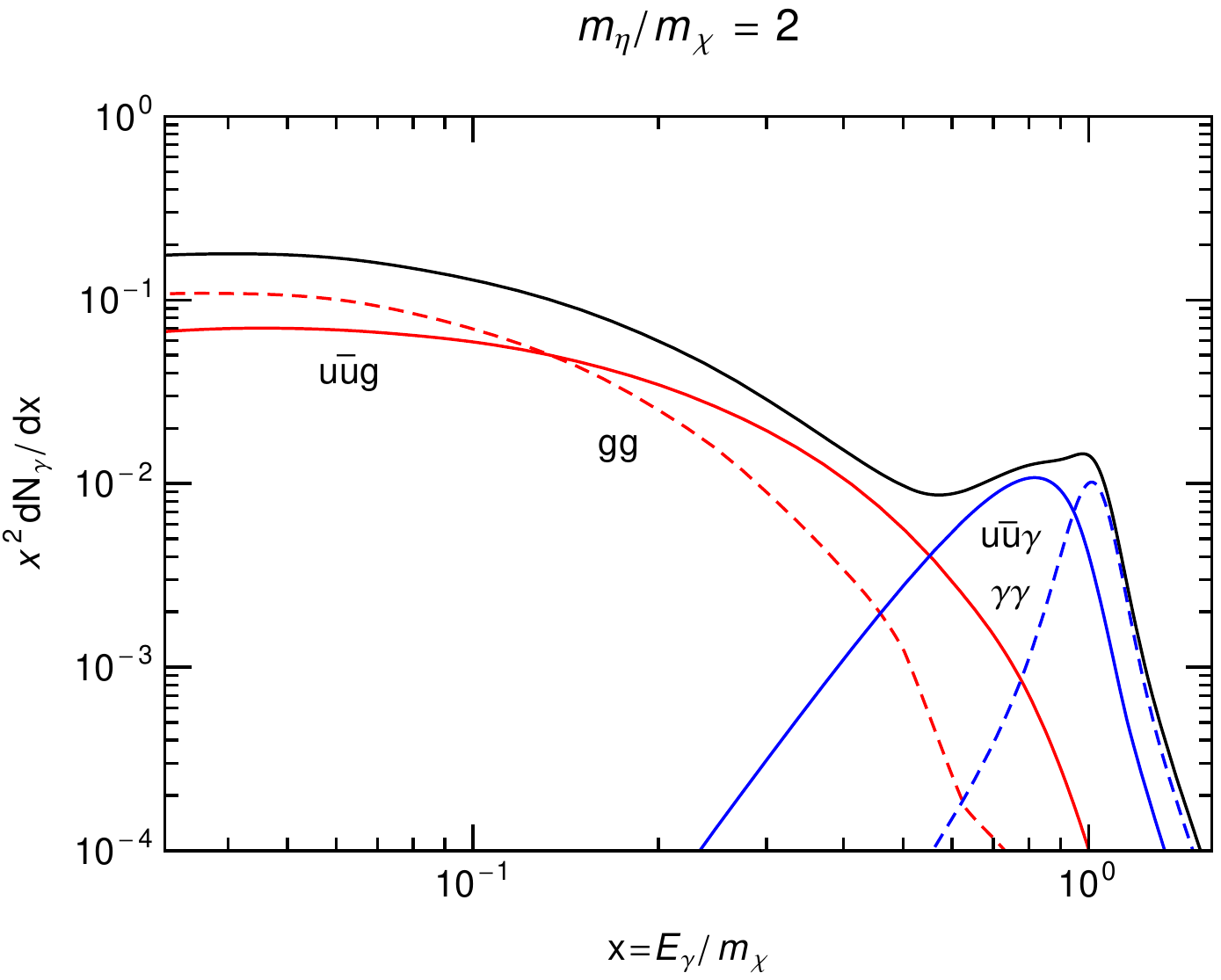}
    \includegraphics[width=0.49\textwidth]{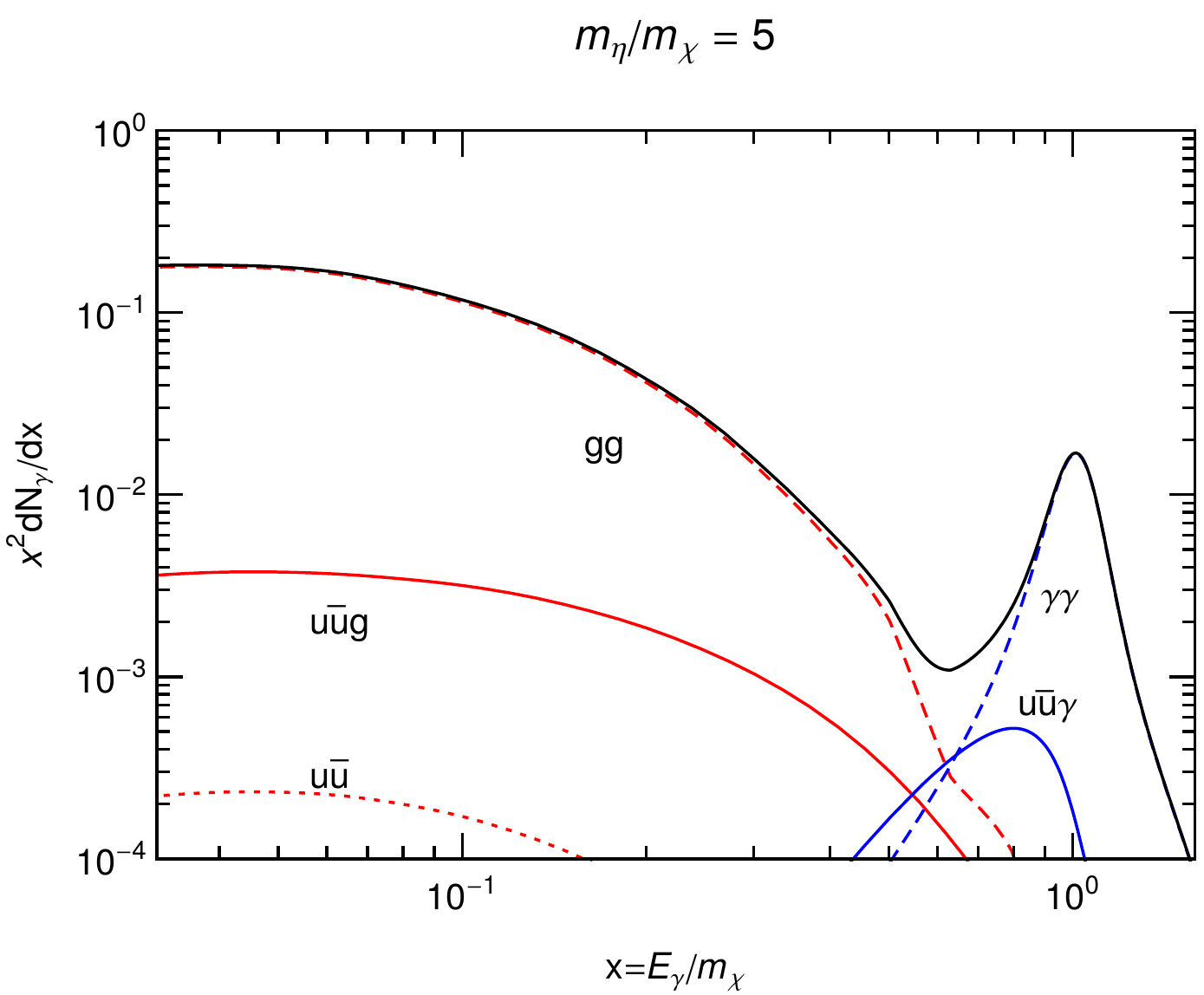}
  \end{center}
  \caption{\label{fig:spectrum} Gamma ray spectrum from dark matter annihilation for a colored mediator coupling to right-handed up-quarks, for $m_\eta/m_\chi=1.01$ (left upper plot), $m_\eta/m_\chi=1.1$ (right upper plot), $m_\eta/m_\chi=2$ (left lower plot), $m_\eta/m_\chi=5$ (right lower plot).}
\end{figure}

The total gamma ray spectrum consists of three contributions: the monochromatic gamma-ray lines from two-body annihilation with one or two photons in the final state, the sharp spectral feature from electromagnetic internal bremsstrahlung, and the soft spectrum of secondary gamma rays produced in other annihilation channels. Some exemplary spectra are shown in Fig.\,\ref{fig:spectrum}, for the cases $m_\eta/m_\chi=1.01$ (upper left plot),  1.1 (upper right plot), 2 (lower left plot) and 5 (lower right plot), assuming coupling to the right-handed up-quark and an energy resolution of 10\%, which is typical for current gamma-ray telescopes. In the plot, we have neglected the contribution from annihilations into $\gamma Z$ which, as mentioned above, gives a subdominant contribution to the total gamma-ray flux.  It is noticeable from the figure the change in the relative weight of the internal bremsstrahlung processes and the one-loop processes in the photon spectrum, and which follows from the dependence of the branching ratios on $m_\eta/m_\chi$, {\it cf.} Fig.~\ref{fig:BRs}. Namely, in the degenerate limit the internal bremsstrahlung dominates the total spectrum of annihilations at the highest energies, while in the hierarchical case, it is the one loop annihilation into $\gamma \gamma$. Besides, the relative importance of the annihilations into $u\bar u g$ and $u\bar u \gamma$ on one hand, and $g g$ and $\gamma \gamma$ on the other hand, is fixed (see Fig.~\ref{fig:BRs} and Table \ref{tab:ratios}). For coupling to down-type quarks, the feature from internal bremsstrahlung is suppressed compared to the continuum by a factor of 4, and the monochromatic contribution by a factor of 16, due to the smaller electric charge of the down-quark. For coupling to leptons, on the other hand, the contribution from continuum gamma rays is much smaller and the sharp spectral features are even more salient. 

The gamma-ray flux produced in dark matter annihilations is dominated close to the kinematical endpoint by the channels $\chi\chi\rightarrow f \bar f \gamma$ and $\chi\chi\rightarrow \gamma\gamma$ (as mentioned above, we neglect in this discussion the contribution from $\gamma Z$, which is suppressed with respect to $\gamma\gamma$). More specifically, the source term Eq.~\ref{eqn:source} reads in this case:
\begin{align}
Q(E_\gamma, r)= \frac{1}{2} \frac{\rho^2_{\chi}(r)}{m^2_\chi} \left( \frac{d(\sigma v)_{q\bar q\gamma}}{dE_\gamma} + 2(\sigma v)_{\gamma\gamma}\delta(E_\gamma-m_\chi) \right)  \,.
\end{align}
Hence, the differential gamma-ray flux as seen from the Earth under the angle $\xi$ with respect to the Galactic Center is given by,
\begin{align}
  \frac{d \Phi}{dE_\gamma d\Omega } = \frac{1}{4\pi} \, \int_0^\infty ds\, Q(E_\gamma,r) \,,
\end{align}
where $r=\sqrt{(r_0-s \cos\xi)^2+(s\sin\xi)^2}$ and $r_0=8.5$\,kpc.  

\begin{figure}[t]
\begin{tabular}{cc}
\hspace*{-14mm}
\includegraphics[width=0.58\textwidth]{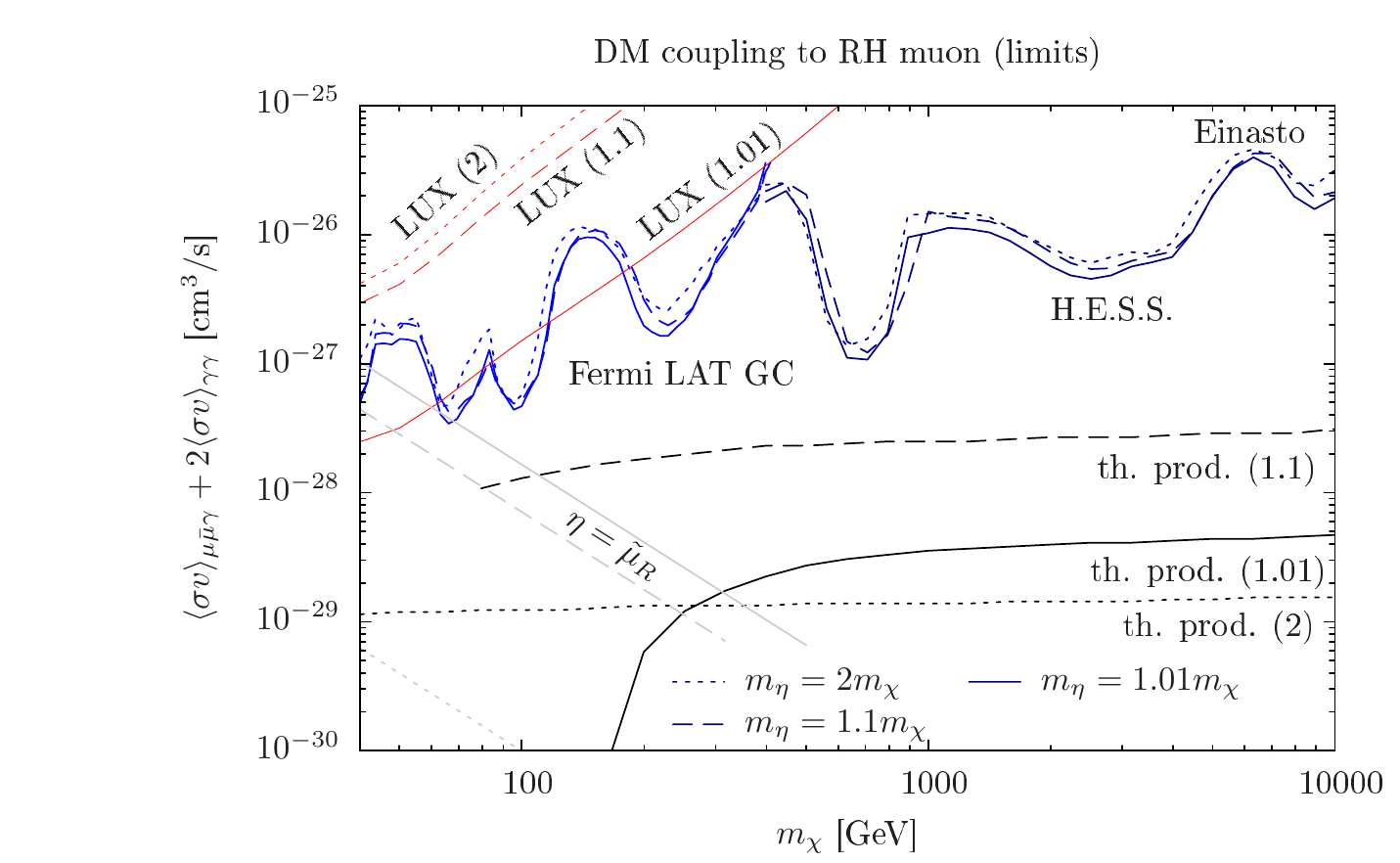} &
\hspace*{-14mm}
\includegraphics[width=0.58\textwidth]{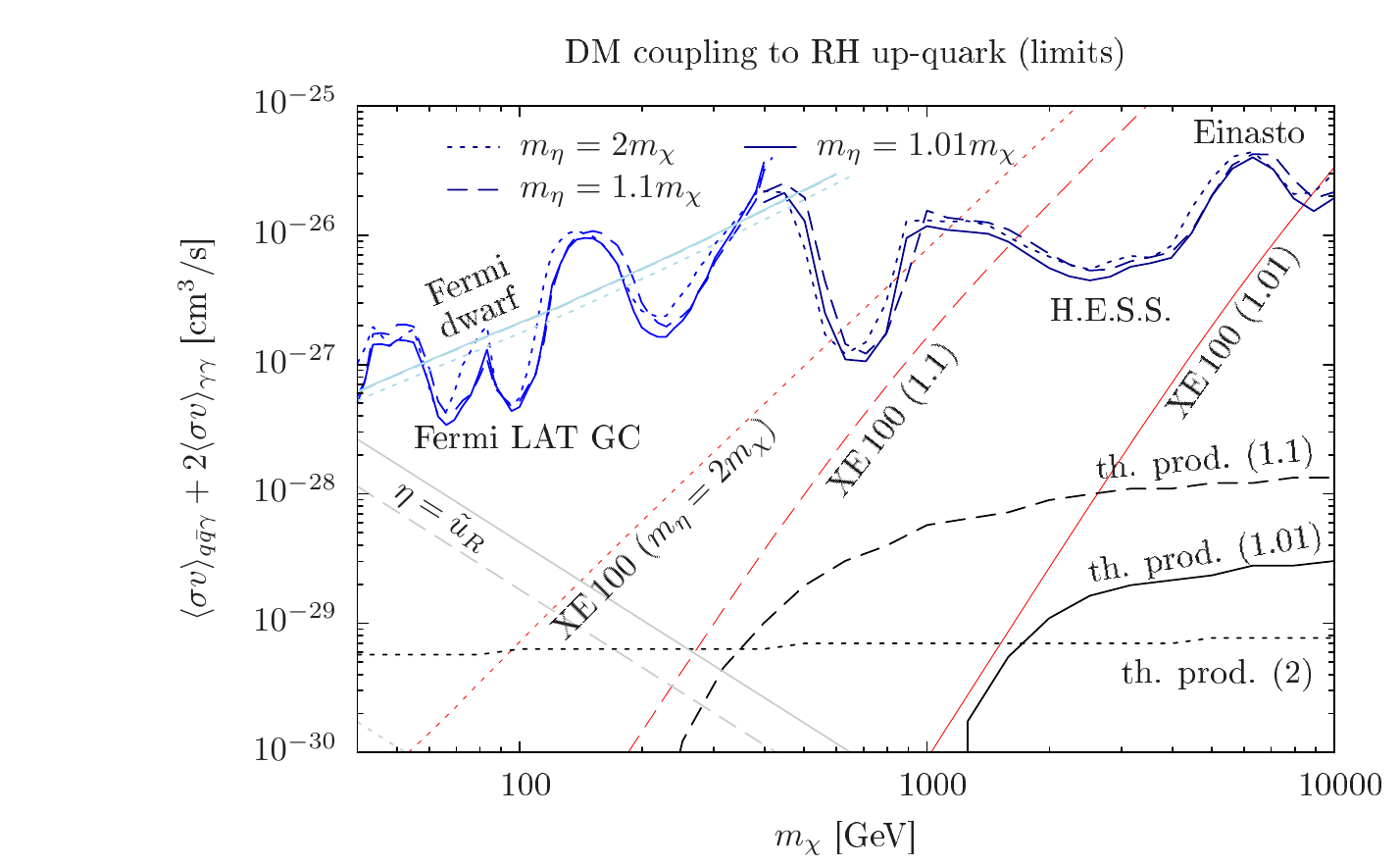} \\
\hspace*{-14mm}
\includegraphics[width=0.58\textwidth]{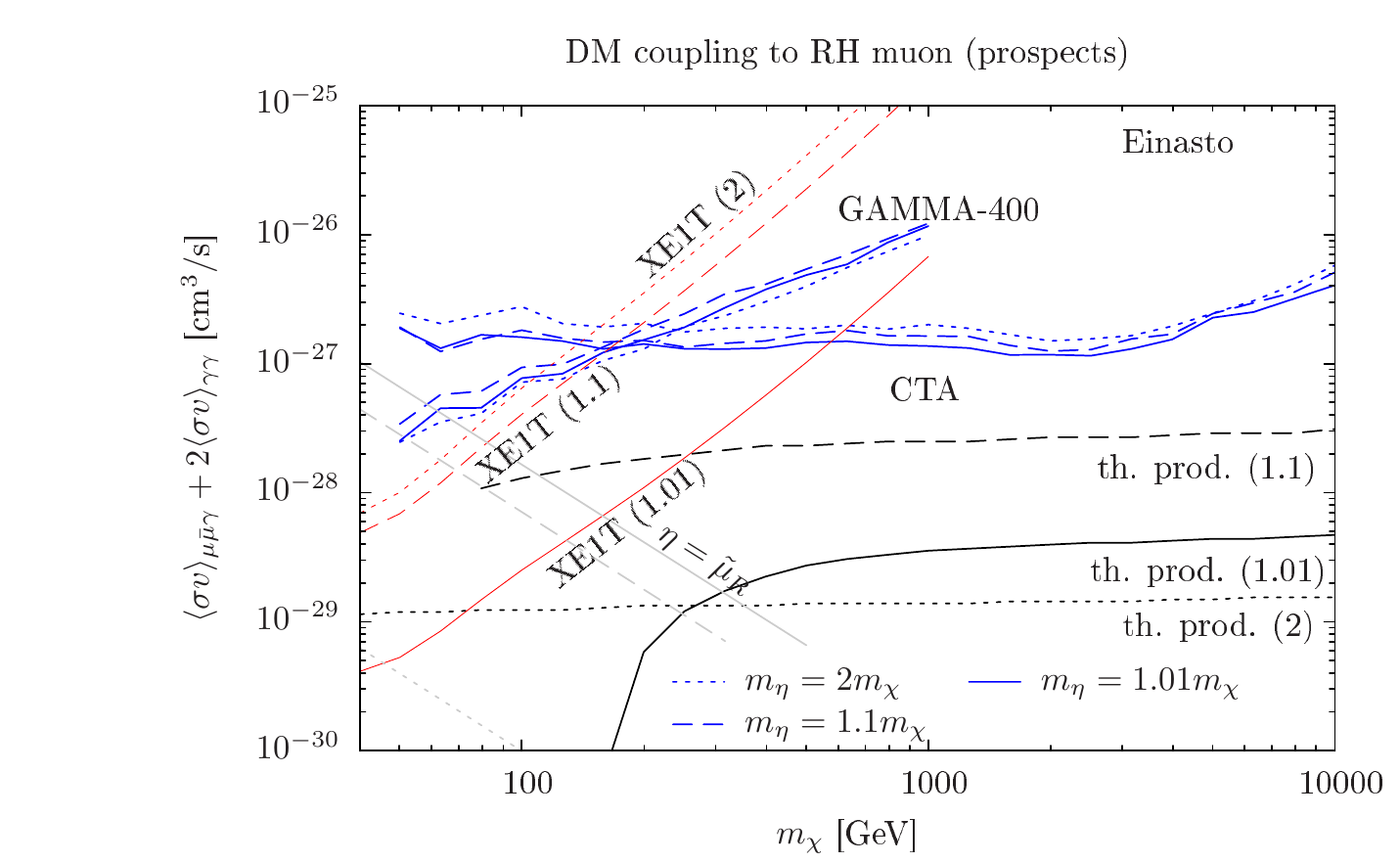} &
\hspace*{-14mm}
\includegraphics[width=0.58\textwidth]{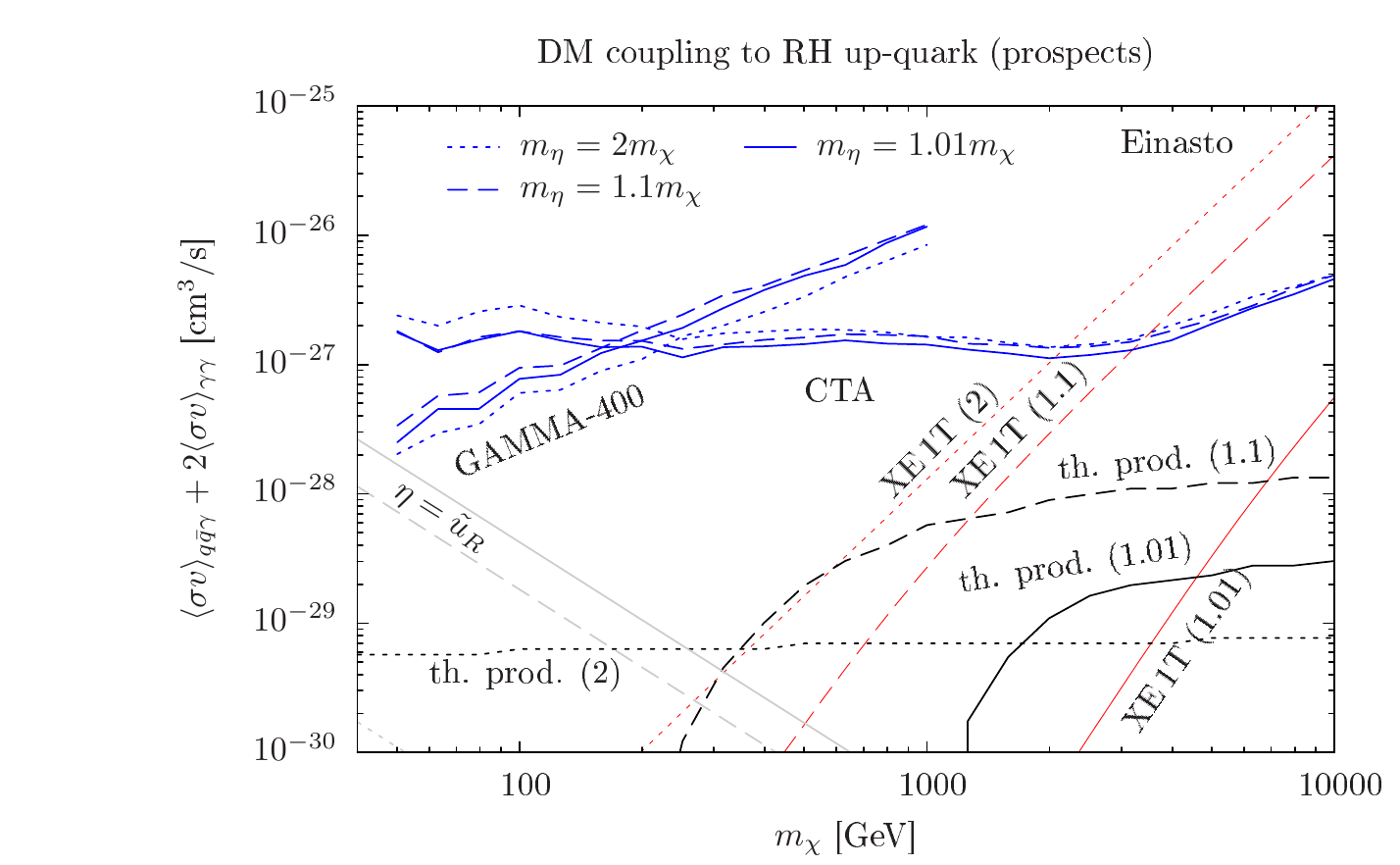} \\
\end{tabular}
\caption{\small Current constraints from Fermi LAT and H.E.S.S. (upper row) and future prospects for CTA and GAMMA-400 (lower row) on the spectral gamma ray feature resulting from the superposition of internal bremsstrahlung and
loop-induced annihilation into monochromatic photons, for $m_\eta/m_\chi=1.01$ (solid), $1.1$ (dashed) and $2$ (dotted). The left column corresponds to a charged mediator which couples to leptons (here $\mu_R$), and the right column to a colored mediator which couples to quarks (here $u_R$). For comparison we also show the cross section expected for a thermal relic for each value of $m_\eta/m_\chi$ (black), the cross-section for right-handed squark or slepton mediators (gray), as well as current constraints from XENON100/LUX and prospects for XENON1T (red, see Sec.\,\ref{sec:DD}).}
\label{fig:gamma}
\end{figure}

Searches for line-like features  have been conducted with data collected by the Fermi-LAT~\cite{Bringmann:2012vr,Fermi-LAT:2013uma} in the energy range 5-300 GeV and by the H.E.S.S. collaboration~\cite{Abramowski:2013ax} in the energy range 500 GeV-25 TeV. 
The $95\%$ C.L. upper limits on the annihilation cross section $\sigma v_{f\bar f\gamma}+2\sigma v_{\gamma\gamma}$ obtained from the combined contribution of internal bremsstrahlung and lines were calculated in Ref.~\citen{Garny:2013ama} and are shown as a blue line in  Fig.\,\ref{fig:gamma}, upper plots, for dark matter particles coupling to the right-handed muons (left plot) or to the right-handed up quarks (right plot), and for a mass ratio with the scalar mediator $m_\eta/m_\chi=1.01$, 1.1 or 2. The limits range from about $5\times 10^{-28}$cm$^3/$s to $10^{-26}$cm$^3/$s for dark matter masses from $40$\,GeV up to $10$\,TeV, and depend relatively weakly on the mass ratio $m_\eta/m_\chi$. The plot also shows the corresponding value of the cross section into sharp gamma-ray spectral features expected for a thermally produced dark matter particle, which lags well below the current limits, although for some choices of parameters it is only necessary a boost in the flux from annihilations by a factor of 5-10 to produce an observable signal. The next-generation of gamma ray telescopes GAMMA-400~\cite{Galper:2012fp} and especially CTA~\cite{Consortium:2010bc}, will improve the limits by a factor of a few, depending on the energy, as shown in  Fig.\,\ref{fig:gamma}, lower plots. 

We also show in the figures for comparison the present limit and future reach in the annihilation cross section into sharp spectral features that can be derived from the limits on the model parameters from current and projected direct detection experiments, which depend strongly on the mass difference between the mediator and the dark matter particle. The derivation of these limits will be discussed in more detail in  Sec.\,\ref{sec:DD}. It is notable from the plot the strong impact of direct detection experiments on the searches for sharp spectral features, especially in the  mass degenerate scenario when the dark matter particle couples to the right-handed up quark.

Finally, complementary limits on the model can be derived from the search for the continuum emission of gamma-rays in the annihilation. Among the existing targets, dwarf spheroidal galaxies are particularly suitable  for the search, due to their very large mass-to-light ratio and their moderate astrophysical activity, which translates into a negligible gamma-ray emission from astrophysical sources. Current data show no significant excess of gamma-rays in the direction of dwarf galaxies, thus allowing to set fairly stringent upper limits on the flux~\cite{Abdo:2010ex, Ackermann:2011wa,Ackermann:2013yva, Aleksic:2013xea}. 
The flux limits can in turn be translated into limits on the annihilation cross section in the channel $\chi\chi\rightarrow f\bar f V$ and which can produce a continuum flux of gamma-rays. These limits can be particularly relevant when the dark matter particle couples to a quark, due to the large branching fraction of the annihilation $\chi\chi\rightarrow q \bar q g$. In this case, the continuum flux of gamma-rays is generated in the fragmentation and decay of the gluon and the quarks. The impact of the dwarf limits on the model is illustrated in Fig.\,\ref{fig:gamma}, upper left plot, for the particular case $q=u_R$ (the limits for $q=d_R$ are a factor of four stronger, due to the different quark charges). Here, the dwarf limits on the model parameters have been translated into limits on the cross section for the annihilations producing sharp spectral features. As apparent from the plot, for a colored mediator dwarf galaxy observations provide limits on the model parameters which can be competitive to the limits obtained from the search for sharp spectral features. For an uncolored mediator, on the other hand, the continuum part of the annihilation spectrum is much fainter due to the absence of gluon bremsstrahlung, therefore, in this case the strongest limits stem from the search for sharp spectral features.

\subsection{Antimatter}

Dark matter annihilations in general produce antimatter particles that could be detected at the Earth as an excess over the expected astrophysical backgrounds, constituted by a secondary component from spallation of high energy cosmic rays on the interstellar medium, and possibly a primary component from sources, such as the positrons produced by the interactions of high-energy photons in the strong magnetic fields of pulsars~\cite{Sturrock:1971zc,Atoian:1995ux,Chi:1995id} or the positrons and antiprotons produced by hadronic interactions inside the same sources that accelerate galactic cosmic rays~\cite{Blasi:2009hv,Blasi:2009bd}. 

Antimatter particles have electric charge and propagate through the Milky Way in a complicated way before reaching the Earth. The propagation of charged particles in the Galaxy is commonly modeled by a stationary two-zone diffusion model with cylindrical boundary conditions. In this scheme, the number density of antiparticles at the position $\vec r$ per unit kinetic energy, $f(T,\vec{r},t)$, satisfies the following transport equation~\cite{ACR}:
\begin{align}
0=\frac{\partial f}{\partial t}=
\nabla \cdot [K(T,\vec{r})\nabla f] +
\frac{\partial}{\partial T} [b(T,\vec{r}) f]
-\nabla \cdot [\vec{V_c}(\vec{r})  f]
-2 h \delta(z) \Gamma_{\rm ann} f+Q(T,\vec{r})\;.
\label{eq:transport}
\end{align}
The boundary conditions require the solution $f(T,\vec{r},t)$ to vanish at the boundary of the diffusion zone, which is approximated by a cylinder with half-height $L = 1-15~\rm{kpc}$ and radius $ R = 20 ~\rm{kpc}$.

The first term on the right-hand side of the transport equation is the diffusion term, which accounts for the propagation through the tangled Galactic magnetic field. The diffusion coefficient $K(T,\vec{r})$ is assumed to be constant throughout the diffusion zone and is parametrized by $K(T)=K_0 \;\beta\; {\cal R}^\delta$,  where $\beta=v/c$ and ${\cal R}$ is the rigidity of the particle, which is defined as the momentum in GeV per unit charge, ${\cal R}\equiv p({\rm GeV})/Z$. The second term accounts for energy losses due to inverse Compton scattering (ICS) on the interstellar radiation field (ISRF) as well as synchrotron radiation and ionization. The third term is the convection term, which accounts for the drift of charged particles away from the disk induced by the Milky Way's Galactic wind. It has axial direction and is also assumed to be constant inside the diffusion region: $\vec{V}_c(\vec{r})=V_c\; {\rm sign}(z)\; \vec{k}$. The fourth term accounts for antimatter annihilations with rate $\Gamma_{\rm ann}$ when it interacts with ordinary matter in the Galactic disk, which is assumed to be an infinitely thin disk with half-width $h=100$ pc. Lastly, $Q(T,\vec{r})$ is the source term of antimatter particles, given in Eq.~\ref{eqn:source}. In this equation, reacceleration effects and non-annihilating interactions of antimatter in the Galactic disk are neglected.

\begin{table}[t]
  \tbl{Astrophysical parameters compatible with the B/C ratio that yield the
  minimal (MIN), median (MED) and maximal (MAX) antiproton fluxes from dark
  matter annihilations; taken from Ref.~\citen{Maurin:2001sj}.}
  {
    \begin{tabular}{ccccc}
      \toprule
      Model & $\delta$ & $K_0\,({\rm kpc}^2/{\rm Myr})$ & $L\,({\rm kpc})$
      & $V_c\,({\rm km}/{\rm s})$ \\
      \colrule
      MIN & 0.85 & 0.0016 & 1 & 13.5 \\
      MED & 0.70 & 0.0112 & 4 & 12 \\
      MAX & 0.46 & 0.0765 & 15 & 5 \\
      \botrule
    \end{tabular}
  }
  \label{tab:param-propagation}
\end{table}

The transport equation, using the parametrizations of the different terms given above, has a number of free parameters which are inferred from measurements of flux ratios of primary and secondary cosmic-ray species, mainly the Boron-to-Carbon (B/C) ratio. Unfortunately, the model parameters cannot be determined uniquely from current observations, due to degeneracies among the parameters entering in the calculation of the cosmic-ray fluxes, for instance  between the diffusion coefficient and the height of the magnetic diffusion zone. The ranges of the astrophysical parameters that are consistent with the B/C ratio and that produce the minimal (MIN), median (MED) and maximal (MAX) antiproton fluxes from dark matter annihilations were calculated in Ref.~\citen{Maurin:2001sj} and are listed in Table \ref{tab:param-propagation}. 

Finally, the flux of  primary antiparticles at the Solar System from dark matter annihilations is given by:
\begin{align}
\Phi^{\rm{prim}}(T) = \frac{v}{4 \pi} f(T),
\label{flux}
\end{align}
where $v$ is the velocity of the antimatter particle.

At energies smaller than $\sim 10$ GeV the antimatter fluxes at the top of the Earth's atmosphere can differ considerably from the interstellar fluxes due to solar modulation effects. One frequently used parametrization of the effect of solar modulation, which can be derived from the full diffusion and convection equations describing the solar wind, is the force-field approximation~\cite{solarmodulation1,solarmodulation2}. The fluxes at the top of the atmosphere in this approximation are related to the interstellar fluxes  by the following relation~\cite{perko}:
\begin{align}
  \Phi^{\rm TOA}(T_{\rm TOA})=
  \left(
    \frac{2 m T_{\rm TOA}+T_{\rm TOA}^2}{2 m T_{\rm IS}+T_{\rm IS}^2}
  \right)
  \Phi^{\rm IS}(T_{\rm IS}),
  \label{eq:solar-modulation}
\end{align}
where $m$ is the mass of the cosmic-ray antimatter particle and $T_{\rm IS}=T_{\rm TOA}+\phi_F$, with $T_{\rm IS}$ and $T_{\rm TOA}$ being the kinetic energies of the antimatter particles at the heliospheric boundary and at the top of the Earth's atmosphere, respectively, and $\phi_F$ being the Fisk potential, which varies between 500 MV and 1.3 GV over the eleven-year solar cycle.

For antiprotons, due to their comparatively large mass, energy losses are negligible and therefore the general transport equation, Eq.~(\ref{eq:transport}), can be simplified. The transport equation for the antiproton density, $f_{\bar p}(T_{\bar p},\vec{r},t)$ then reads
\begin{align}
  0=\frac{\partial f_{\bar p}}{\partial t}=
  \vec{\nabla} \cdot (K(T_{\bar p},\vec{r})\vec{\nabla} f_{\bar p})
  -\vec{\nabla} \cdot (\vec{V_c}(\vec{r})  f_{\bar p})
  -2 h \delta(z) \Gamma_{\rm ann} f_{\bar p}+Q(T_{\bar p},\vec{r})\;,
  \label{transport-antip}
\end{align}
where the annihilation rate, $\Gamma_{\rm ann}$, is given by
\begin{align}
  \Gamma_{\rm ann}=(n_{\rm H}+4^{2/3} n_{\rm He})
  \sigma^{\rm ann}_{\bar p p} v_{\bar p}\;.
\end{align}
In this expression it has been assumed that the annihilation cross-section between an antiproton and a helium nucleus is related to the annihilation cross-section between an antiproton and a proton by the simple geometrical factor $4^{2/3}$.  Furthermore, $n_{\rm H}\sim 1\;{\rm cm}^{-3}$ is the number density of Hydrogen nuclei in the Milky Way disk, $n_{\rm He}\sim 0.07 ~n_{\rm H}$ the number density of Helium nuclei and $\sigma^{\rm ann}_{\bar p p}$ is the proton--antiproton annihilation cross-section, which is given {\it e.g.} in Refs.~\citen{Tan:1983de, Protheroe:1981gj}.

The spectrum of cosmic antiprotons has been measured by WiZard/CAPRICE~\cite{Boezio:1997ec}, between 0.62 and 3.19 GeV, AMS \cite{Aguilar:2002ad} between 0.2 and 4 GeV, and more recently by the PAMELA satellite~\cite{Adriani:2010rc} between 60 MeV and 180 GeV and BESS-Polar II~\cite{Abe:2011nx} between 0.17 and 3.5 GeV.  In the near future, data from the AMS-02 experiment are expected to yield further information on the cosmic antiproton flux~\cite{Malinin:2007zz,Pato:2010ih}. The measured flux as well as the antiproton-to-proton ratio agree well with the expectations from secondary production of antiprotons from spallation of cosmic ray nuclei, mainly protons and Helium, on the interstellar medium \cite{Bringmann:2006im,Evoli:2011id}. This allows to set stringent upper limits on a possible primary contribution generated from dark matter annihilations. 

\begin{figure}[t]
\begin{center}
\includegraphics[width=0.49\textwidth]{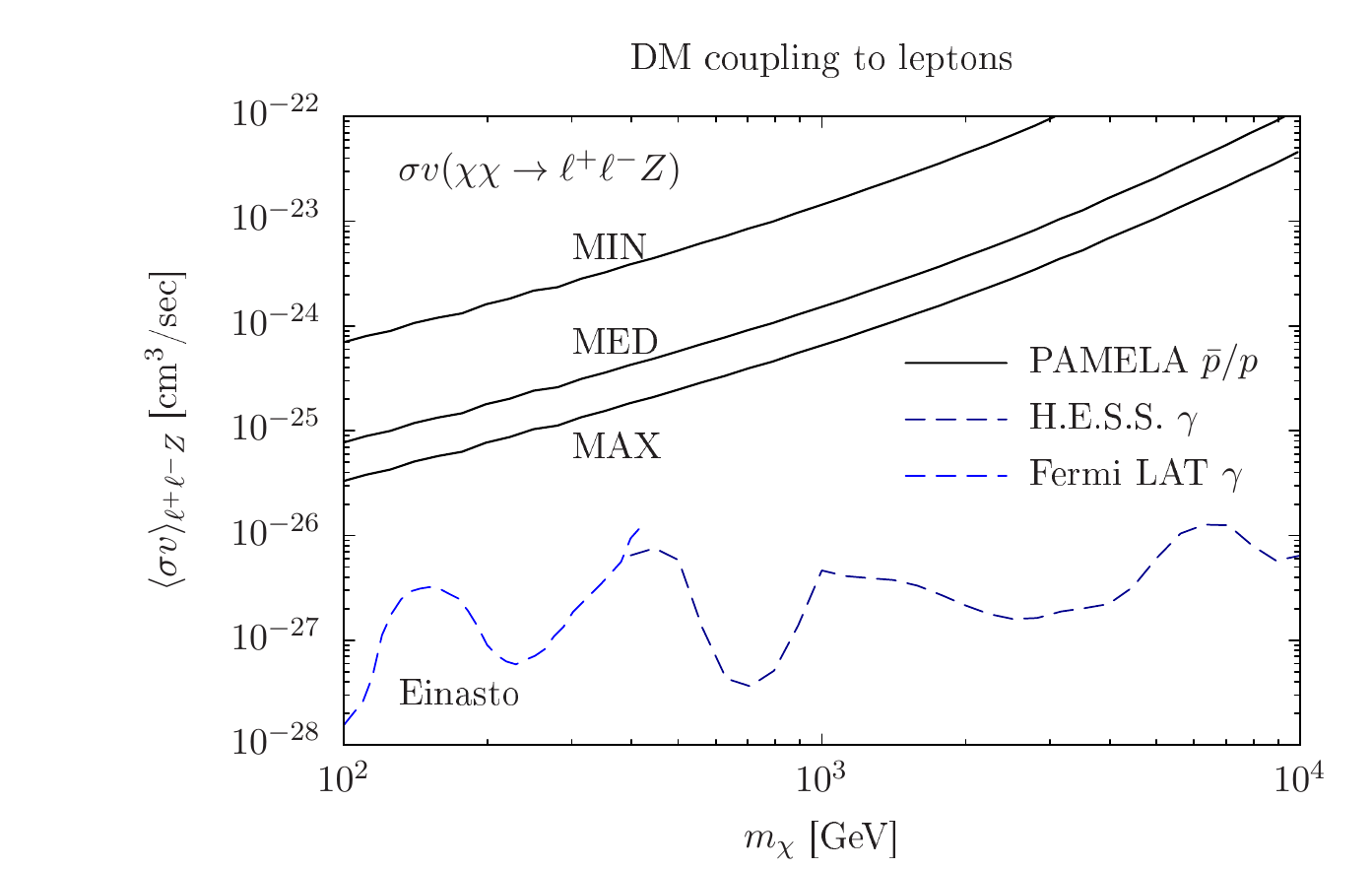} \\
\includegraphics[width=0.49\textwidth]{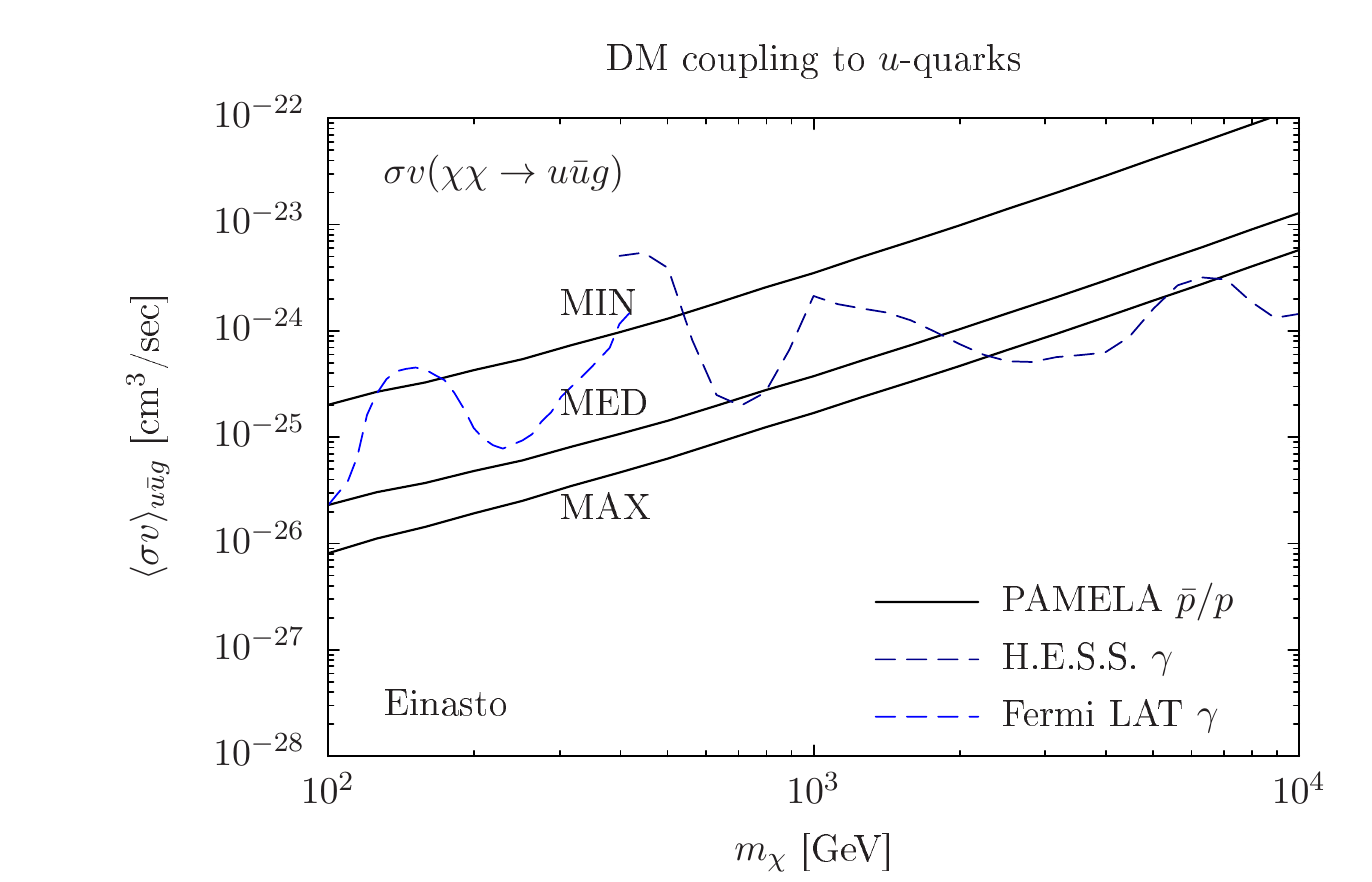}
\hspace{-0.5cm}
\includegraphics[width=0.49\textwidth]{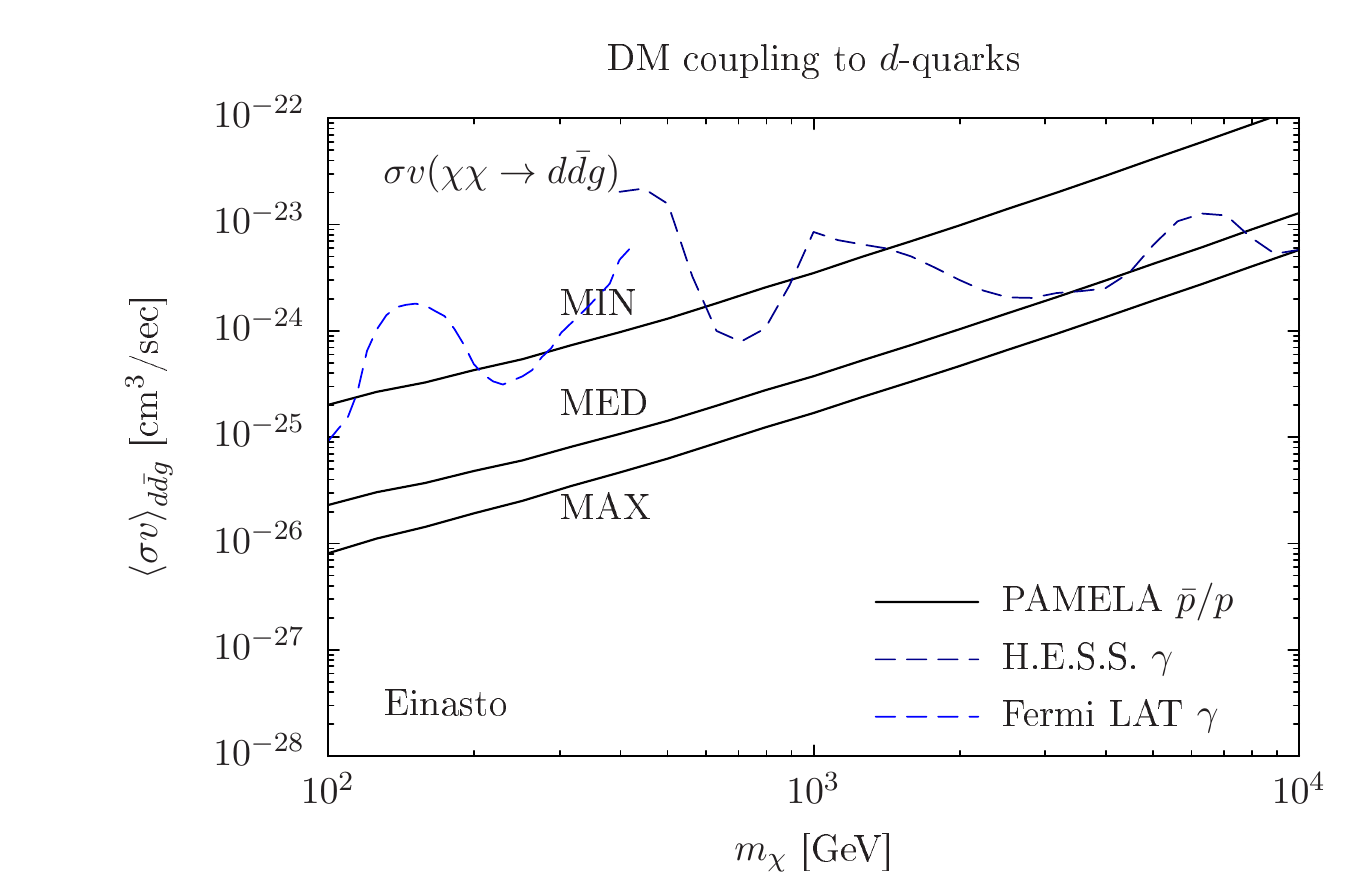}
\end{center}
 \caption{\label{fig:exclusionPlot1} 95\% C.L. upper bounds on the annihilation cross section for the higher order processes $\chi\chi\to \ell^+ \ell^- Z$ (upper plot) $\chi\chi\to u \bar u g$ (lower left plot) and $\chi\chi\to d \bar d g$ (lower right plot) derived in Ref.~\citen{Garny:2011ii} from the PAMELA data on the antiproton-to-proton fraction \cite{Adriani:2010rc}, adopting the MIN, MED and MAX propagation models defined in Table \ref{tab:param-propagation}. In the figures, it was assumed the Einasto profile and a dark matter particle quasi-degenerate in mass with the scalar mediator. For comparison, we also show as dashed lines the limits on the cross sections for the corresponding channels from gamma-ray data from the Fermi-LAT~\cite{Bringmann:2012vr} and H.E.S.S.~\cite{Abramowski:2013ax} data.}
\end{figure}

Annihilations of Majorana dark matter particles coupling to leptons do not produce antiprotons in the lowest order process $\chi\chi\rightarrow \ell^+ \ell^-$,with $\ell=e,\mu,\tau$. Nevertheless, the higher order process $\chi\chi\rightarrow \ell^+ \ell^- Z$ produces antiprotons in the fragmentation of the $Z$ boson, and therefore the antiproton measurements can constrain this model, despite its leptophilic nature. This process has been studied in various works, {\it e.g.} Refs. \citen{Ciafaloni:2011sa,Bell:2011if,Garny:2011cj,Barger:2011jg,Ciafaloni:2011gv,Garny:2011ii,Ciafaloni:2012gs,Bell:2012dk,Ibarra:2013eba,Shudo:2013lca,Cavasonza:2014xra}. Fig. \,\ref{fig:exclusionPlot1} shows the 95\% C.L. upper limits on the annihilation cross section for the higher order process $\chi\chi\rightarrow \ell^+ \ell^- Z$ calculated in Ref. \citen{Garny:2011ii} from the non-observation of a significant excess of antiprotons in the  PAMELA $\bar p/p$ data \cite{Adriani:2010rc} over the background flux calculated in Ref.~\citen{Donato:2001ms}. In the plot it was assumed the Einasto profile and $m_{\eta}/m_\chi=1.1$, however the  limits are rather insensitive to the choice of halo profile and to the concrete value of the mass of the scalar mediator as long as $m_{\eta}/m_\chi\sim\mathcal{O}(1)$. Besides, in order to bracket the astrophysical uncertainties, the plot shows limits for the three propagation models discussed above. For the MED propagation model the limits are rather weak and lie in the range $\sigma v\lesssim 10^{-25}\cm^3/\s$ for $m_\chi=100\GeV$ and $10^{-24}\cm^3/\s$ for $m_\chi=1\TeV$, which approximately translate into an upper limit on the Yukawa coupling $y\lesssim 5.4$ for $m_\chi=100\GeV$ and $y\lesssim 26$ for $m_\chi=1\TeV$; these limits are very sensitive to the choice of the propagation parameters, as apparent from the figure. The plot also shows for comparison the limits on that same channel inferred from the non-observation of a significant excess of gamma-rays in the Fermi-LAT~\cite{Bringmann:2012vr} and H.E.S.S.~\cite{Abramowski:2013ax} data, which are independent of the propagation model. It follows from the plot that for leptophilic models the limits on the cross section derived from gamma-ray observations are more than one order of magnitude stronger than those derived from antiprotons. 

On the other hand, the lower left and right plots show, respectively, the corresponding limits for the annihilation channel $\chi\chi\rightarrow u \bar u g$ and $\chi\chi\rightarrow d \bar d g$. The limits on the cross section are only a factor of $\sim 3$ better than for $\chi\chi\rightarrow \ell^+ \ell^- Z$, however the limits on the Yukawa coupling $y$ are significantly better for dark matter particles coupling to quarks, due to the larger cross section of the three body final state involving one gluon. More specifically,  for the MED propagation model, it follows that $y\lesssim 0.8$ for $m_\chi=100\GeV$ and $y\lesssim 5.4$ for $m_\chi=1\TeV$. The plot also shows the limits inferred from the Fermi-LAT and H.E.S.S. data, and which are a factor of four weaker for couplings to down-quarks compared to the couplings to up-quarks due to the different electric charges. As apparent from the plot, for couplings to quarks the limits from antiprotons are competitive, and in some instances better, than the limits from the search for gamma-ray spectral features.

The production of antiprotons and antineutrons in dark matter annihilations leads to the possibility of also producing antinuclei, such as antideuterons~\cite{Donato:1999gy} or antihelium~\cite{Carlson:2014ssa,Cirelli:2014qia}. These channels are particularly interesting due to the very small background flux of antideuterons~\cite{Donato:2008yx,Ibarra:2013qt} and antihelium~\cite{Carlson:2014ssa,Cirelli:2014qia} expected at the energies relevant for experiments. In fact, the expected flux from spallations of cosmic rays on the interstellar medium lies more than three orders of magnitude below the current limit by BESS~\cite{Fuke:2005it} and more than one order of magnitude below the expected sensitivity of AMS-02 and GAPS~\cite{Doetinchem}, therefore, the observation of a few antideuterons in experiments will constitute a strong hint for their exotic origin. The antideuteron flux at the Earth from dark matter annihilations is, on the other hand, strongly constrained by the non-observation of an excess in antiprotons~\cite{Ibarra:2012cc}. Given the stringent limits existing from PAMELA, and despite the various sources of uncertainty in the modeling of the antideuteron production, the observation of an antideuteron flux at AMS-02 or GAPS from dark matter annihilations will be challenging~\cite{Ibarra:2012cc}. 

Dark matter annihilations also produce positrons. In this case galactic convection and annihilations in the disk can be neglected in the transport equation, which is then simplified to:
\begin{align}
\nabla \cdot [K(E_e,\vec{r})\nabla f_{e^+}] +
\frac{\partial}{\partial E_e} [b(E_e,\vec{r}) f_{e^+}]+Q(E_e,\vec{r})=0\;,
\label{transport-positron}
\end{align}
where we have approximated the kinetic energy of the positron $T_e$ by its total energy $E_e$. Positrons with energies in the GeV-TeV range lose energy in their propagation through the galactic diffusive halo mainly via the inverse Compton scattering (ICS) on the interstellar radiation field (ISRF) and via the synchrotron losses on the Galactic $B$-field: $b=b_{\rm ICS}+b_{\rm syn}$. The rate of energy loss due to ICS is given by
\begin{align}
  b_{\rm ICS}(E_e,\vec{r})=\int_0^\infty d\epsilon
  \int_{\epsilon}^{E_\gamma^{\rm max}} dE_\gamma\,
  (E_\gamma-\epsilon)\, \frac{d\sigma^{\rm  IC}(E_e,\epsilon)}{dE_\gamma}f_{\rm ISRF}(\epsilon,\vec{r})\;.
  \label{eq:b-ICS}
\end{align}
Here $d\sigma^{\rm IC}/d E_\gamma$ denotes the differential cross section of inverse Compton scattering of a positron with energy $E_e$, where an ISRF photon with energy $\epsilon$ is up-scattered to energies between $E_\gamma$ and $E_\gamma+dE_\gamma$. In the limit $\epsilon, m_e\ll E_e$, kinematics and the neglect of down-scattering require that $\epsilon \leq E_\gamma \leq  (1/E_e+1/4\gamma_e^2 \epsilon)^{-1} \equiv E_\gamma^{\rm max}$, with $\gamma_e\equiv E_e/m_e$. Besides, $f_{\rm ISRF}(\epsilon,\vec{r})$ is the number density of photons of the interstellar radiation field, which includes the cosmic microwave background, thermal dust radiation and starlight. An explicit model of the interstellar radiation field can be found, \emph{e.g.}, in Ref.~\citen{Porter:2005qx}.  For a positron energy of $E_e = 1\,{\rm GeV}$, $b_{\rm ICS}$ ranges between $4.1 \times 10^{-17}\,{\rm GeV}{\rm s}^{-1}$ and $1.9 \times 10^{-15}\,{\rm GeV}{\rm s}^{-1}$, depending on the position in the Galaxy. On the other hand, the rate of energy loss due to synchrotron emission is given by
\begin{align}
  b_{\rm syn}(E_e,\vec{r})=\frac{4}{3}\sigma_{\rm T}
  \gamma_e^2 \frac{B^2}{2}\;,
  \label{eq:b-syn}
\end{align}
where $\sigma_{\rm T}=0.67$ barn denotes the Compton scattering cross section in the Thomson limit and $B^2/2$ is the energy density of the Galactic magnetic field, which is conventionally taken as $B\simeq 6\,\mu{\rm G}\exp(-|z|/2\,{\rm kpc}-r/10\,{\rm kpc})$~\cite{Strong:1998fr}, although the size and spatial dependence are not precisely known. At the position of the Sun this magnetic field yields a synchrotron loss rate given by $b_{\rm syn}\simeq 4.0\times10^{-17}(E_e/\,{\rm GeV})^2\,{\rm GeV}\;{\rm s}^{-1}$. In many analyses, the rate of energy loss is approximated by a spatially constant function parametrized by $b(E)=\frac{E^2}{E_0\tau_E}$,  with $E_0=1\;{\rm GeV}$ and $\tau_E=10^{16}\;{\rm s}$.

\begin{figure}[t]
\begin{center}
\includegraphics[width=0.7\textwidth]{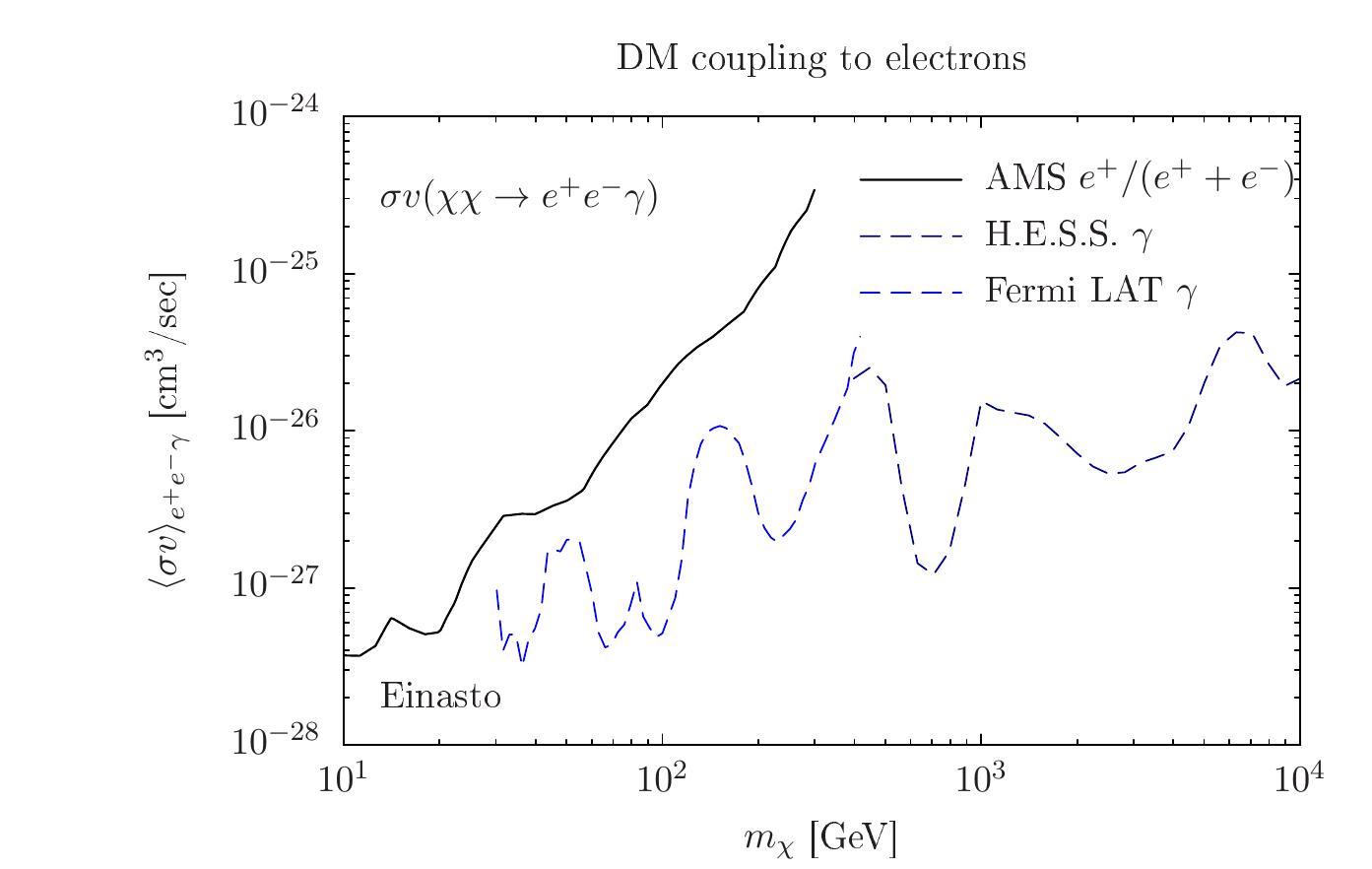}
\end{center}
 \caption{\label{fig:exclusionPlotAMS} 95\% C.L. upper bounds on the annihilation cross section for the higher order processes $\chi\chi\to e^+ e^- \gamma$ derived in Ref.~\citen{Bergstrom:2013jra} using the AMS-02 data on the positron fraction.\cite{Accardo:2014lma}. In the figures, it was assumed the Einasto profile and a dark matter particle quasi-degenerate in mass with the scalar mediator. For comparison, we also show as dashed lines the limits on the cross sections for the corresponding channels from gamma-ray data from the Fermi-LAT \cite{Bringmann:2012vr} and H.E.S.S. data \cite{Abramowski:2013ax}.}
\end{figure}

Most experiments measure the positron fraction, defined as the flux of positrons divided by the total flux of electrons plus positrons, which is less susceptible to systematics since most sources of systematic error, such as detector acceptance or trigger efficiency, cancel out when computing the ratio of particle fluxes. This is the case of CAPRICE~\cite{Boezio:2000zz}, HEAT~\cite{DuVernois:2001bb}, AMS-01~\cite{Aguilar:2007yf}, PAMELA~\cite{Adriani:2008zr,Adriani:2013uda} and AMS-02~\cite{Aguilar:2013qda,Accardo:2014lma}. Some experiments also measure the positron flux itself, such as HEAT~\cite{DuVernois:2001bb}, PAMELA~\cite{Adriani:2013uda} and AMS-02~\cite{Aguilar:2014mma}. The most striking feature in the data is the rise in the positron fraction at energies larger than $\sim 10\GeV$, first observed by PAMELA ~\cite{Adriani:2008zr} an recently confirmed by the AMS-02 collaboration~\cite{Aguilar:2013qda,Accardo:2014lma}. While dark matter annihilations or decays provide a possible explanation for this observation, the cosmic positron backgrounds are still poorly understood and it is not possible at the moment to make a definite statement about the origin of this excess.  Nevertheless, the exquisite quality of the positron data can be used to set stringent limits on the dark matter annihilation cross section~\cite{Bergstrom:2013jra,Ibarra:2013zia}. Fig.\,\ref{fig:exclusionPlotAMS} shows the limits on the annihilation cross section for the annihilation $\chi\chi\rightarrow e^+ e^-\gamma$ which follow from the non-observation of a sharp feature in the positron fraction, and which approximately read  $\sigma v\lesssim 3\times 10^{-28} (3\times 10^{-26})\,{\rm cm}^3\,{\rm s}^{-1}$ for $m_\chi=10\,(100)$ GeV~\cite{Bergstrom:2013jra}. The plot also shows for comparison the limits on the same annihilation channel from the non-observation of a sharp gamma ray feature in the Fermi-LAT \cite{Bringmann:2012vr} and H.E.S.S. data \cite{Abramowski:2013ax}. As apparent from the plot, the current limits from gamma-rays are stronger than those that follow from the positron fraction data. 

\subsection{Neutrinos}
\label{sec:neutrinos}

Dark matter particles traversing the Sun could scatter-off a nucleus in the solar interior and lose energy. After several scatterings, the dark matter particles eventually sink to the solar core where they accumulate~\cite{Press:1985ug}. The subsequent annihilation of the dark matter particles produces, depending on the final state, a high energy neutrino flux in the direction of the center of the Sun which could be detected in a neutrino telescope~\cite{Silk:1985ax,Gould:1987ir}. Cosmic ray interactions with the solar corona constitute an irreducible background in this search, however, the predicted flux at the Earth is far below the sensitivity of current neutrino telescopes~\cite{Ingelman:1996mj}, therefore, the observation of an excess of high energy neutrino events correlated to the direction of the Sun would constitute a strong hint for dark matter annihilations. 
 
The time evolution of the number of dark matter particles $N$ in the solar core is described by the following differential equation~\cite{Griest:1986yu}:
    \begin{align}
        \frac{dN}{dt} = \sub{\Gamma}{C} - \sub{C}{A} N^2 - \sub{C}{E} N \, , \label{eq:DEDMDensitySun} 
    \end{align}
where $\sub{\Gamma}{C}$ is the capture rate, $\sub{C}{A}$ the annihilation constant and $\sub{C}{E}$ the evaporation constant.  For dark matter masses above $\sim 10\GeV$ the evaporation of dark matter particles can be safely neglected \cite{Jungman:1995df,Busoni:2013kaa}. Then, after solving Eq.\eqref{eq:DEDMDensitySun}, one finds an annihilation rate as a function of time given by
  \begin{align}
      \sub{\Gamma}{A}(t) &=\frac{1}{2} \sub{C}{A} N(t)^2= \frac{1}{2} \sub{\Gamma}{C} \tanh^2 \left( t / \tau \right) \, , \label{eq:SolutionDMDensitySun}
  \end{align}
where 
  \begin{align}
       \tau &= \frac{1}{\sqrt{\sub{\Gamma}{C} \sub{C}{A}}} \, . \label{eq:EquilibrationTime} 
  \end{align}
The annihilation rate reaches a maximum when $t\gg \tau$. In this regime, captures and annihilations are in equilibrium and the annihilation rate reaches its maximum value, which is determined only by the capture rate: $\sub{\Gamma}{A}= \sub{\Gamma}{C}/2$. For thermally produced dark matter particles, this regime is never reached in the interior of the Earth\cite{Ibarra:2013eba}. Instead, the annihilation rate is very suppressed due to the small number of dark matter particles trapped inside the Earth and,  correspondingly, the expected high energy neutrino flux. Therefore, the detection of  a signal from dark matter annihilations in the direction of the center of the Earth becomes very challenging, if not impossible in practice.  On the other hand, for dark matter particles which have a sizable coupling to quarks, equilibration is generically reached inside the Sun and hence the high energy neutrino flux is not suppressed, thus making the center of the Sun a prime target to search for dark matter annihilations. 

In the class of scenarios under consideration the annihilation channels with largest branching fraction are  $\chi \chi \rightarrow f  \bar{f} V$  an $\chi\chi\rightarrow V V'$, with $V, V'$ gauge bosons, which produce in their decay and hadronization a neutrino flux. It is important to note that the annihilations take place in dense medium of the solar core, therefore the neutrino spectrum generated can significantly differ from the corresponding spectrum in vacuum, due to the interactions with the medium of the charged leptons and the hadrons before they decay. Concretely, the muons and light hadrons, such as pions and kaons, are stopped in the Sun before decaying and hence produce neutrinos with energies below 1 GeV. On the other hand, taus and heavy hadrons, such as charmed or beauty hadrons, decay in flight after losing a fraction of their kinetic energy, hence producing energetic neutrinos, while weak gauge bosons decay promptly producing neutrinos which typically dominate the energy spectrum at the highest energies. Finally, the neutrinos produced in the annihilation propagate from the solar interior to the surface, undergoing flavor oscillations and scatterings off solar matter, and from the solar surface to a neutrino telescope at the Earth, undergoing flavor oscillations in vacuum. 

\begin{figure}[t]
\begin{center}
\includegraphics[width=0.49\textwidth]{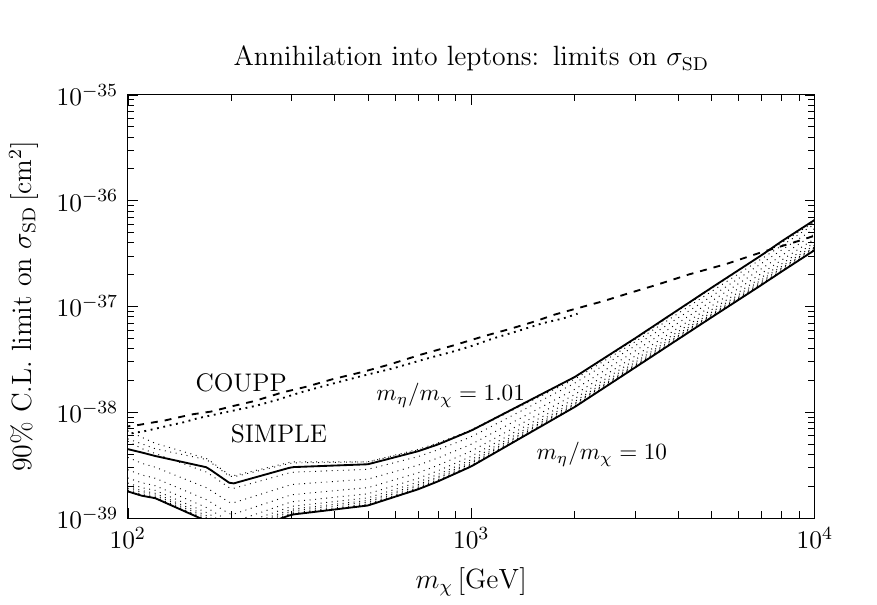}
\includegraphics[width=0.49\textwidth]{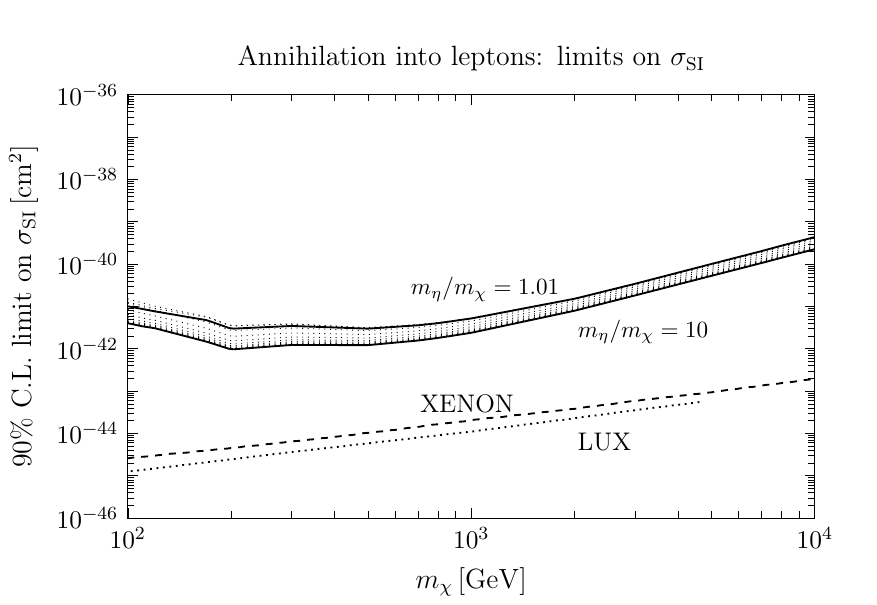} \\
\includegraphics[width=0.49\textwidth]{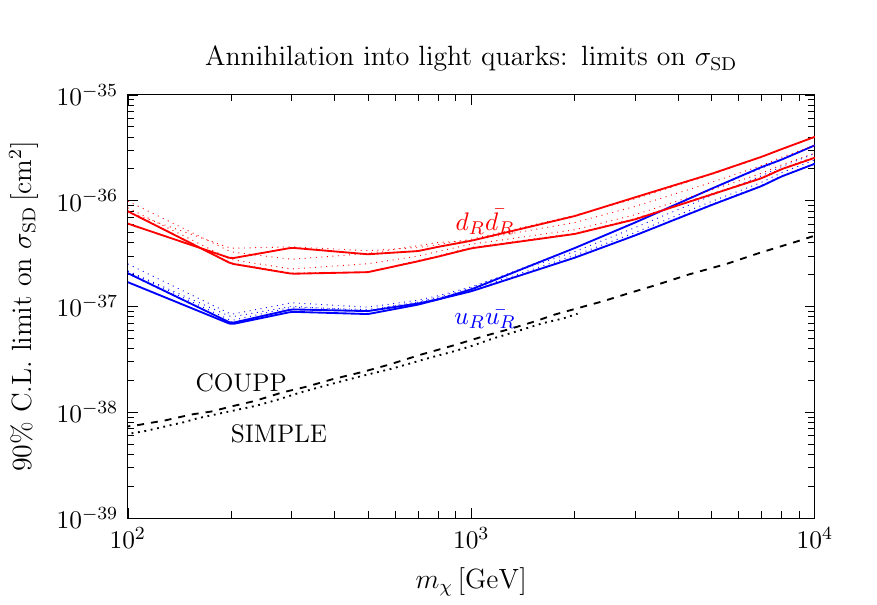}
\includegraphics[width=0.49\textwidth]{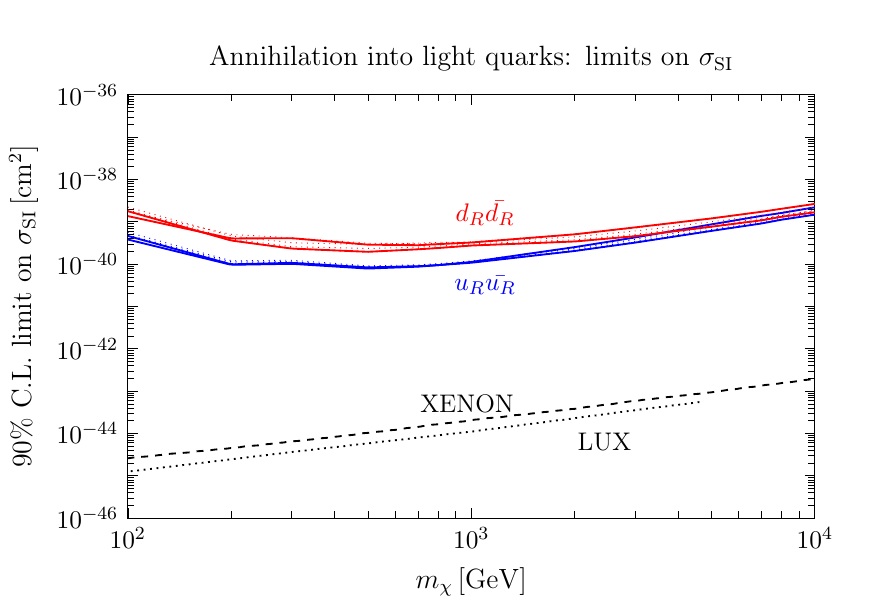} \\
\caption{\small 90\% C.L. limits on the spin-dependent (left plots) and spin-independent (right plots) interaction cross sections as a function of the dark matter mass  for couplings to the right-handed electron (top plots) or to the right-handed first generation quark (lower plots), for various values of the mass ratio $m_\eta/m_\chi$. The best limits on the scattering cross section from direct detection experiments are also shown for comparison. } 
\label{fig:limits-toy-model}
\end{center}
\end{figure}

The non-observation at IceCube of a significant neutrino excess in the direction of the Sun with respect to the expected atmospheric background then allows to set constraints on the capture cross section in scenarios where the dark matter particle couples to a light fermion~\cite{Ibarra:2014vya,Ibarra:2013eba,Bell:2012dk}. The resulting limits on the spin-dependent (left panel) and spin-independent (right panel) cross section are shown in Fig.~\ref{fig:limits-toy-model}, for dark matter annihilations mediated by couplings to right-handed electrons (upper plots) and right-handed up- or down-quarks (lower plots) as a function of the dark matter mass, for eight different values of the parameter $m_\eta/m_\chi$ ranging between $1.01$ and 10, where the dominant annihilation channel is $f_R \bar f_R Z $ and  $\gamma Z$ respectively. The figure also shows  the limits on the spin-dependent interaction cross sections from direct detection experiments COUPP~\cite{Behnke:2012ys} and SIMPLE~\cite{Felizardo:2011uw}, and on the spin-independent interaction cross sections from XENON100~\cite{Aprile:2012nq} and LUX~\cite{Akerib:2013tjd}.For couplings to leptons, the limits on the spin-dependent cross section are significantly stronger than the direct detection limits, while for quarks they are weaker, since in this case the dark matter particle annihilates mostly into final states involving gluons which have a much larger branching fraction than those involving weak gauge bosons. On the other hand, the limits on the spin-independent cross section from IceCube are much weaker, at least one order of magnitude than the direct detection limits for couplings to $e_R$ and at least two orders of magnitude for couplings to $u_R$.  For lighter dark matter particles, complementary constraints on the cross-section can be obtained from the non-observation at Super-Kamiokande of the neutrino flux produced by the stopped muons and charged pions and which peaks at energies below 1 GeV~\cite{Rott:2012qb,Bernal:2012qh}.

A search for a high energy neutrino flux from dark matter annihilations has also been conducted in other regions of the Universe with an overdensity of dark matter particles, such as the Milky Way center~\cite{Abbasi:2012ws, Aartsen:2013mla,Hernandez-Rey:2014ssa, Zornoza:2014cra, Wendell:2014dka}, dwarf galaxies~\cite{Aartsen:2013dxa}, the Andromeda galaxy~\cite{Aartsen:2013dxa}, the Coma Cluster~\cite{Aartsen:2013dxa} and the diffuse extragalactic background~\cite{Beacom:2006tt,Moline:2014xua}. For annihilations into $W^+ W^-$, the upper limit on the cross-section derived from neutrino data is ${\cal O}(10^{-22})\,{\rm cm}^3\,{\rm s}^{-1}$~\cite{Wendell:2014dka}, which is weaker than the corresponding limits derived from gamma-ray or antiproton data, unless the dark matter mass is in the multi-TeV range. Similar conclusions are expected for the annihilation channels into $\gamma Z$ and $f\bar f Z$, and which are the dominant ones in this scenario.

\section{Direct detection}\label{sec:DD}

\begin{figure}[thb]
  \begin{center}
    \includegraphics[width=0.9\textwidth]{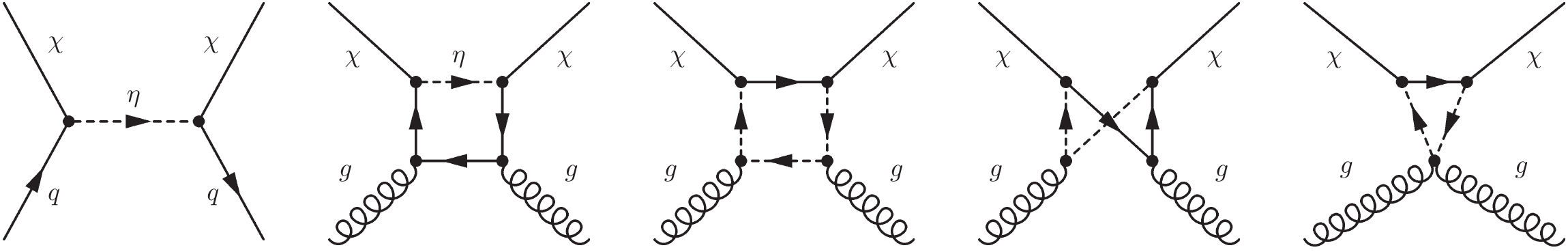}\\[2ex]
    \includegraphics[width=0.35\textwidth]{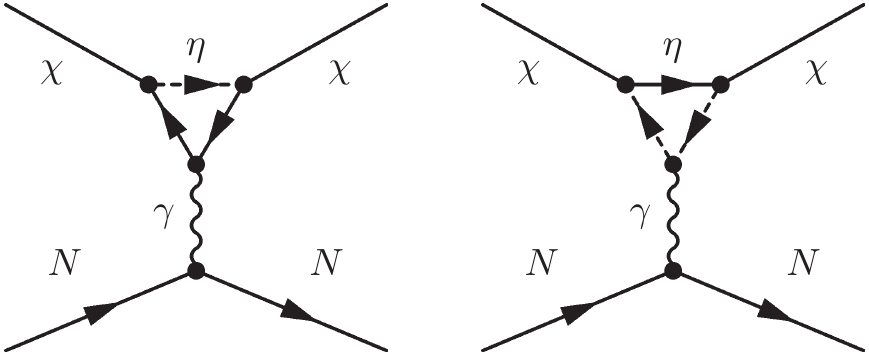}
  \end{center}
  \caption{\label{fig:feynDD} Feynman diagrams for the processes contributing to the scattering of the dark matter particle off nucleons for a colored mediator (upper row) and an uncolored mediator (lower row). Diagrams with flipped initial and final state $\chi$ are not shown for brevity.}
\end{figure}

The rapid experimental development in direct detection makes it possible to probe even scenarios where the expected scattering rate of dark matter on nuclei is relatively small. It turns out that this is indeed the case for the simplified models of Majorana dark matter considered here. Nevertheless, one should keep in mind that the constraints depend on the local density and velocity distribution of dark matter particles, and therefore are subject to large uncertainties which are complementary to those affecting indirect constraints. In the following we assume a standard dark matter halo model characterized by a truncated Maxwellian distribution with $v_0=220$\,km/s, $v_{esc}=544$\,km/s and local dark matter abundance $\rho_0=0.4$\,GeV/cm$^3$ (see {\it e.g.} Refs.~\citen{Catena:2009mf,Aprile:2012nq}).

\subsection{Colored mediator}

The formalism for computing direct detection event rates is well developed and we do not repeat it here (see {\it e.g.} Refs.~\citen{Jungman, Garny:2012eb}). Instead we briefly discuss the effective WIMP-nucleon interaction for Majorana dark matter which couples to the light Standard Model quarks via a scalar mediator with purely chiral coupling, and collect the quantities entering the cross section. The relevant Feynman diagrams are shown in Fig.\,\ref{fig:feynDD}. For spin-independent interactions, the leading dimension six operators $\bar\chi \chi\bar q q$ and $\chi\gamma^\mu\chi \bar q\gamma_\mu q$ both vanish, due to the chiral interaction and the Majorana condition, respectively. Consequently, the leading contribution arises from dimension eight operators \cite{Hisano:2010ct},
\begin{equation}
{\cal L}_{\rm eff, tree}^{\rm SI} = -\frac{y^2}{2(m_\eta^2-(m_\chi+m_q)^2)^2}(\bar\chi\gamma^\mu D_\nu\chi) (\bar q_R \gamma_\mu D_\nu q_R-(D_\nu\bar q_R) \gamma_\mu q_R)\;.
\end{equation} 
Additionally, a loop-induced interaction with the gluon content gives rise to an effective dimension seven interaction \cite{Gondolo:2013wwa}. For spin-dependent scattering, the axial dimension six operator gives the dominant contribution,
\begin{equation}
{\cal L}_{\rm eff, tree}^{\rm SD} = -\frac{y^2}{2(m_\eta^2-(m_\chi+m_q)^2)}\bar\chi\gamma^\mu\gamma_5\chi \bar q_R \gamma_\mu\gamma_5 q_R\;.
\end{equation} 
Note that the tree-level contributions are enhancement for small mass splitting. The above expressions are applicable for $m_\eta-m_\chi \gg m_p$ (see \ref{app:DD}).

The corresponding scattering cross sections on nucleons are given by
\begin{equation}
  \sigma_{\rm SI}^N = \frac{4}{\pi}\mu_N^2 f_N^2, \qquad 
  \sigma_{\rm SD}^N = \frac{12}{\pi}\mu_N^2 \left(\sum_{q=u,d,s} \Delta q^N a_q\right)^2, 
\end{equation} 
where $\mu_N$ is the WIMP-nucleon reduced mass, $\Delta q^N$ is the spin-content of the nucleon \cite{Ellis:2008hf}, and
\begin{equation}
  a_q = \frac{y^2}{8}\frac{1}{m_\eta^2-(m_\chi+m_q)^2}\,,
\end{equation} 
is the coefficient of the axial-vector effective interaction. The effective nucleon coupling for spin-independent scattering can be parametrized as
\begin{equation}
  \frac{f_N}{m_N} = - \frac{m_\chi}{2}\sum_{q=u,d,s}\left(f_{Tq}^N+3q(2)+3\bar q(2)\right)g_q - \frac{8\pi}{9\alpha_s} f_{TG}^N b + \frac34 G(2) g_G\,,
\end{equation}
where $f_{Tq}^N \propto \langle N | \bar q q | N\rangle$ and $f_{TG}^N=1-\sum_{q=u,d,s} f_{Tq}^N$ are related to the integrated quark and gluon content \cite{Ellis:2008hf},
$G(2)$, $q(2)$, $\bar q(2)$ are the second moments of the gluon, quark and antiquark distribution \cite{Hisano:2010ct}, respectively, and
\begin{equation}
  g_q = -\frac{y^2}{8}\frac{1}{(m_\eta^2-(m_\chi+m_q)^2)^2}\,,
\end{equation} 
is the coefficient of the effective dimension eight operator contributing to spin-independent scattering. The loop-induced interactions with the gluon content are proportional to the loop functions
\begin{align}
  b &= \frac{\alpha_s}{4\pi} \frac{m_\chi}{24} \frac{y^2}{2} \left( 3I_2 - I_5 -2m_\chi^2 I_4 \right) \,,\\
  g_G &= \frac{\alpha_s}{3\pi} \frac{m_\chi}{8} \frac{y^2}{2} \left( I_5 + 2m_\chi^2 I_4 \right) \,,
\end{align}
where the loop integrals $I_n$ are given in \cite{DreesNojiri} (we use the corrected expressions from Ref.~\citen{Gondolo:2013wwa}). As was emphasized in Ref.~\citen{Gondolo:2013wwa}, the loop-induced effective couplings are regular for $m_\eta\to m_\chi+m_q$, while the effective couplings $a_q$ and $g_q$ which arise from tree-level exchange of the mediator are resonantly enhanced.

Due to the absence of dimension six operators for spin-independent scattering, limits on spin-dependent scattering yield the dominant constraint over a wide region of the parameter space. Nevertheless, since limits on spin-independent scattering are much stronger, the latter are important  especially for small mass splitting, due to the less severe suppression of the dimension eight operators.

In our analysis we consider spin-independent limits from XENON100 \cite{Aprile:2012nq} and from LUX \cite{Akerib:2013tjd}, which are most sensitive at present within the mass range $m_\chi \gtrsim 40$\,GeV considered here. Although xenon is not an optimal material for spin-dependent searches, the presence of the isotopes ${}^{129}$Xe and ${}^{131}$Xe
together with the large exposure and small background allow to set stringent limits on the spin-dependent WIMP-neutron scattering cross section, {\it e.g.} based on data collected by XENON100 \cite{Garny:2012it,Aprile:2013doa}. For the simplified models considered here, the coupling strength to neutrons and protons is comparable. Therefore, also experiments probing the spin-dependent proton cross section ({\it e.g.} SIMPLE \cite{SIMPLE11}, COUPP \cite{COUPP12}) yield comparable constraints to those derived from the XENON100 limits. Additionally, these interactions are probed by observations of the neutrino flux from the sun (cf. Sec.\,\ref{sec:neutrinos}).

\begin{figure}[t]
\begin{center}
\includegraphics[width=0.49\textwidth]{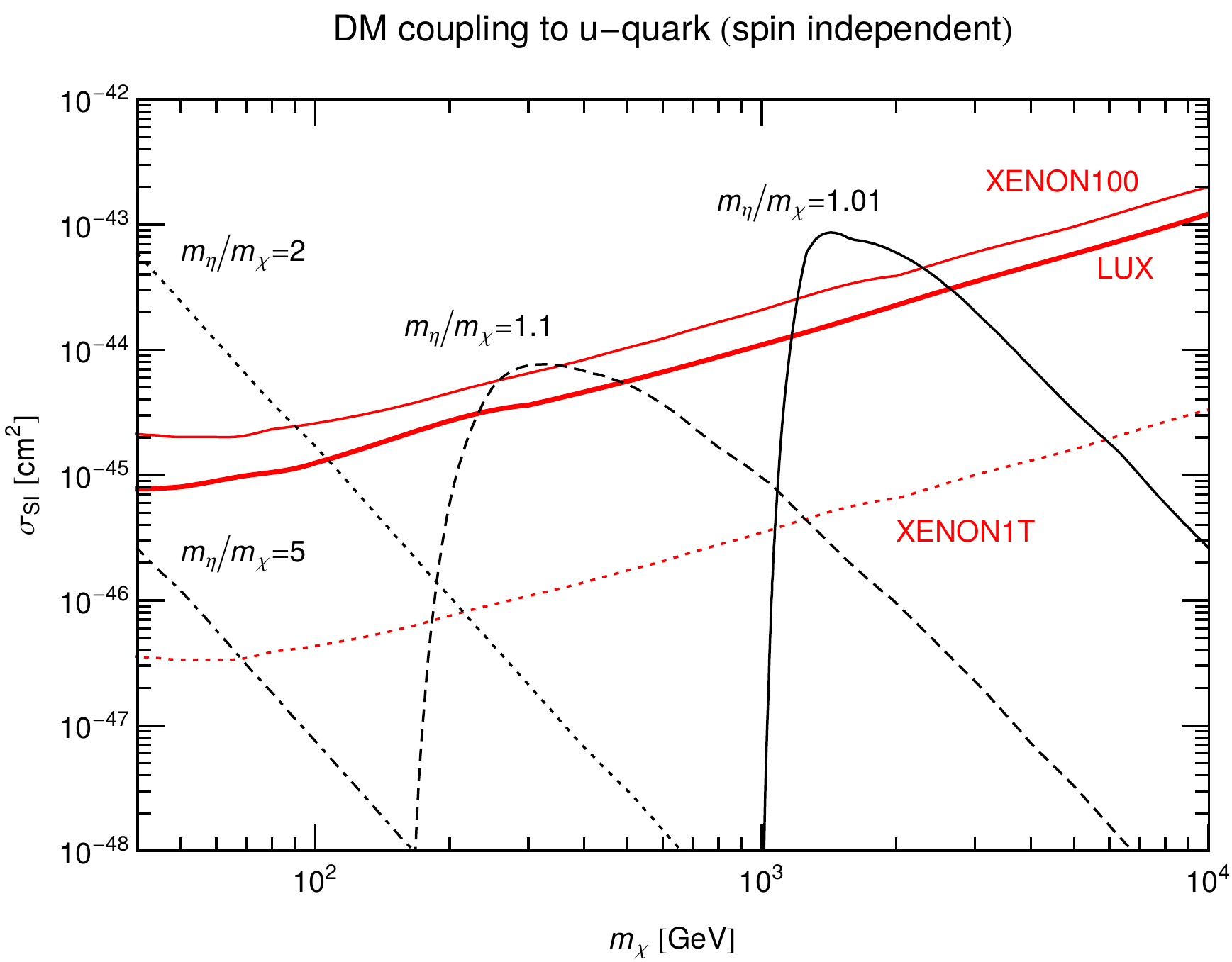}\\
\includegraphics[width=0.49\textwidth]{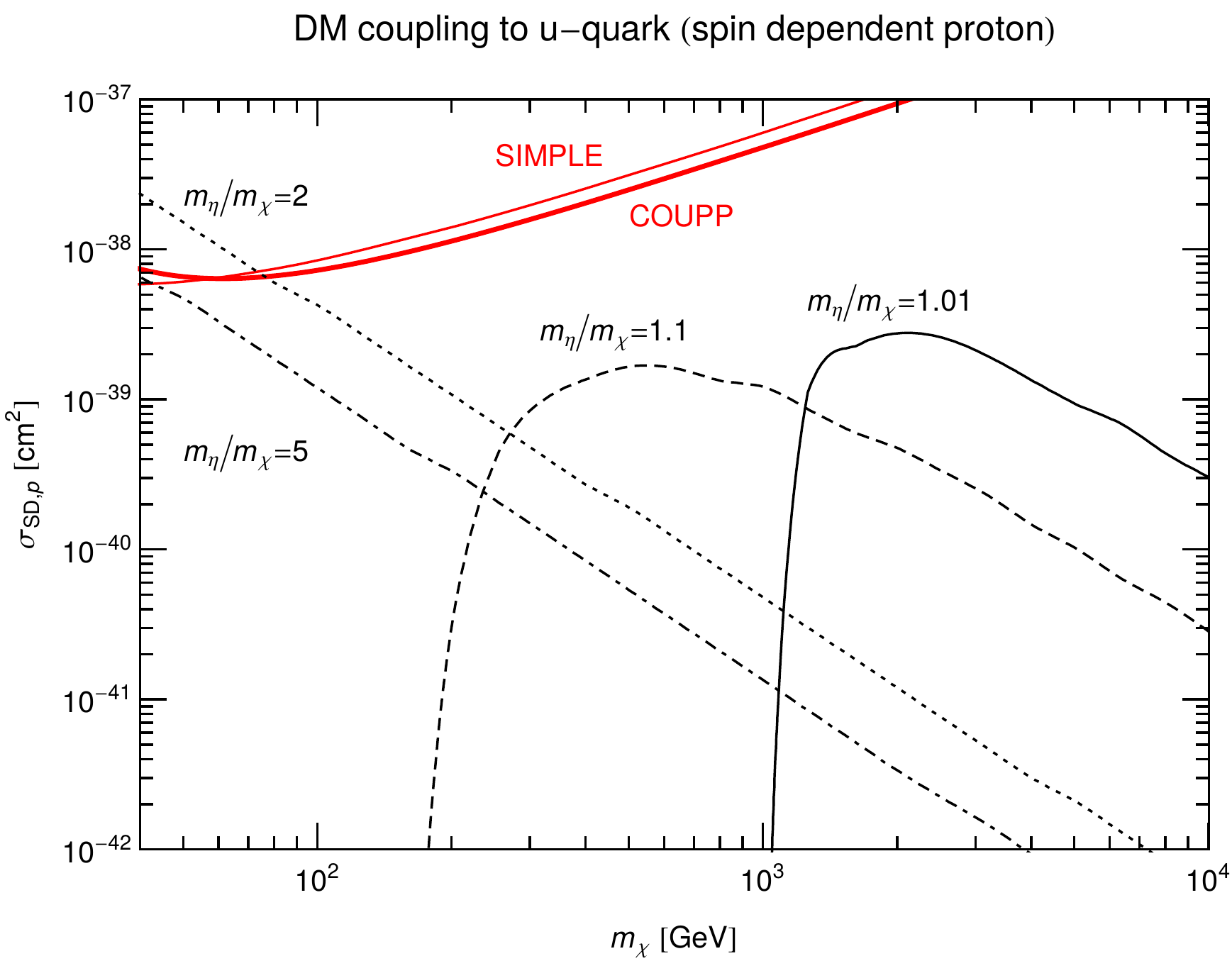}
\includegraphics[width=0.49\textwidth]{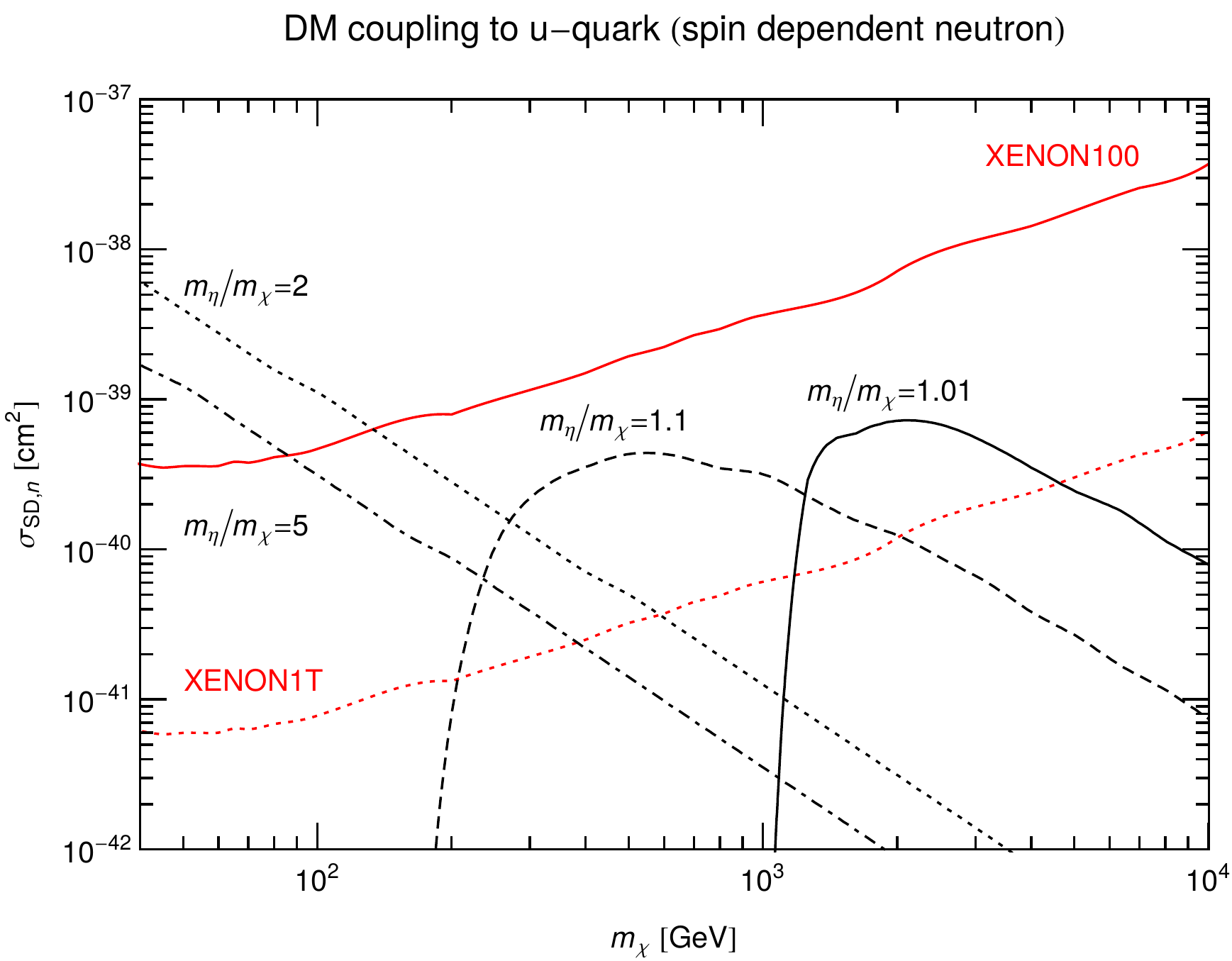}
\caption{\small   $90\%$C.L. direct detection limits for spin-independent scattering (upper plot), spin-dependent scattering off protons (lower left plot), and off neutrons (lower right plot) for $m_\eta/m_\chi=1.01, 1.1, 2, 5$. Black lines correspond to the scattering cross section expected for thermal production, and red lines to experimental constraints from XENON100~\cite{Aprile:2012nq}, LUX~\cite{Akerib:2013tjd}, COUPP~\cite{Behnke:2012ys} and SIMPLE~\cite{Felizardo:2011uw}. Also shown is a projection for XENON1T~\cite{Aprile:2012zx,Garny:2012eb}.}
\label{fig:sigDD}
\end{center}
\end{figure}

As an example, we show in Fig.\,\ref{fig:sigDD} the spin-independent as well as spin-dependent scattering cross section that corresponds to a thermally produced dark matter particle for a scenario where the mediator couples to right-handed up-quarks. The cross section depends strongly on the mass splitting and also on the dark matter mass.
For small mass splitting $m_\eta/m_\chi = 1.1 (1.01)$, the cross section is suppressed below $m_\chi\lesssim 200$\,GeV ($1$\,TeV) due to strong coannihilations, while
being sizable for larger masses due to the resonant enhancement of the scattering cross section. For ${\cal O}(1)$ mass splitting, the qualitative behavior changes
due to the absence of coannihilations. In addition, spin-dependent scattering is more sensitive compared to spin-independent scattering, since the latter is induced by dimension eight operators for Majorana dark matter. For the standard halo model, thermally produced dark matter can be excluded from constraints on spin-independent scattering from LUX~\cite{Akerib:2013tjd} for $230$\,GeV$\lesssim m_\chi \lesssim 550\,$GeV if $m_\eta/m_\chi=1.1$, and from spin-dependent scattering constraints from XENON100~\cite{Aprile:2012nq} for $m_\chi \lesssim 150\,$GeV if $m_\eta/m_\chi=2$.

The expected increase in sensitivity in XENON1T  \cite{Aprile:2012zx}, and later LUX-ZEPLIN (LZ), will allow us to probe a significant fraction of the accessible parameter space for thermally produced Majorana dark matter that couples to first-generation quarks. If dark matter couples to heavy quarks, the constraints are significantly weaker, \emph{e.g.} about an order of magnitude in $\sigma_{\rm SI}\propto y^4$ for bottom quarks \cite{Garny:2012eb}. Nevertheless, even if dark matter couples only to top quarks, LZ and XENON1T are sensitive to scattering cross sections expected for thermal production for mass splittings in the few$\times {\cal O}(10\%)$ range \cite{Ibarra:2015nca}. 

\subsection{Uncolored mediator}

\begin{figure}[t]
\begin{center}
\includegraphics[width=0.6\textwidth]{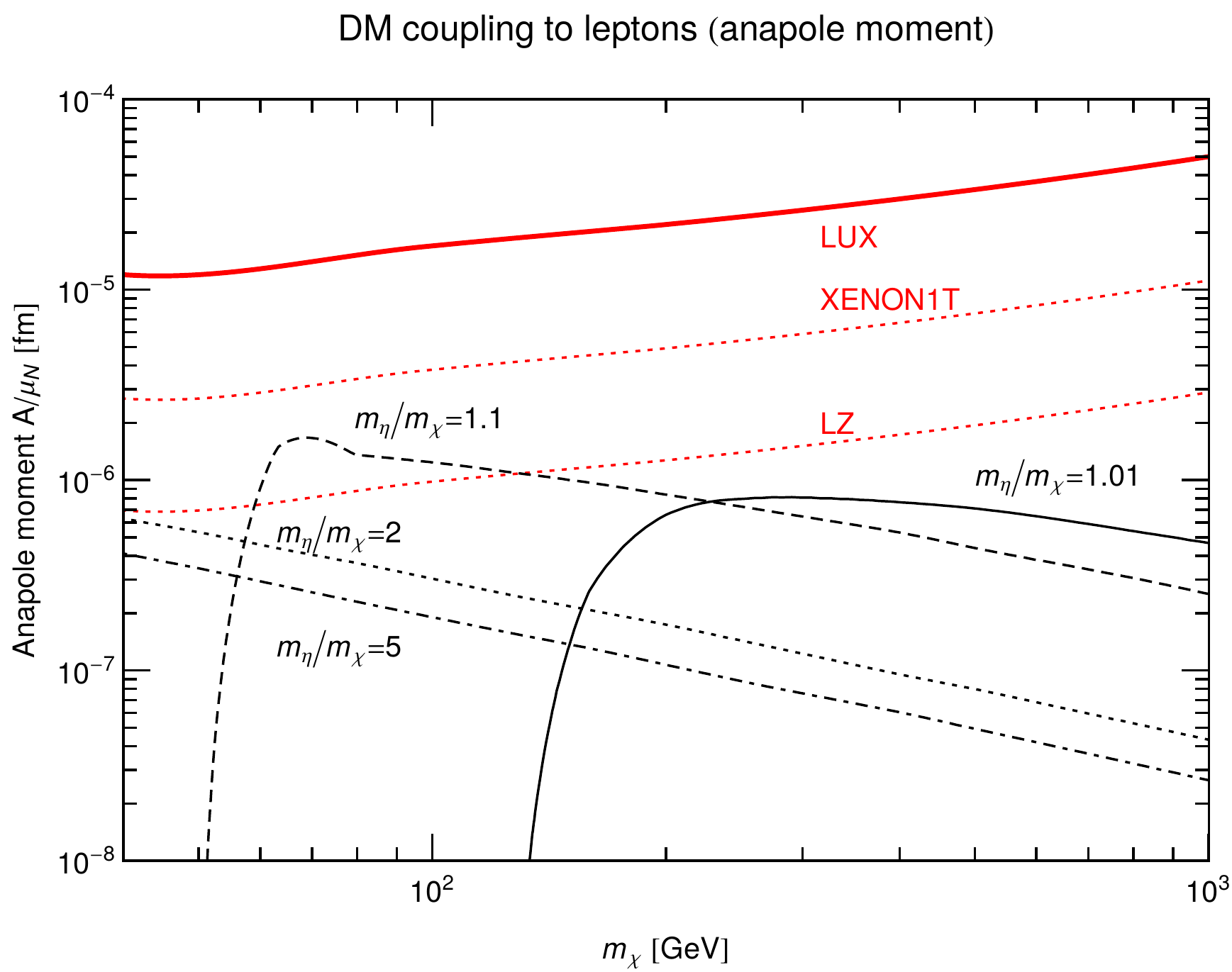}
\caption{\small $90\%$C.L. direct detection limits on the anapole moment ${\cal A}/\mu_N$ derived from LUX limits~\cite{Kopp:2014tsa}. Black lines correspond to the anapole moment expected for thermal production for $m_\eta/m_\chi=1.01, 1.1, 2, 5$, respectively, if dark matter couples to right-handed muons. Also shown is a projection for XENON1T~\cite{Aprile:2012zx,Garny:2012eb} and for LUX-ZEPLIN~\cite{Kopp:2014tsa}.} 
\label{fig:sigDDana}
\end{center}
\end{figure}

It is interesting to note that even for a pure coupling to leptons, dark matter interactions can be potentially probed in next-generation direct detection experiments due to the loop-induced nucleon coupling \cite{Kopp:2014tsa,Chang:2014tea,Bai:2014osa}. In particular, the charged mediator gives rise to an effective electromagnetic coupling. However, for Majorana dark matter, the loop-induced dipole moments vanish, and the leading contribution is the electromagnetic anapole moment,
\begin{equation}
  {\cal L}_{eff} = {\cal A} \, \bar \chi \gamma^\mu \gamma^5\chi \partial^\nu F_{\mu\nu} \,,
\end{equation}
with  \cite{Kopp:2014tsa}
\begin{align}
  {\cal A} &= - \frac{y^2e}{96\pi^2 m_\chi^2}\Bigg\{ \frac32\ln\left(\frac{m_\eta^2}{m_\ell^2}\right) - \frac{m_\chi^2+3m_\eta^2-3m_\ell^2}{\sqrt{(m_\eta^2-m_\chi^2-m_\ell^2)^2-4m_\chi^2m_\ell^2}} \nonumber\\
 & \times \mbox{arctanh}\left(\frac{\sqrt{(m_\eta^2-m_\chi^2-m_\ell^2)^2-4m_\chi^2m_\ell^2}}{m_\eta^2-m_\chi^2+m_\ell^2}\right) \Bigg\}\;.
\end{align}
This expression is valid if the momentum transfer is smaller than the mass of the particles in the loop, in particular for $\ell=\mu,\tau$.
For $\ell=e$, a form factor depending on the momentum should be included  \cite{Kopp:2014tsa}.

Although the anapole moment is enhanced logarithmically for very small mass splitting, it is difficult to probe thermally produced dark matter in this region due to efficient coannihilations, which largely reduce the value of the coupling $y$ required for thermal production. For a coupling to muons, the anapole moment corresponding to thermal production is shown in Fig.\,\ref{fig:sigDDana}, together with constraints inferred from LUX~\cite{Kopp:2014tsa} as well as prospects for XENON1T~\cite{Aprile:2012zx,Garny:2012eb} and for LZ~\cite{Kopp:2014tsa}. Taking coannihilations into account, the expected sensitivity of LZ may be large enough to detect thermally produced leptophilic dark matter in a small portion of the parameter space.
The LUX constraints are significantly weaker compared to Fermi LAT and H.E.S.S. limits except for very fine mass splittings $m_\eta/m_\chi\sim 1.01$, where they are comparable (\emph{cf.} Fig.\,\ref{fig:gamma}). Prospects for XENON1T are competitive with GAMMA-400 and CTA for similarly small mass splitting and low dark matter masses.

\section{Collider signatures}\label{sec:collider}

The non-observation at LEP or the LHC of signals of new physics allows to set limits on the parameter of the dark matter model we consider in this review. The search strategies crucially depend on the mass difference between the dark matter particle and the scalar mediator. When the mass difference is very large,  only the dark matter particle can be directly produced at the collider experiment. In this case, the most relevant experimental signature is a monojet, monophoton or mono-W/Z boson plus missing transverse momentum, which is generated by the pair production of dark matter particles which recoil against a jet or gauge boson. This signature has been studied in detail by ATLAS\cite{Aad:2015zva,atlasMono} and CMS\cite{Khachatryan:2014rra}, mostly within an effective operator description \cite{Rajaraman:2011wf,Fox:2011pm} (see Refs.\citen{Buchmueller:2013dya,Busoni:2013lha,Busoni:2014haa,Racco:2015dxa} for some recent discussions of the validity of this approach). On the other hand, when the mass difference is moderate, the scalar mediators can be directly pair-produced\cite{Garny:2013ama,Chang:2013oia,An:2013xka,Bai:2013iqa,DiFranzo:2013vra,Papucci:2014iwa,Garny:2014waa,Chang:2014tea,Bai:2014osa,Abdallah:2014hon}. The subsequent decay of the mediator into the dark matter particle and the corresponding quark or lepton leads to signatures with missing transverse momentum and two jets, or with two same-flavor, opposite-charged leptons, respectively. Signatures of this type are being studied extensively in connection to SUSY searches  \cite{Chatrchyan:2014lfa,Aad:2014wea,Aad:2014vma,Khachatryan:2014qwa}. 
Lastly, when the mass difference is small, the scalar mediator is also directly produced in the collider. However, the jets or charged leptons produced in their decay are too soft to be detected,\cite{Dreiner:2012sh,Aad:2014tda,Aad:2014nra}. therefore the final state is completely invisible to the detector and the monojet or monophoton constraints give again the strongest limits on the model parameters.

The production rate of mediators at the LHC, and correspondingly the limits on the model, crucially depends on whether they carry color charge or not. We discuss in what follows each case separately.

\subsection{Colored mediator}

\begin{figure}[t]
  \begin{center}
    \includegraphics[width=0.9\textwidth]{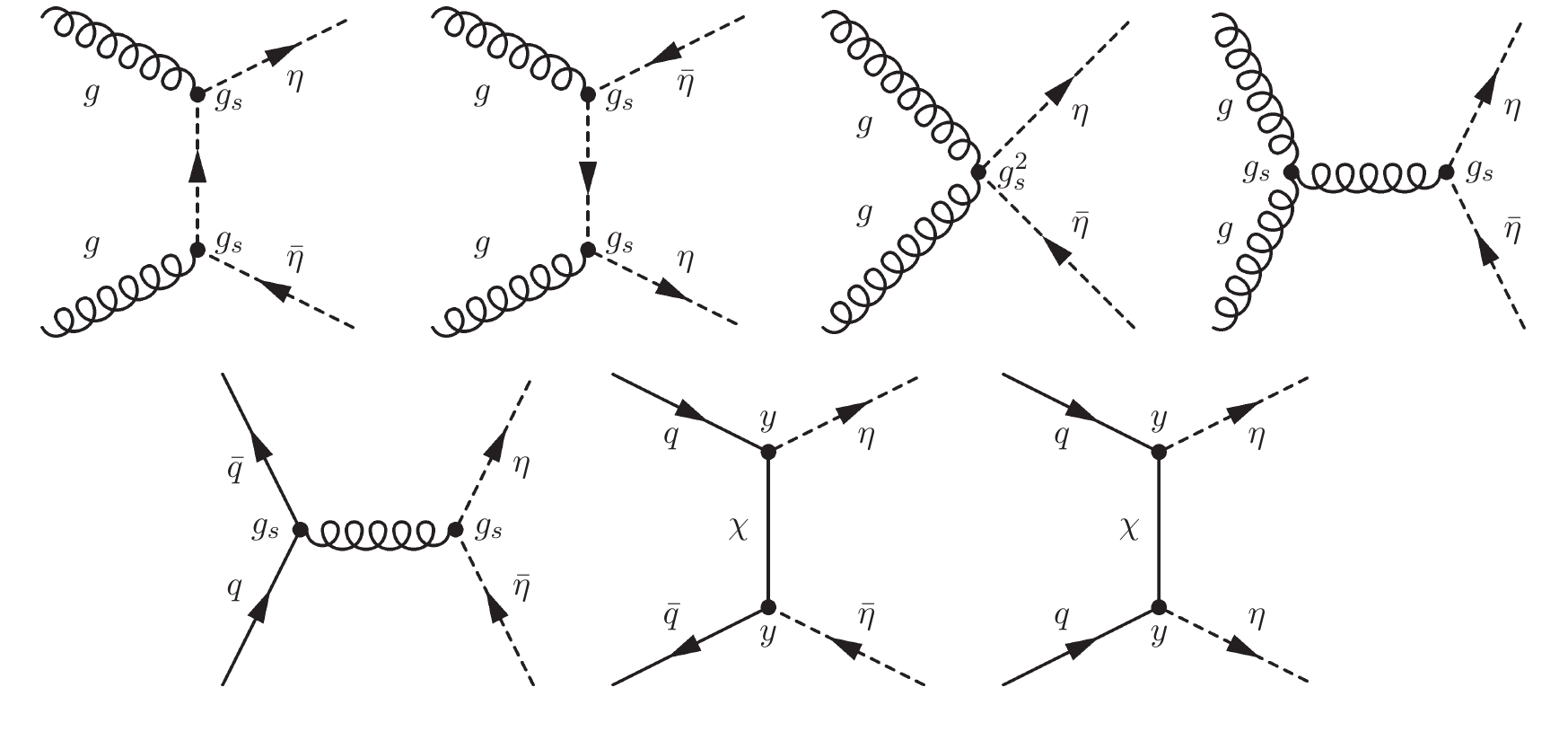}
    \includegraphics[width=0.45\textwidth]{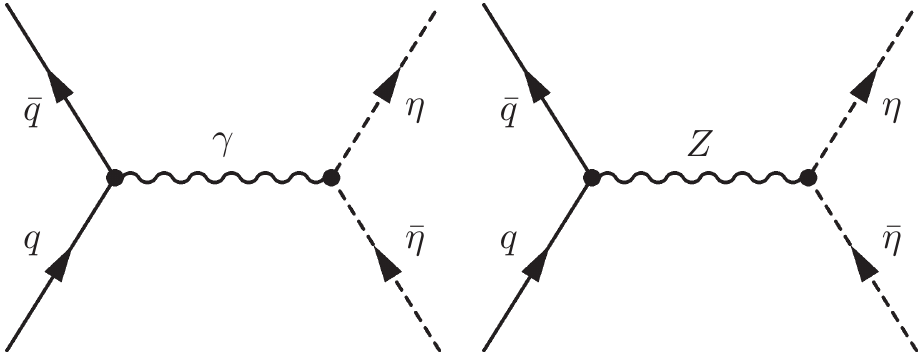}
  \end{center}
  \caption{\label{fig:prodLHC} Feynman diagrams for the processes contributing to the direct production of a colored mediator and an uncolored mediator (lower row) at the LHC.}
\end{figure}

\begin{figure}[t]
  \begin{center}
    \includegraphics[width=0.49\textwidth]{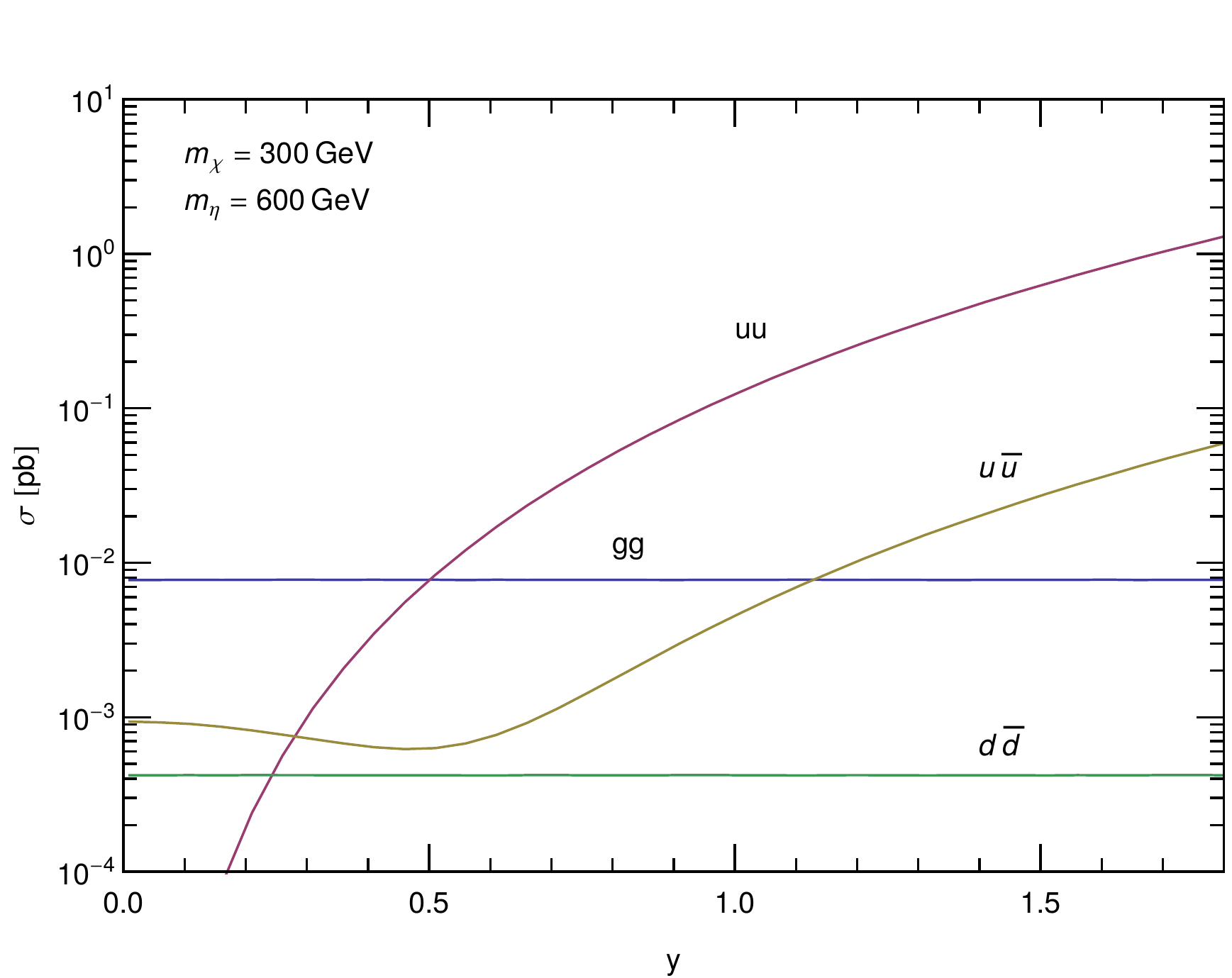}
    \includegraphics[width=0.49\textwidth]{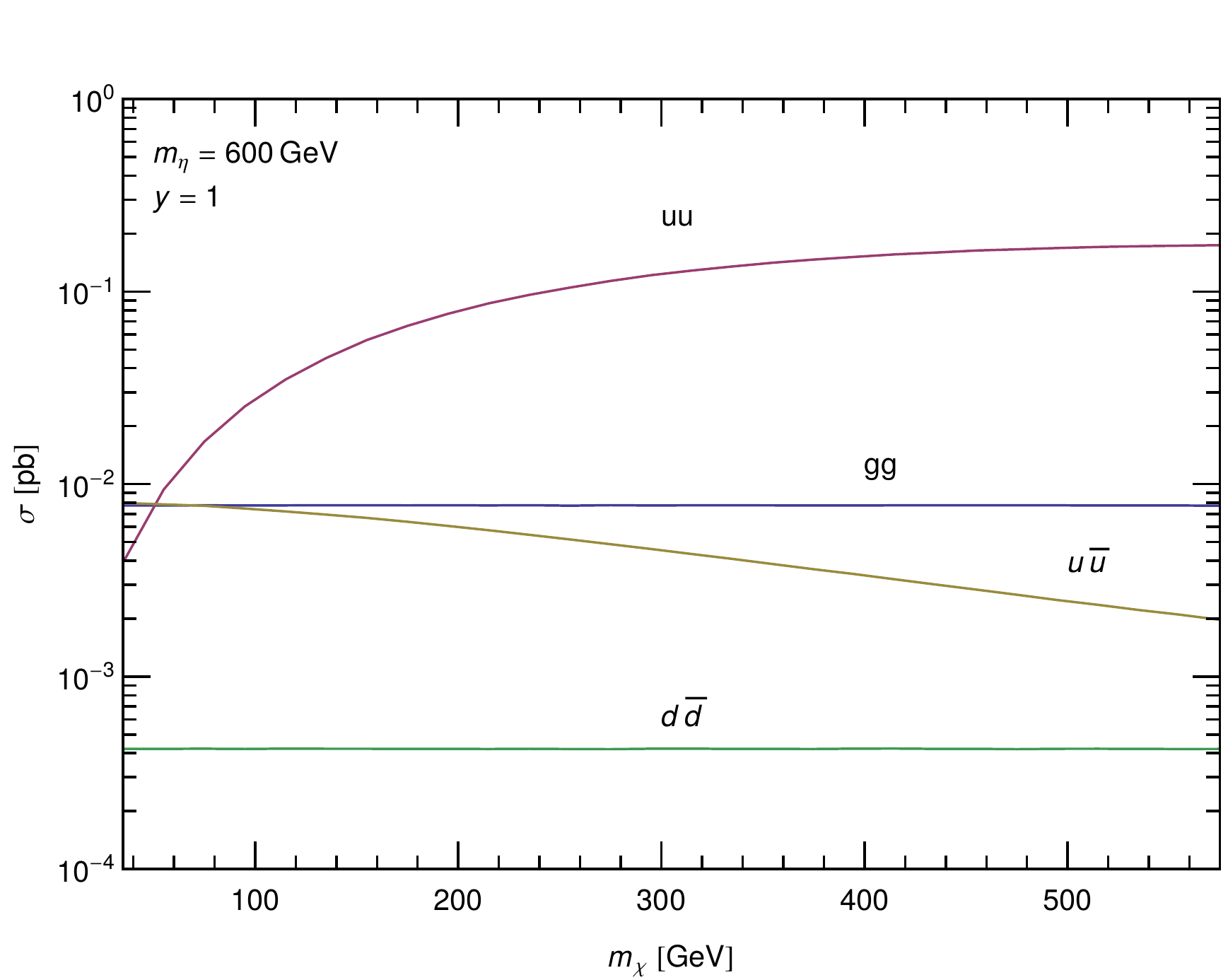}
  \end{center}
  \caption{\label{fig:prodvsy} Production cross section of pairs of the colored mediator at the LHC as a function of the Yukawa coupling $y$ (left plot) and the dark matter mass (right plot), for different production channels.}
\end{figure}

We consider first the case of a colored scalar mediator which couples to the dark matter particle and to the right-handed up or down quark. The Feynman diagrams contributing to the production at the LHC  of the mediator are shown in Fig.\,\ref{fig:prodLHC}. The first five diagrams are mediated by strong interactions and contribute to the subprocesses $gg\rightarrow \eta \bar\eta$, for the first four, and $q\bar q\rightarrow \eta \bar \eta$, for the fifth. The sixth and seventh diagrams are important when the dark matter particle couples to a first generation quark, and correspond to the subprocesses $q \bar q\rightarrow \eta \bar \eta$ and $q  q\rightarrow \eta \eta$ via the exchange of a dark matter particle in the $t$-channel (note that the latter process is a consequence of the Majorana nature of our dark matter candidate). Finally, the last two diagrams correspond to the subprocess $q\bar q\rightarrow \eta \bar \eta$ via electroweak interactions and give a negligible contribution at the LHC for a colored mediator. The production cross sections for the different subprocesses can be found in Fig.\,\ref{fig:prodvsy}, for  some specific choices of the parameters and
$\sqrt{s}= 8$ TeV. It follows from the plot that for small Yukawa couplings the dominant production process at the LHC is $gg\rightarrow \eta\bar\eta$, while for larger Yukawas ($y\gtrsim 0.5$ for the parameters of the figure)  the dominant process is $uu\rightarrow \eta\eta$, due to the enhancement of the rate by the parton distribution functions. This process is specific for a Majorana dark matter particle, and the corresponding cross section scales with the squared of the dark matter Majorana mass (\emph{cf.} right plot in Fig.\,\ref{fig:prodvsy}). Thermal dark matter production typically requires a sizable Yukawa coupling, {\it cf.} Fig.~\ref{fig:relic}, therefore for this particularly interesting case the dominant production mechanism of colored scalar mediators at the LHC is due to the exchange of a dark matter particle in the $t$-channel. This is in contrast to the squark production in the Minimal Supersymmetric Standard Model, which is dominated by channels involving the strong interaction. As a consequence, the ATLAS and CMS limits on simplified SUSY scenarios  cannot be straightforwardly applied to our toy model, since the efficiency of the search  depends on the underlying hard process. A critical discussion can be found in Ref.~\citen{Garny:2014waa}.

\begin{figure}[t]
\begin{center}
\includegraphics[width=0.49\textwidth]{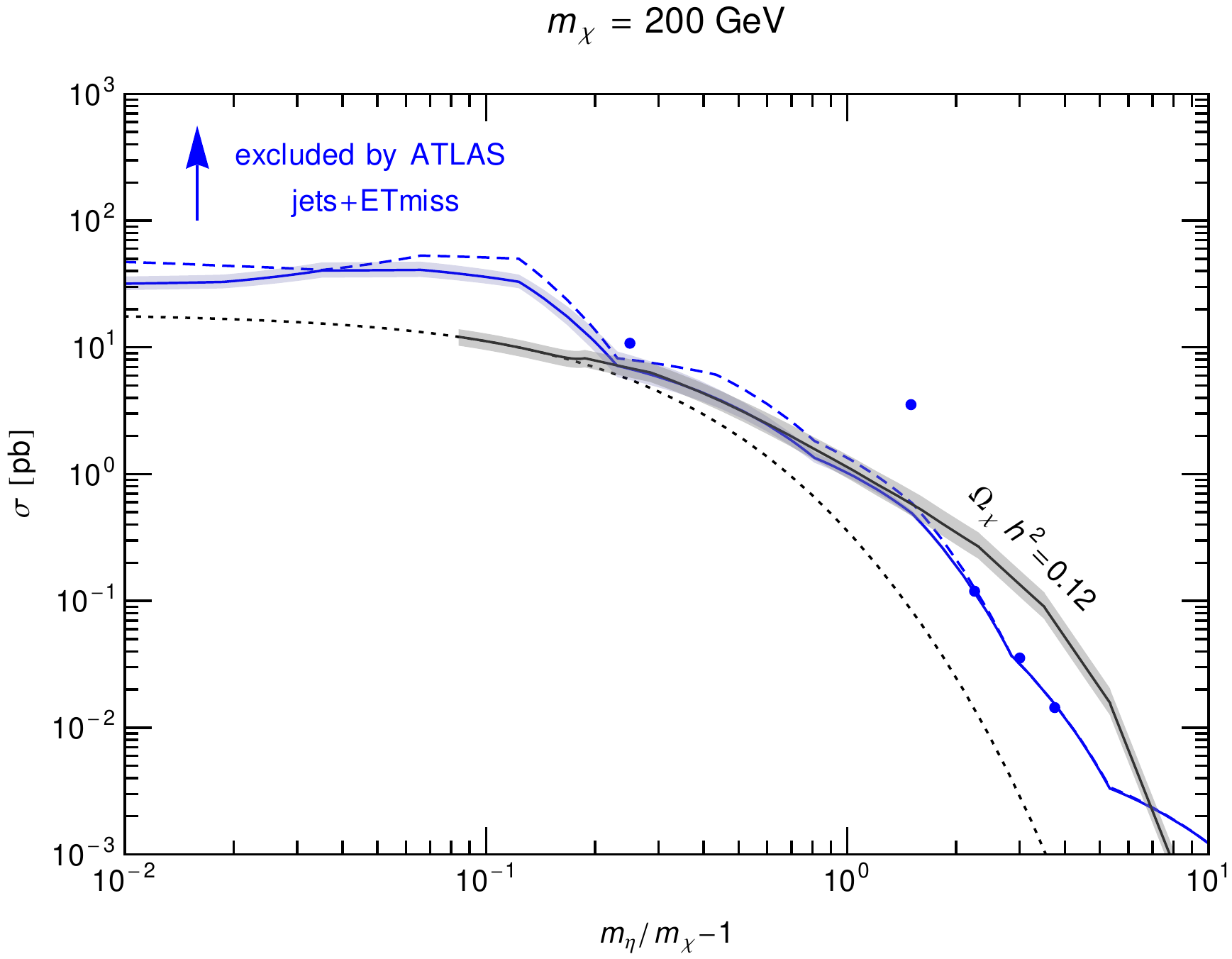}
\includegraphics[width=0.49\textwidth]{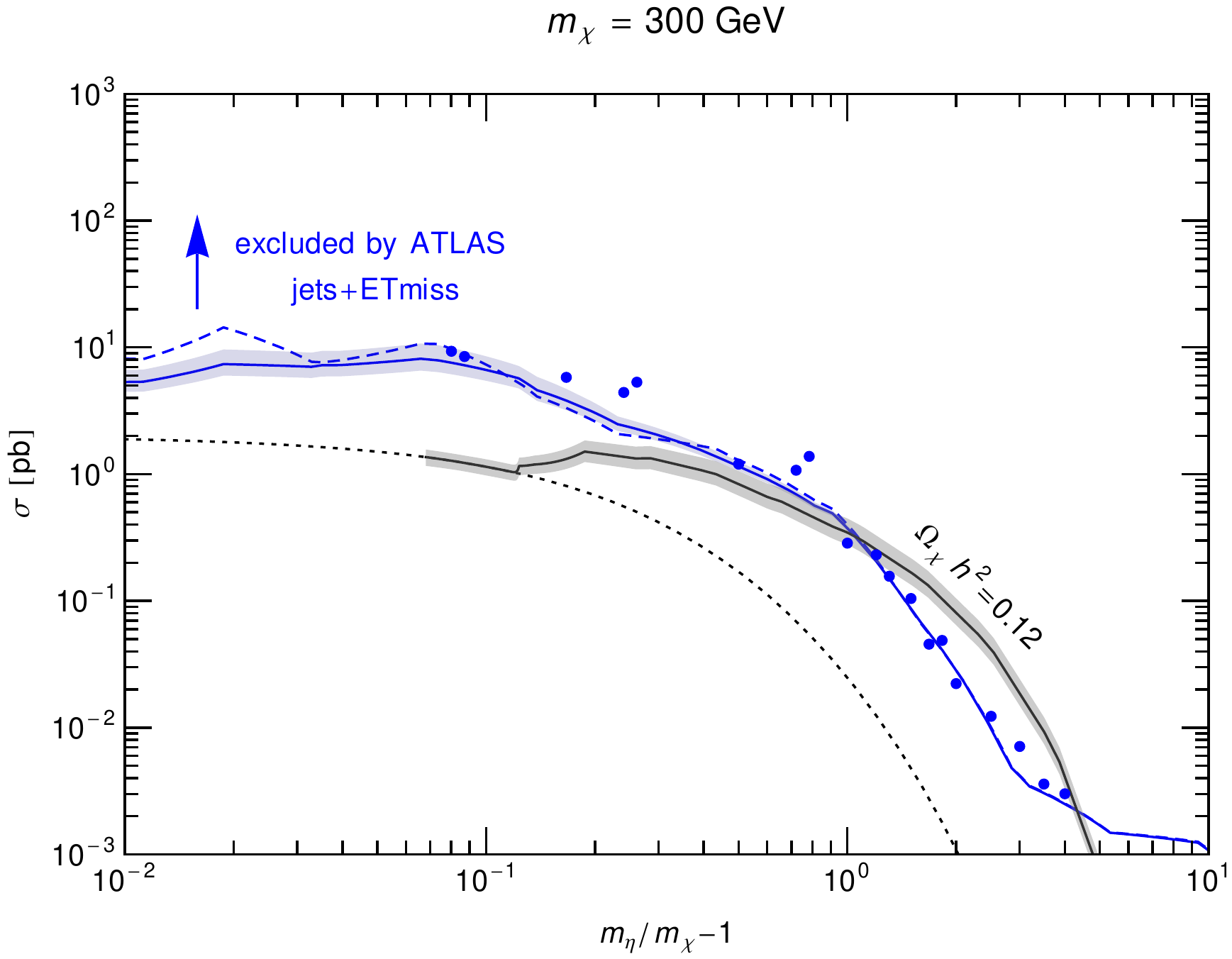}\\
\includegraphics[width=0.49\textwidth]{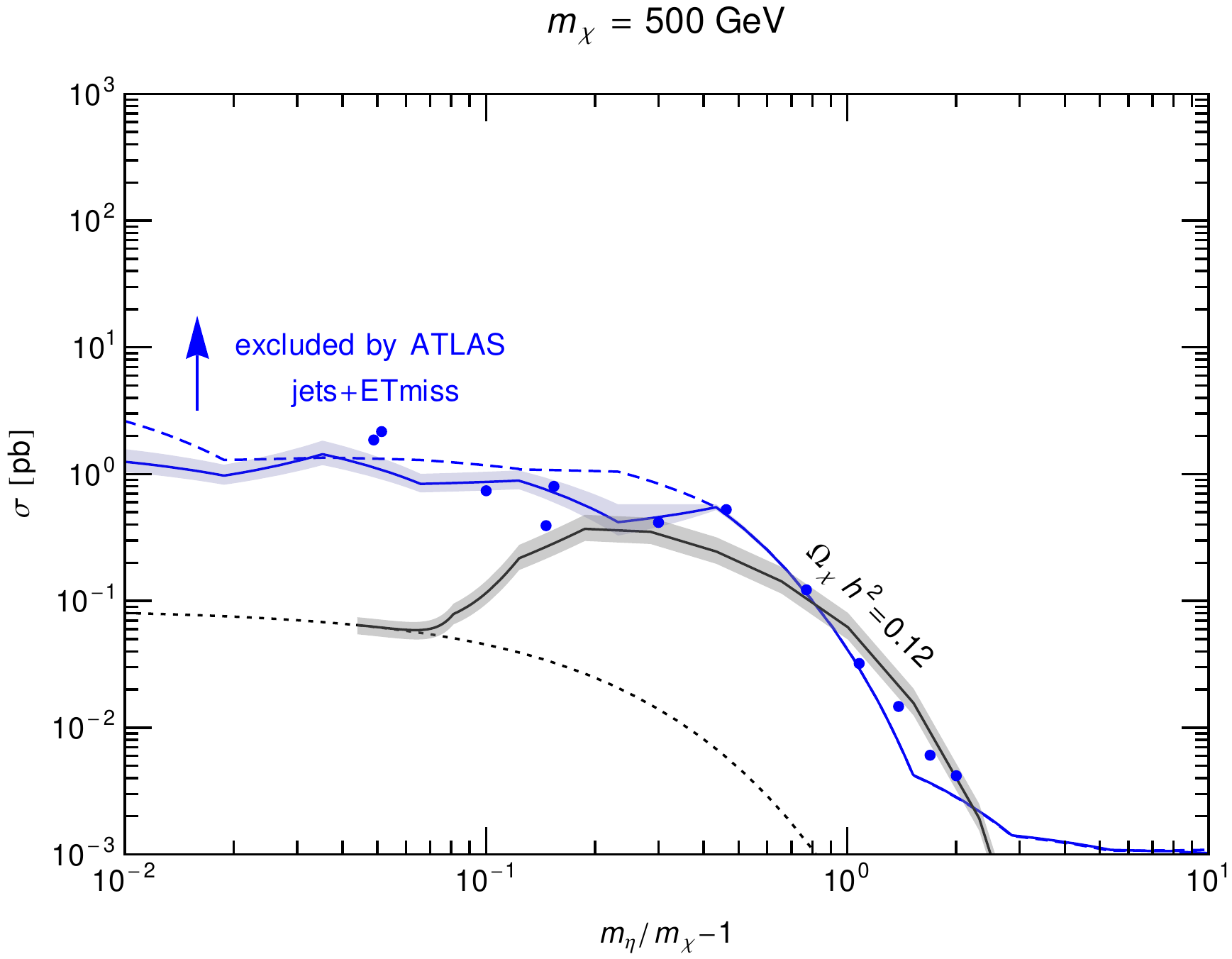}
\includegraphics[width=0.49\textwidth]{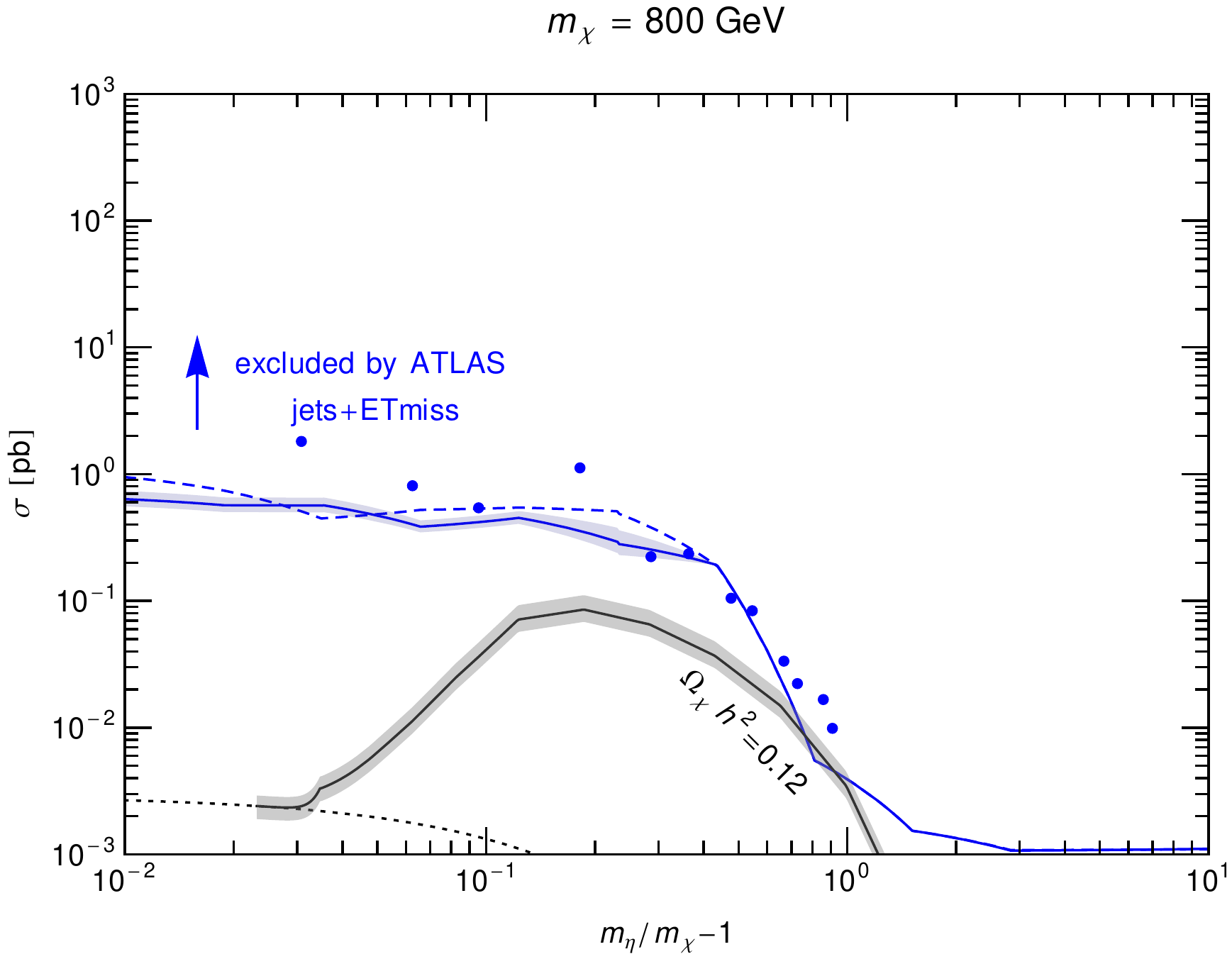}
\caption{\small $95\%$C.L. constraints on the production cross section for direct production of the colored mediator for $m_\chi=200, 300, 500, 800$\,GeV, obtained
from a reinterpretation of the ATLAS search~\cite{Aad:2014wea} for jets and missing energy. Blue solid lines show the upper limit when using two hard jets in the
matching procedure, and blue dashed lines when using only one jet. The black dotted line corresponds to the theoretical prediction for $y=0$, and the black solid line for $y=y_{th}$. The shaded regions correspond to an estimate of uncertainties (see Ref.~\citen{Garny:2014waa} for details), and the blue dots are the upper limits obtained by ATLAS for a simplified supersymmetric model.} 
\label{fig:sigEtaEta}
\end{center}
\end{figure}

A dedicated search for the signals of Majorana dark matter coupling to a light quark has been performed in Ref.~\citen{Garny:2014waa}, from a re-analysis of the ATLAS search \cite{Aad:2014wea} for jets and missing energy based on ${\cal L}=20.3$fb$^{-1}$ at $\sqrt{s}=8$ TeV (see also Refs.~\citen{Chang:2013oia,An:2013xka,Bai:2013iqa,DiFranzo:2013vra} for related works). The resulting upper limits on the pair-production cross section of the colored mediator are shown in Fig.\,\ref{fig:sigEtaEta}, as a function of the mass splitting $m_\eta/m_\chi-1$ and various dark matter masses. For small mass splitting, the constraint reaches a plateau given by $\sigma\simeq 0.6-30$\,pb, depending on the dark matter mass $m_\chi=800-200\,$GeV. For ${\cal O}(1)$ mass splitting, the constraints are more stringent, of the order $1-0.01$\,pb. The theoretical expectation for a thermally produced dark matter particle is also shown in Fig.\,\ref{fig:sigEtaEta}. The coupling strength required for thermal production can be excluded for example for $1\lesssim m_\eta/m_\chi-1\lesssim 7$ if $m_\chi=200\,$GeV and for $1\lesssim m_\eta/m_\chi-1\lesssim 2$ if $m_\chi=500\,$GeV.

These exclusion limits are obtained under the assumption that $\eta\to \chi q$ is the only possible decay channel, which is the case as long as the simplified model is a good description. When relaxing this assumption, and considering an additional hypothetical untagged decay channel, the multijet limits degrade due to two effects: first, due to the reduced event rate suppressed by ${\rm BR}(\eta\to \chi q)^2$. Second, for very large total width, the narrow width approximation becomes questionable\cite{Papucci:2014iwa}.

Qualitatively, this picture remains the same when considering multiple colored mediators and/or a coupling to down-type quarks, or left-handed first generation quarks. For example, for the case of two mediators that couple to the right-handed up- and charm-quarks, respectively, the quantitative change in the excluded Yukawa coupling strength is less than about $30\%$ for $m_\chi>300$GeV or $m_\eta/m_\chi>2$, but can be larger otherwise due to the two times larger contribution from $SU(3)_c$-mediated processes to the production cross section \cite{Garny:2014waa}.

Apart from searches for multiple jets, constraints on a colored $t$-channel mediator have also been derived for the monojet channel~\cite{Chang:2013oia,An:2013xka,Bai:2013iqa,DiFranzo:2013vra,Racco:2015dxa}. In addition to the process $q\bar q \to \chi\chi g$, via radiation of the gluon off the initial state or off the mediator in the $t$-channel, also  the processes $qg\to \chi\chi q$, via initial-state radiation or on-shell production of $\chi\eta$, contribute to the monojet channel\footnote{When radiating additional gluons, these processes can also give a certain contribution to multijet searches (\emph{cf.} Ref.\cite{Papucci:2014iwa} for an analysis for Dirac dark matter).  However, for Majorana dark matter, their relative contribution is less important due to the large contribution from $qq\to \eta\eta$.}. For very small mass splitting, the direct pair production of the mediator also contributes to the monojet signal \cite{Dreiner:2012sh}.
For Majorana dark matter with ${\cal O}(1)$ mass splitting of the mediator, the constraints on the Yukawa coupling $y$ inferred from monojet searches have been found to be less stringent compared to the multijet channel~\cite{Chang:2013oia,An:2013xka,Bai:2013iqa,DiFranzo:2013vra}. For nearly degenerate masses, the monojet exclusion derived in Ref.~\citen{Dreiner:2012sh}, taking $\eta\bar\eta$ production via strong interactions into account, covers some parameter space which is complementary to the multijet search discussed above, which is however already excluded by XENON100~\cite{Garny:2013ama}. Similar constraints can be derived also from monophoton searches \cite{Aad:2014tda}.

\subsection{Uncolored mediator}

For a mediator which is singlet under $SU(3)_c$, the production at LHC is clearly much less efficient than for colored mediators, and therefore the corresponding constraints are expected to be significantly weaker. A scalar mediator which couples the Majorana dark matter particle to a Standard Model lepton can be pair-produced via the Drell-Yan process (\emph{cf.} diagrams in the lowest row of Fig.\,\ref{fig:prodLHC}), and the subsequent decay into a pair of opposite-sign, same-flavor leptons as well as missing transverse momentum carried away by the dark matter particles provides a clear signature. Since the production processes as well as the decay is in this case identical to the simplified supersymmetric model containing neutralino and sleptons, the corresponding exclusion limits apply directly in this case, and no re-interpretation is necessary. On the other hand, this also implies that collider constraints depend only on the masses $m_\eta$ and $m_\chi$, and are essentially independent of the coupling $y$ (as long as $y$ is large enough such that the mediator decays promptly, which is the case possibly except for $m_\eta-m_\chi<m_\tau$, see {\it e.g.} Ref.~\citen{Desai:2014uha}). 

The constraints obtained from the LHC are sensitive to $m_\eta-m_\chi\gtrsim 100$GeV, because smaller mass splittings would lead to leptons which are too soft to be efficiently discriminated from backgrounds. The production cross section depends on whether the mediator couples to right- or left-handed leptons, due to the different coupling strength to the $Z$ boson. Searches for direct slepton production have been performed both by the ATLAS and CMS collaborations \cite{Aad:2014vma,Khachatryan:2014qwa}. For example, the ATLAS search yields a rather mild constraint ranging up to $m_\eta\gtrsim 250(300)$GeV and $m_\chi\gtrsim 100(150)$GeV in the right-(left-)handed case, for $m_\eta-m_\chi\gtrsim 100$GeV. Nevertheless, these constraints are complementary to LEP limits, $m_\eta, m_\chi\gtrsim 100$GeV, which apply for smaller mass splitting. The current constraints are summarized in Fig.\,\ref{fig:leptons}.

\section{Complementarity of searches}\label{sec:complementarity}

Each strategy to search for traces of non-gravitational interactions of dark matter, via cosmic rays from annihilation, scattering off nuclei, and production of dark matter particles at high-energy colliders, provides pieces of information that are complementary in two different respects: first, their sensitivity depends on different properties of the dark matter particle and its interactions with the Standard Model particles. Second, each channel is affected by different types of systematic uncertainties. These include above all the dark matter density and velocity distribution, both locally at the Earth as well as for the primary targets of indirect searches, including the Galactic Center and dwarf galaxies as prominent examples. In addition, the ability to separate a signal from dark matter annihilation from astrophysical foregrounds and efficiently reject backgrounds in direct detection represent major challenges. The systematic uncertainties can be reduced, for example, by improved dynamical
constraints\cite{Iocco:2011jz, Iocco:2015xga}, progress in the theory of structure formation, as well as by exploiting the different spatial and spectral morphology of astrophysical foregrounds with respect to a dark matter annihilation signal. The spectral feature from internal bremsstrahlung is a particular  example for a characteristic signature for which uncertainties from astrophysical foregrounds are considerably reduced.

In the following, we discuss the complementarity of indirect, direct and collider searches for the simplified Majorana dark matter models subject of this review. To make a quantitative comparison, we assume the standard halo model for direct detection as described in Sec.\,\ref{sec:DD}, and quantify gamma-ray constraints relative to the Einasto profile from Sec.\,\ref{sec:ID}. As discussed before, it is important to keep in mind the different sources of systematic uncertainties in each case.

\subsection{Complementarity of constraints on the coupling}

\begin{figure}[t]
  \begin{center}
    \includegraphics[width=0.49\textwidth]{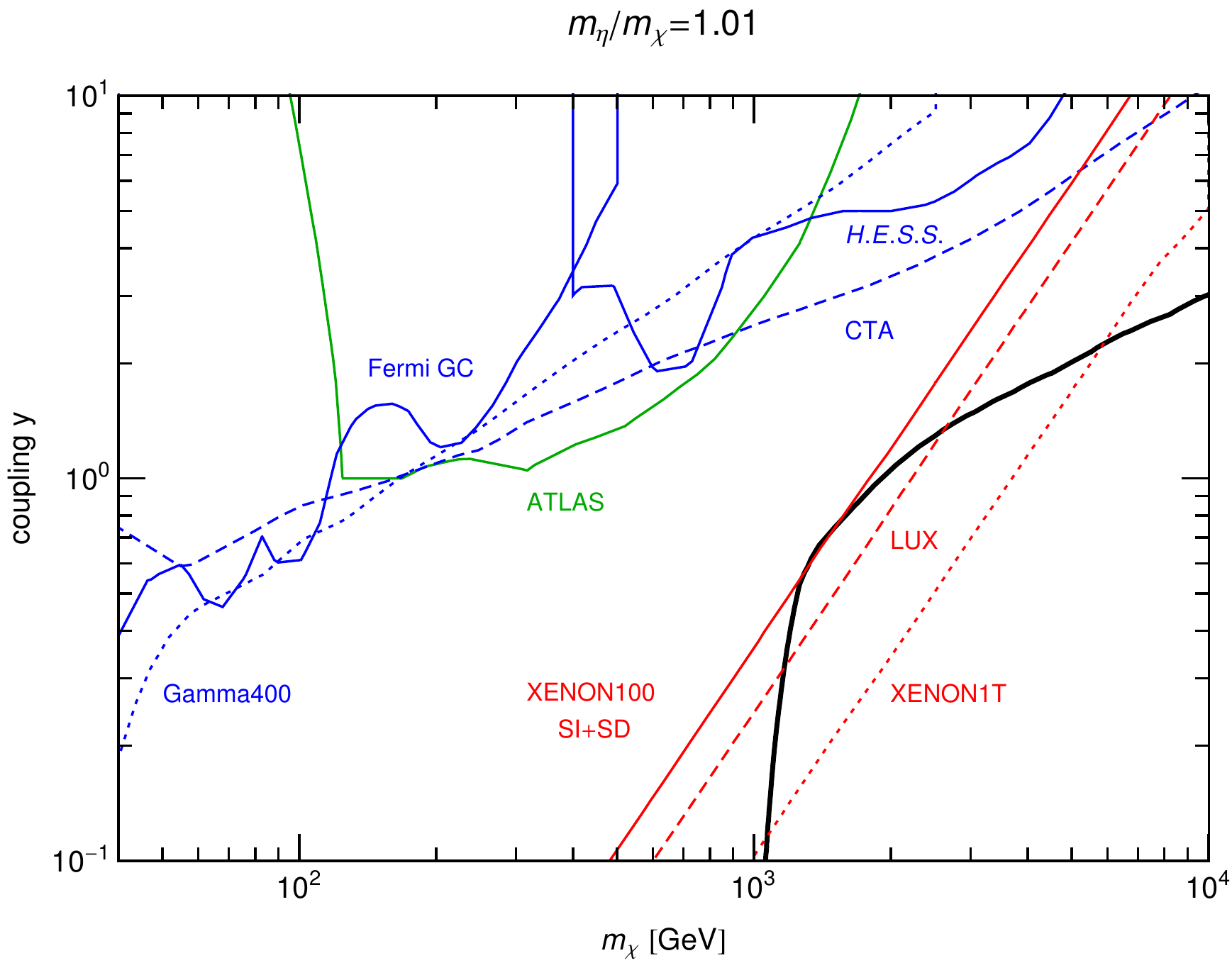}
    \includegraphics[width=0.49\textwidth]{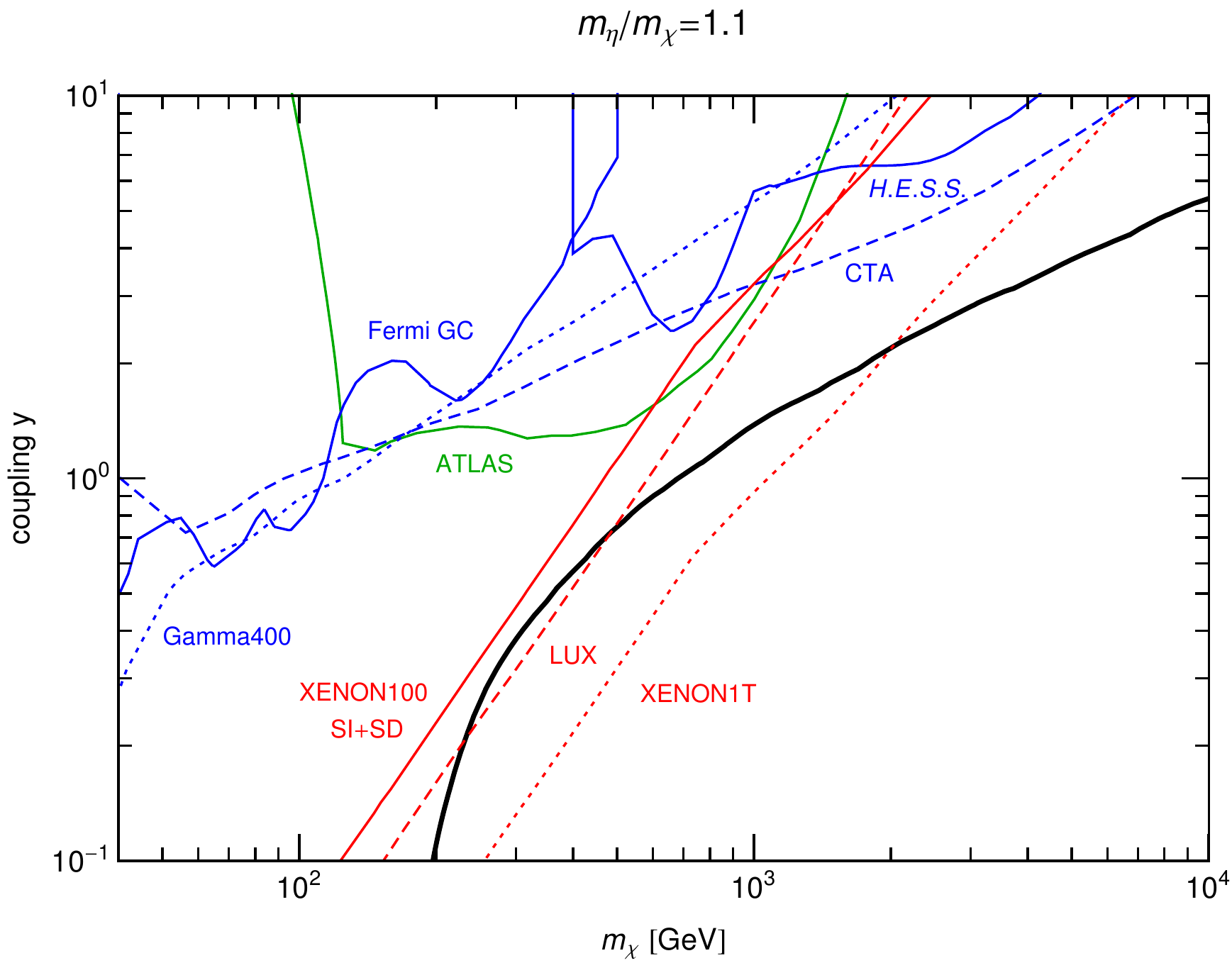}\\[1.5ex]
    \includegraphics[width=0.49\textwidth]{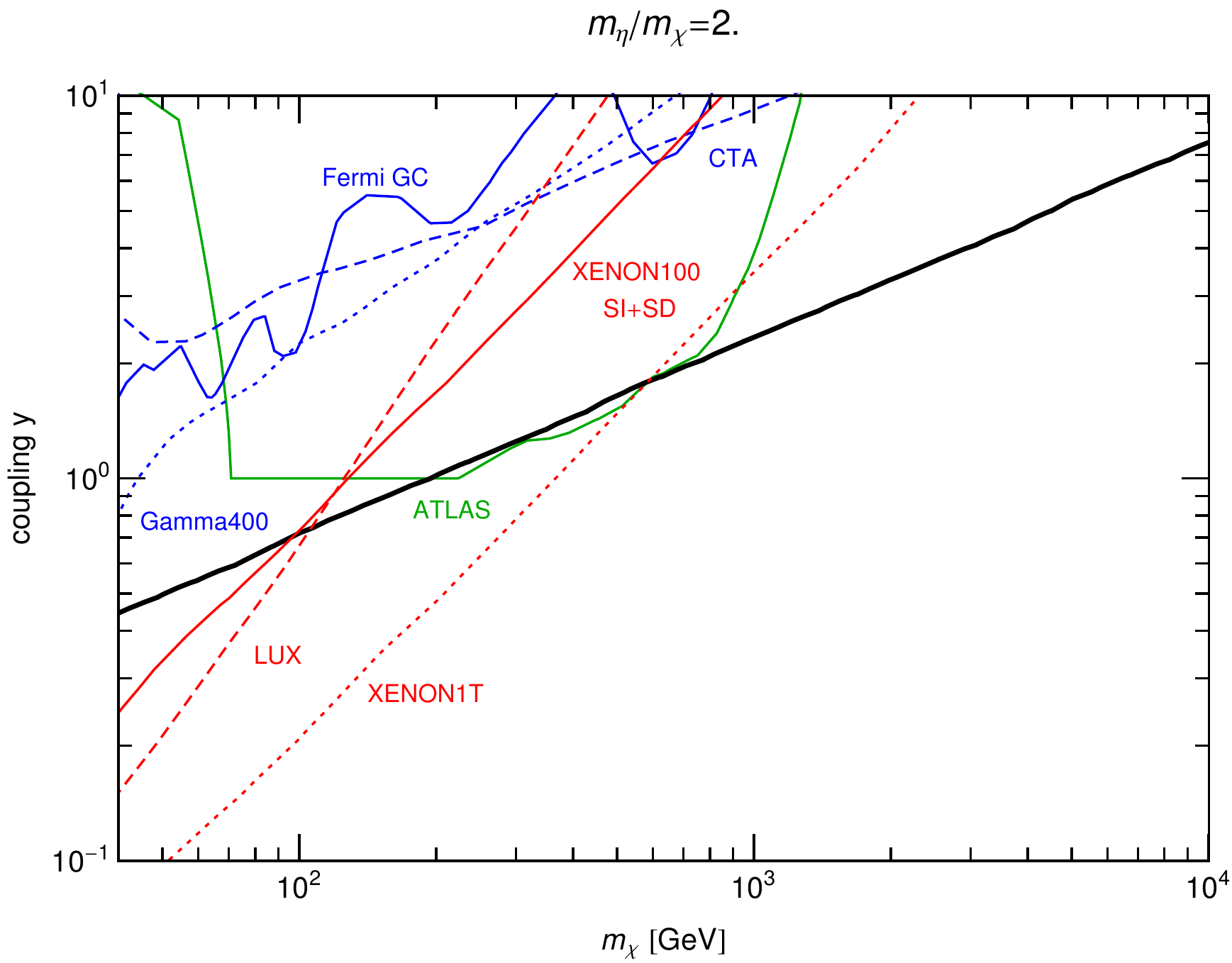}
    \includegraphics[width=0.49\textwidth]{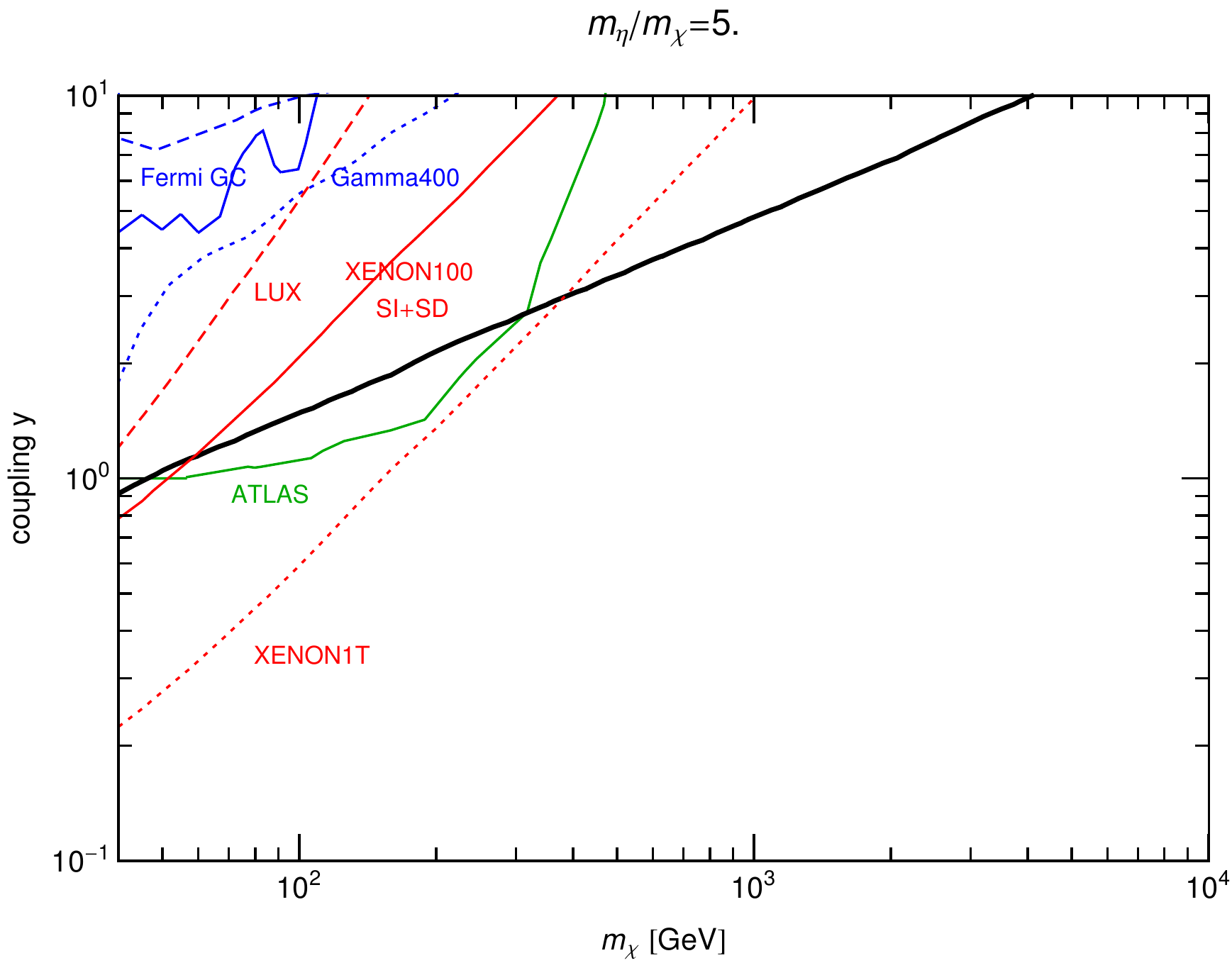}\\[1.5ex]
  \end{center}
  \caption{\label{fig:mDMvsCoupling} Compilation of current constraints on Majorana dark matter coupling to the right-handed up-quark, expressed in terms of the Yukawa coupling $y$, for various values of the mass splitting $m_\eta/m_\chi$. ATLAS constraints on jets and missing energy are shown in green, as well as direct detection (XENON100 in red, LUX red dashed) and indirect detection (Fermi, H.E.S.S. in blue). The red dotted lines shows prospects for XENON1T, blue dotted for GAMMA-400 and blue dashed for CTA, respectively. The thick black lines corresponds to thermal production.}
\end{figure}

\begin{figure}[t]
  \begin{center}
    \includegraphics[width=0.49\textwidth]{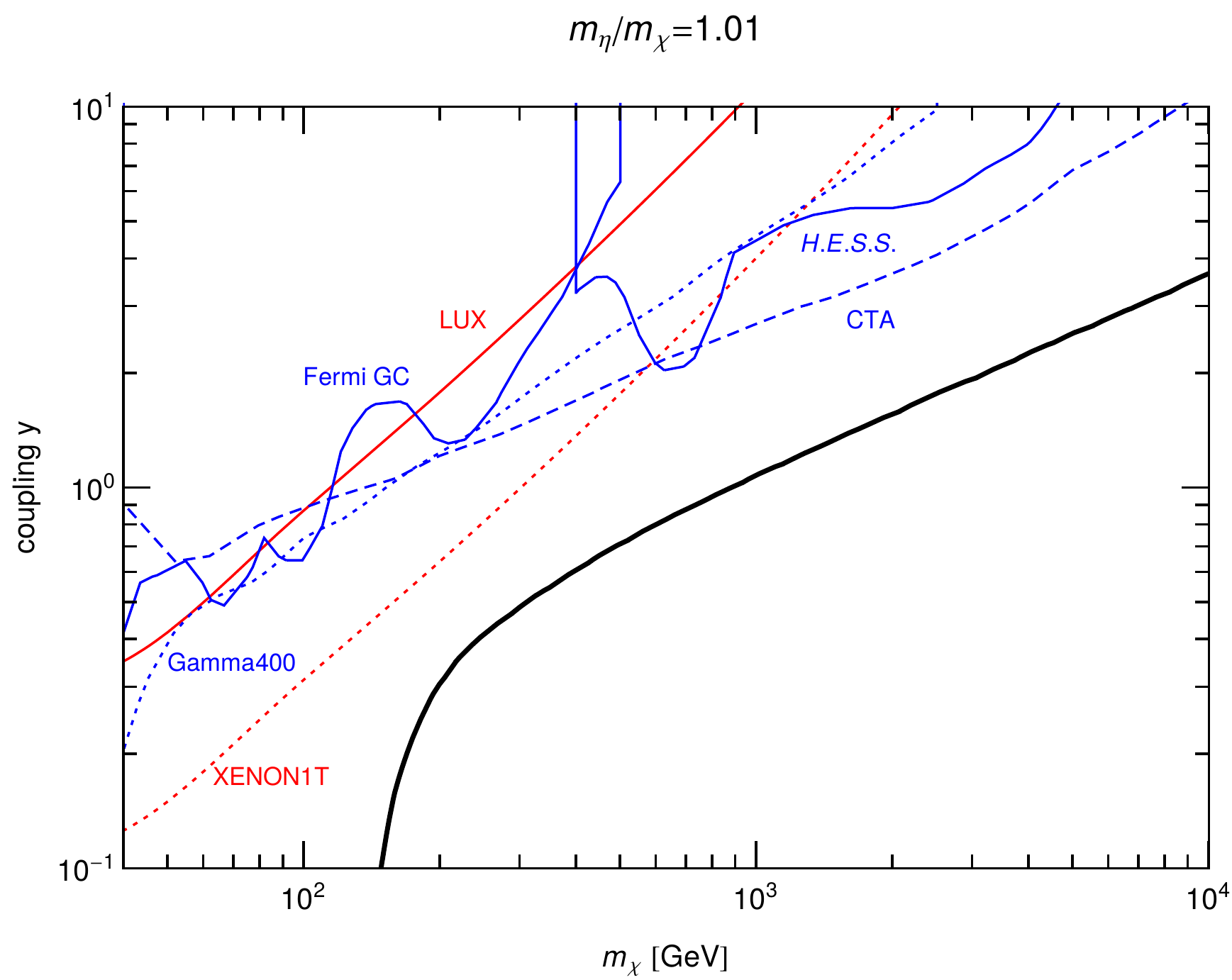}
    \includegraphics[width=0.49\textwidth]{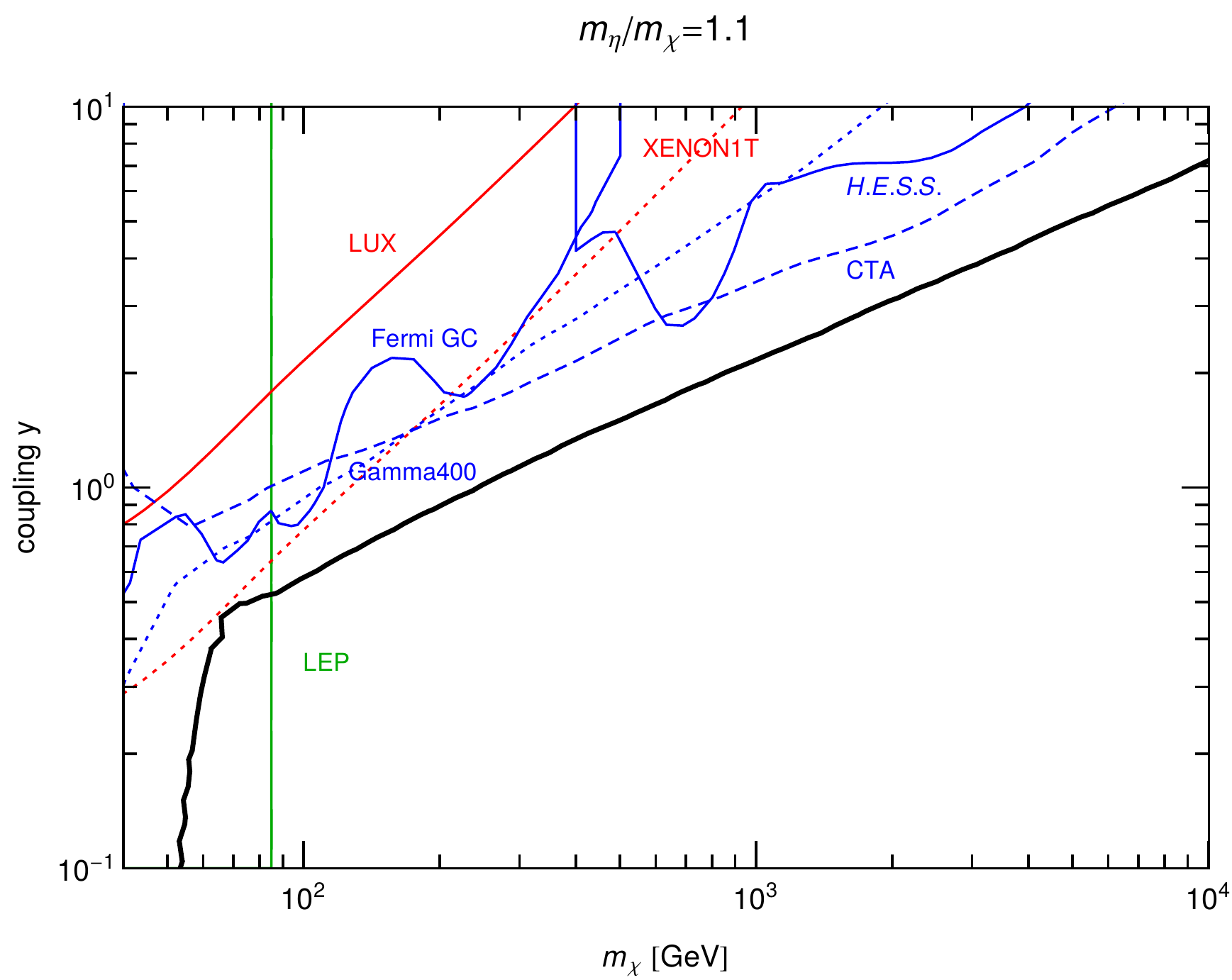}\\[1.5ex]
    \includegraphics[width=0.49\textwidth]{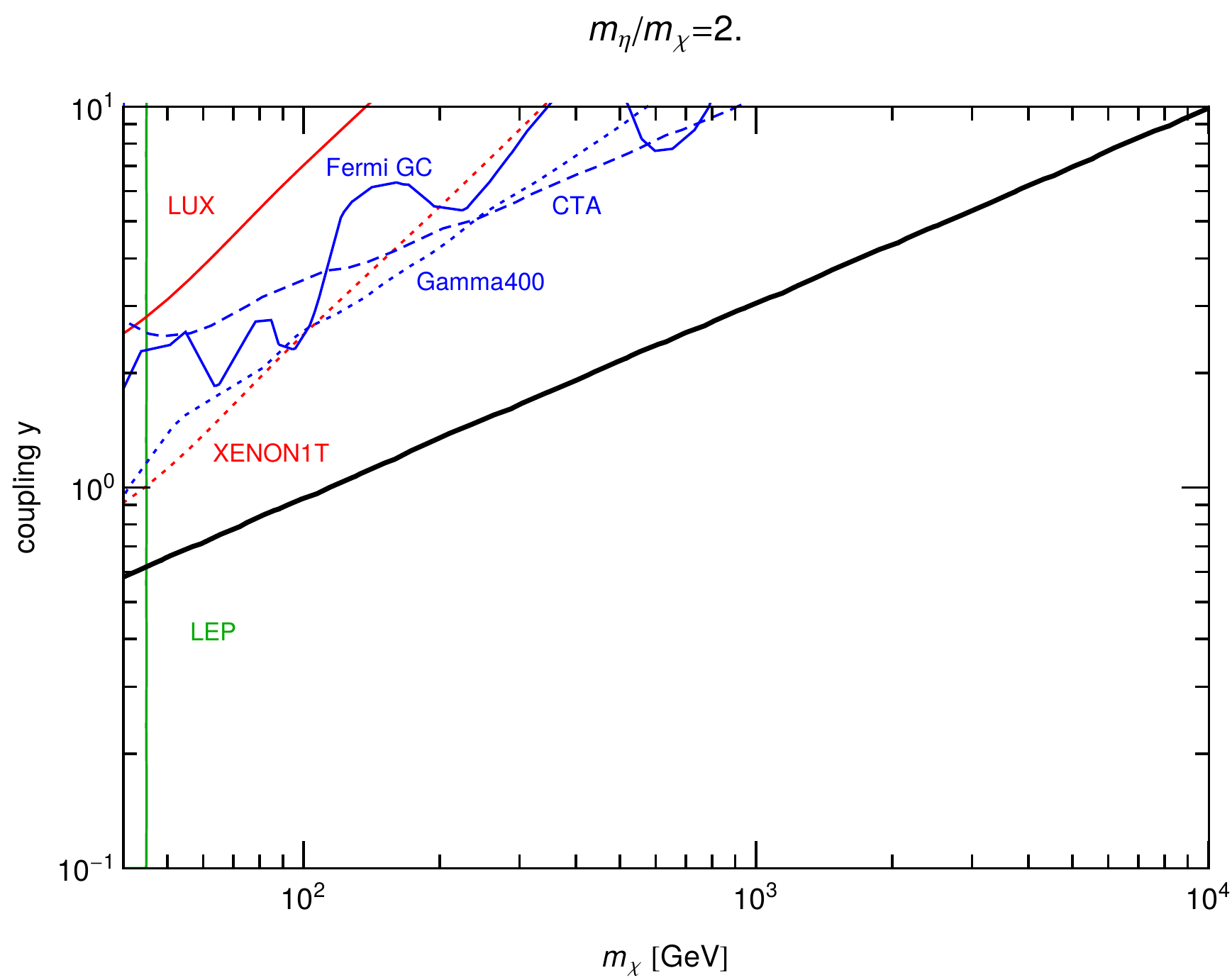}
    \includegraphics[width=0.49\textwidth]{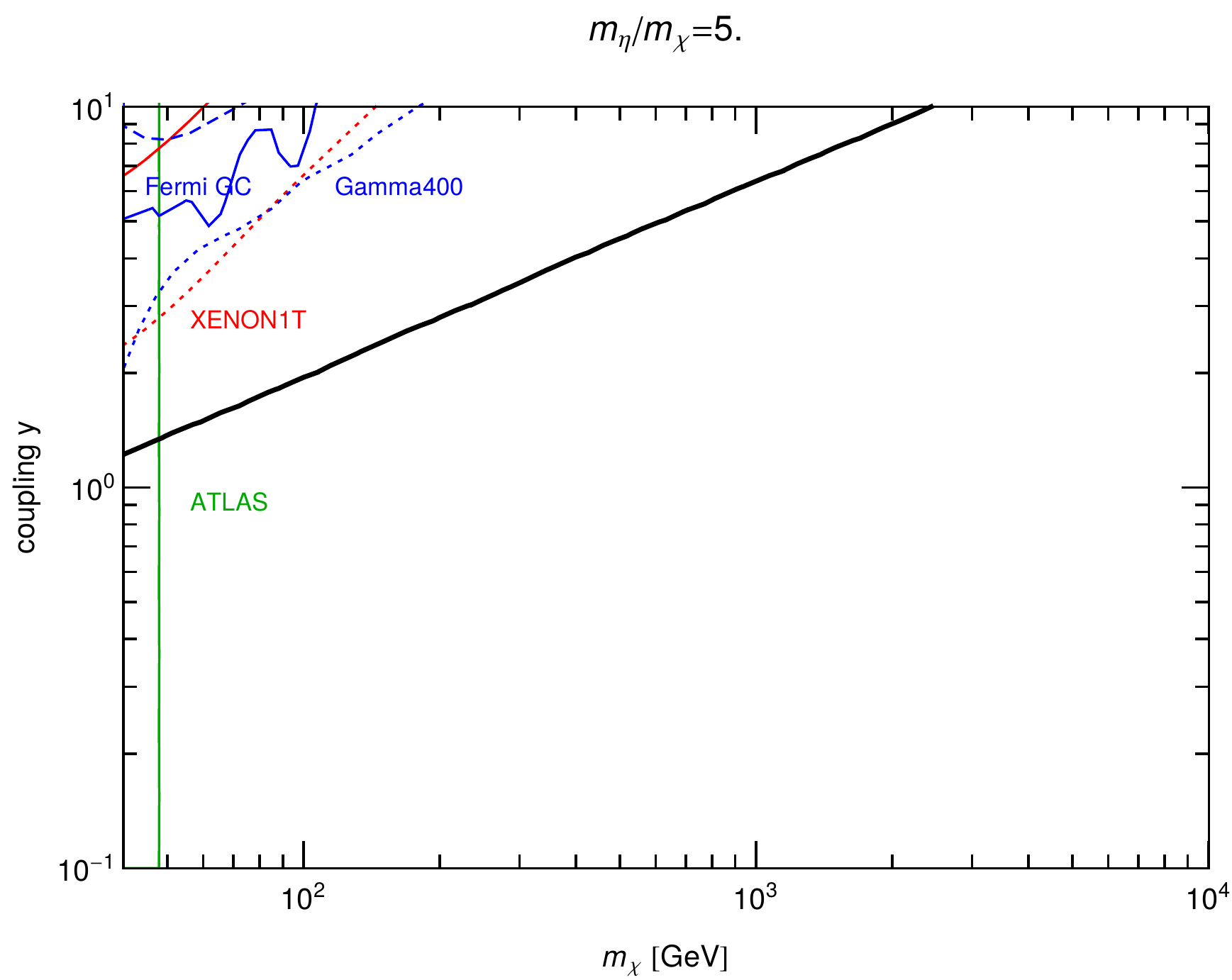}\\[1.5ex]
  \end{center}
  \caption{\label{fig:mDMvsCouplingMuR} Same as Fig.\,\ref{fig:mDMvsCoupling}, but for coupling to the right-handed muon.}
\end{figure}

In Fig.\,\ref{fig:mDMvsCoupling}, we show a compilation of constraints and prospects discussed in the previous sections, for the case of a colored mediator. All cross sections have been converted into limits on the Yukawa coupling strength $y$, and the thick black line shows the coupling $y_{th}(m_\chi,m_\eta)$ for which thermal freeze-out yields a relic abundance $\Omega_\chi h^2\simeq 0.12$. For very small mass splitting $m_\eta/m_\chi=1.01$ direct detection constraints from LUX and XENON100 are by far dominant due to the resonant enhancement of the scattering cross section in this limit. LUX limits exclude thermally produced dark matter for a small range of masses between $1-2$\,TeV, and XENON1T will probe deeply into the region of thermal production. For mass splitting $m_\eta/m_\chi=1.1$, direct detection constraints are much weaker, but still dominant
for $m_\chi\lesssim 1\,$TeV. For higher masses, gamma ray constraints from internal bremsstrahlung are formally the best constraint, although at unrealistically high values of the Yukawa coupling. Constraints from jets and missing energy at LHC are comparable to LUX constraints around $m_\chi \approx 0.5-1\,$TeV. For even larger mass splitting $m_\eta/m_\chi=2$, the overall picture changes and collider constraints become dominant over the whole range of masses $m_\chi\gtrsim 100\,$GeV. This is due to two effects: first, there is no resonant enhancement of the direct detection cross section in this regime. Second, the larger mass splitting leads to more energetic jets in the decay $\eta\to\chi q$ which make the discrimination against background more efficient. The limits inferred from the ATLAS search for jets and missing energy are slightly stronger than the coupling required for thermal production for $m_\chi\approx 200-600\,$GeV. Constraints from internal bremsstrahlung also become weaker, because this process is suppressed for large mass splitting. For very large splitting $m_\eta/m_\chi=5$, collider constraints are dominant as well, and exclude thermally produced dark matter for $m_\chi \lesssim 300\,$GeV.

The constraints shown in Fig.\,\ref{fig:mDMvsCoupling} correspond to a coupling to right-handed up-type quarks. Nevertheless, the qualitative picture remains the same for a coupling to quarks of the first two generations \cite{Garny:2012eb,Garny:2014waa}. For example, for a coupling to down-type quarks the direct detection constraints remain essentially unaffected, the collider constraints weaken only slightly due to the dependence of the production cross section on the parton distribution in the proton, and limits from internal bremsstrahlung weaken by a factor $\sqrt{2}$ due to the smaller electric charge of down-type quarks. Similarly, an extension to a model containing two mediators which couple to the up- and charm-quark, respectively, leaves collider constraints essentially unaffected, because the dominant production process for Majorana dark matter is sensitive to the valence quarks \cite{Garny:2014waa}. Note however that the presence of the second mediator influences thermal freeze-out and therefore affects the coupling for which the density matches the Planck value \cite{Planck:2015xua}. It is also important to note that collider and direct detection constraints are considerably weaker for a coupling to the third generation, while indirect constraints are identical in the limit $m_\chi \gg m_{t/b}$ \cite{Garny:2012eb, Aad:2014vea, Ibarra:2015nca}.

For an uncolored mediator which couples dark matter to leptons, constraints from indirect detection are dominant for $m_\chi\gtrsim 100\,$GeV at present. Therefore, searches for a spectral feature from internal bremsstrahlung represent an important probe of this class of models. The corresponding constraints and prospects are summarized in Fig.\,\ref{fig:mDMvsCouplingMuR}.

\subsection{Complementarity for thermal production}

When assuming that the Majorana particle $\chi$ constitutes the dominant form of cold dark matter, and that it was produced via thermal freeze-out, one parameter of the model can be fixed by requiring $\Omega_\chi h^2=0.12$. Using this constraint to fix the Yukawa coupling strength, $y=y_{th}(m_\chi,m_\eta)$, the parameters are completely specified by the dark matter and the mediator mass, respectively. As discussed in Sec.\,\ref{sec:Relic}, requiring thermal production as well as perturbative couplings implies that only a finite region in this parameter space is theoretically viable. In the following we discuss the current experimental constraints, and to which extent the viable region will be covered in the near future. 

\begin{figure}
  \begin{center}
    \includegraphics[width=0.9\textwidth]{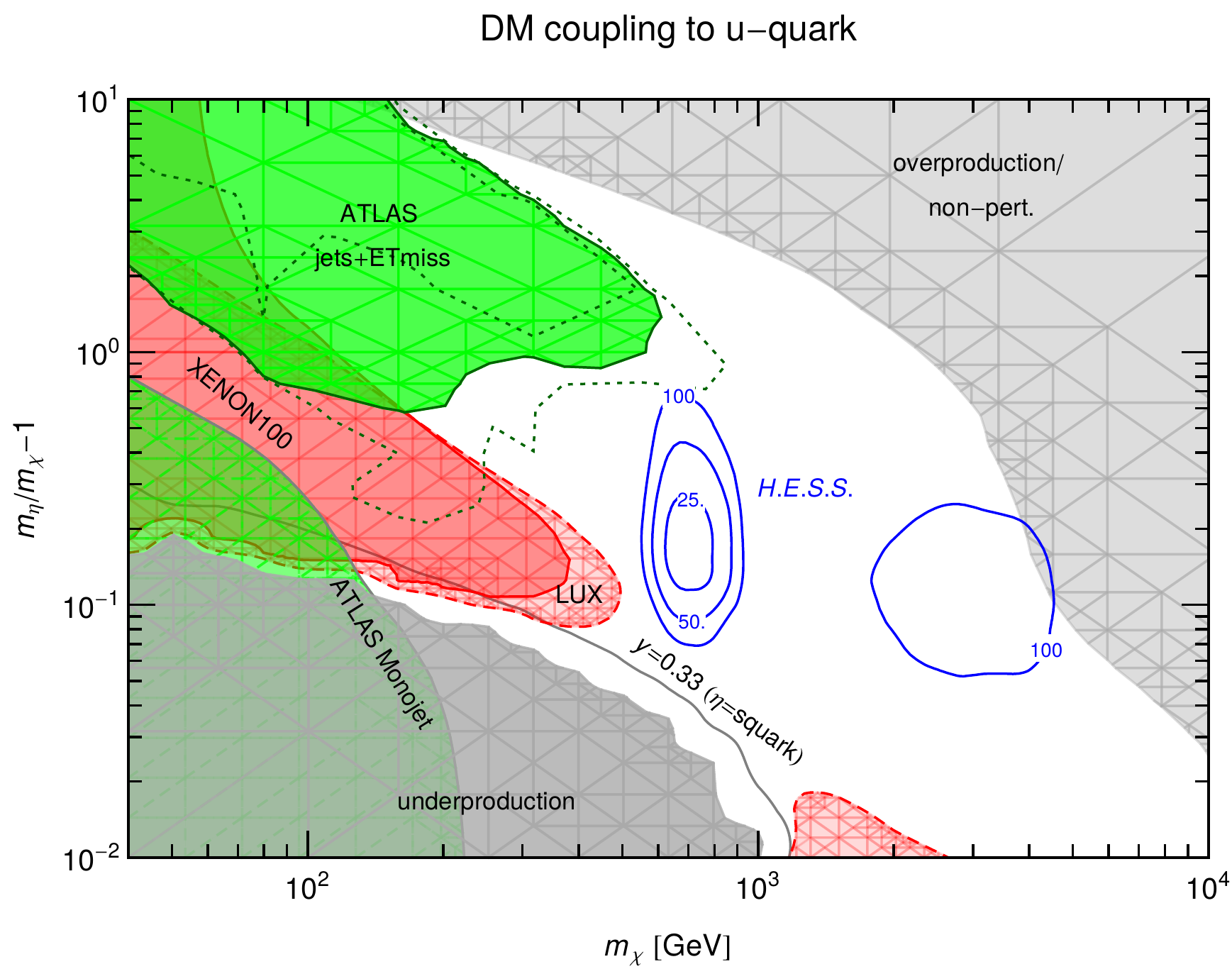}\\[1.5ex]
    \includegraphics[width=0.9\textwidth]{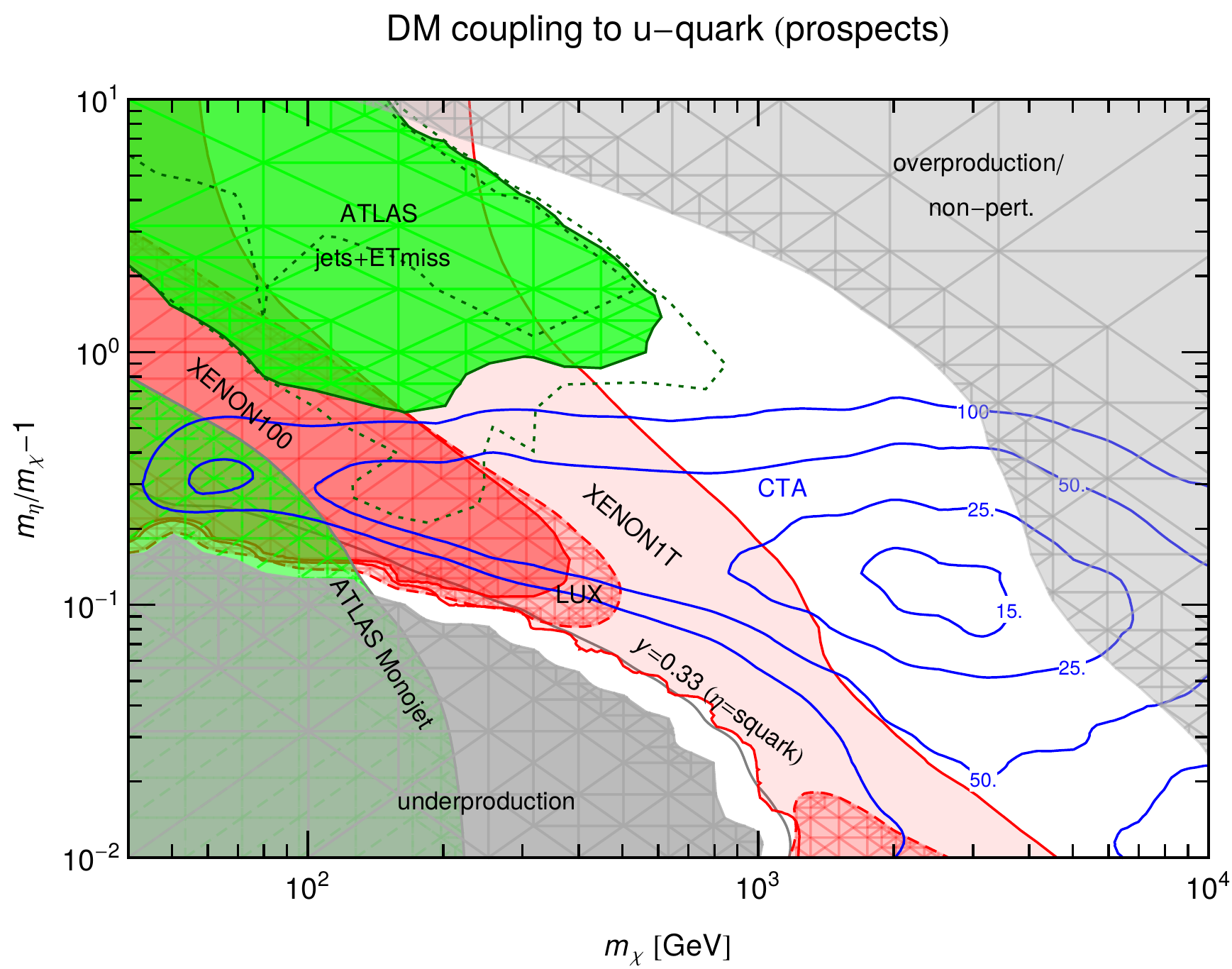}
  \end{center}
  \caption{\label{fig:quarks} Compilation of current constraints from collider searches, direct detection (XENON100, LUX) and indirect detection (H.E.S.S.) on thermally produced Majorana dark matter with colored mediator, coupling to right-handed up quarks. The lower panel shows prospects for direct and indirect constraints achievable with XENON1T and CTA, respectively.}
\end{figure}

For a colored mediator, collider searches as well as direct detection constraints exclude dark matter masses up to $m_\chi=600\,$GeV,
see Fig.\,\ref{fig:quarks} (upper panel). However, the exclusion regions are almost fully complementary: constraints from jets and missing energy are
relevant for mass splitting $m_\eta/m_\chi\gtrsim 2$, and those from direct detection for smaller splittings. As discussed previously, this is due to an
interplay of a resonant enhancement of the scattering cross section off nuclei on the one hand, and the efficiency of producing jets with large transverse momentum
on the other hand. We also show monojet constraints which are sensitive to the quasi-degenerate region \cite{Dreiner:2012sh}. Gamma ray constraints from internal bremsstrahlung at present are sensitive to couplings larger than those required for thermal production, when assuming an Einasto profile. The blue contour lines shown in Fig.\,\ref{fig:quarks} show the ratio of the excluded annihilation cross section to the thermal cross section. The regions inside the contour lines are excluded if the gamma ray flux from annihilation was enhanced relative to the Einasto profile by a corresponding factor, either due to a cuspier profile or due to substructures. 

The lower panel of Fig.\,\ref{fig:quarks} shows prospects for XENON1T and CTA. The former experiment will be able to probe a significant fraction of the theoretically viable parameter space in this model. Depending on the control of systematic uncertainties, CTA will be significantly more sensitive than H.E.S.S.. Furthermore, the region of highest sensitivity, in the multi-TeV region and for small mass splitting of order of tens of percent, will not be covered by direct or collider searches. Therefore, CTA will provide complementary information, although a detection may only be expected for optimistic assumptions on the dark matter distribution in this model.

\begin{figure}
  \begin{center}
    \includegraphics[width=0.9\textwidth]{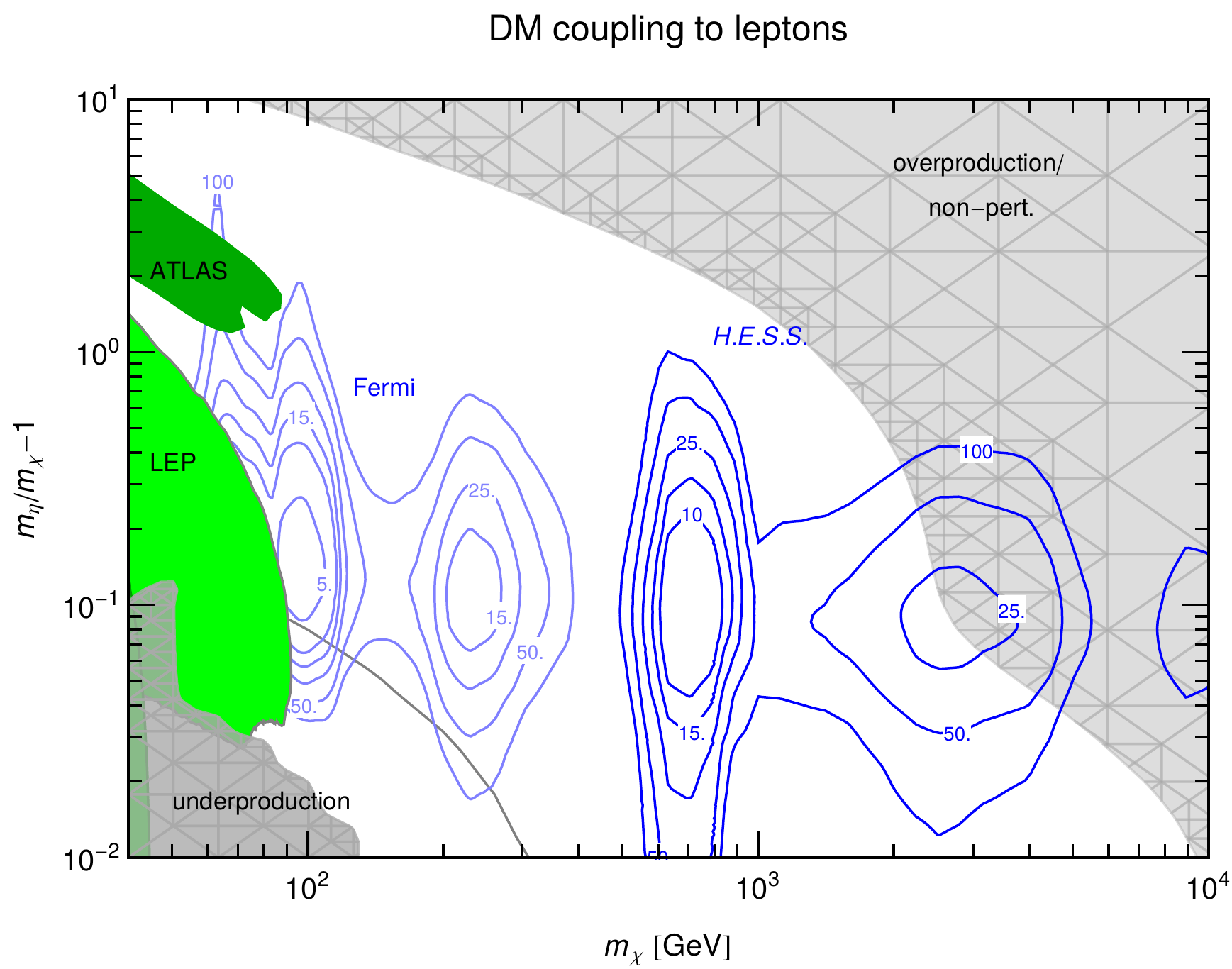}\\[1.5ex]
    \includegraphics[width=0.9\textwidth]{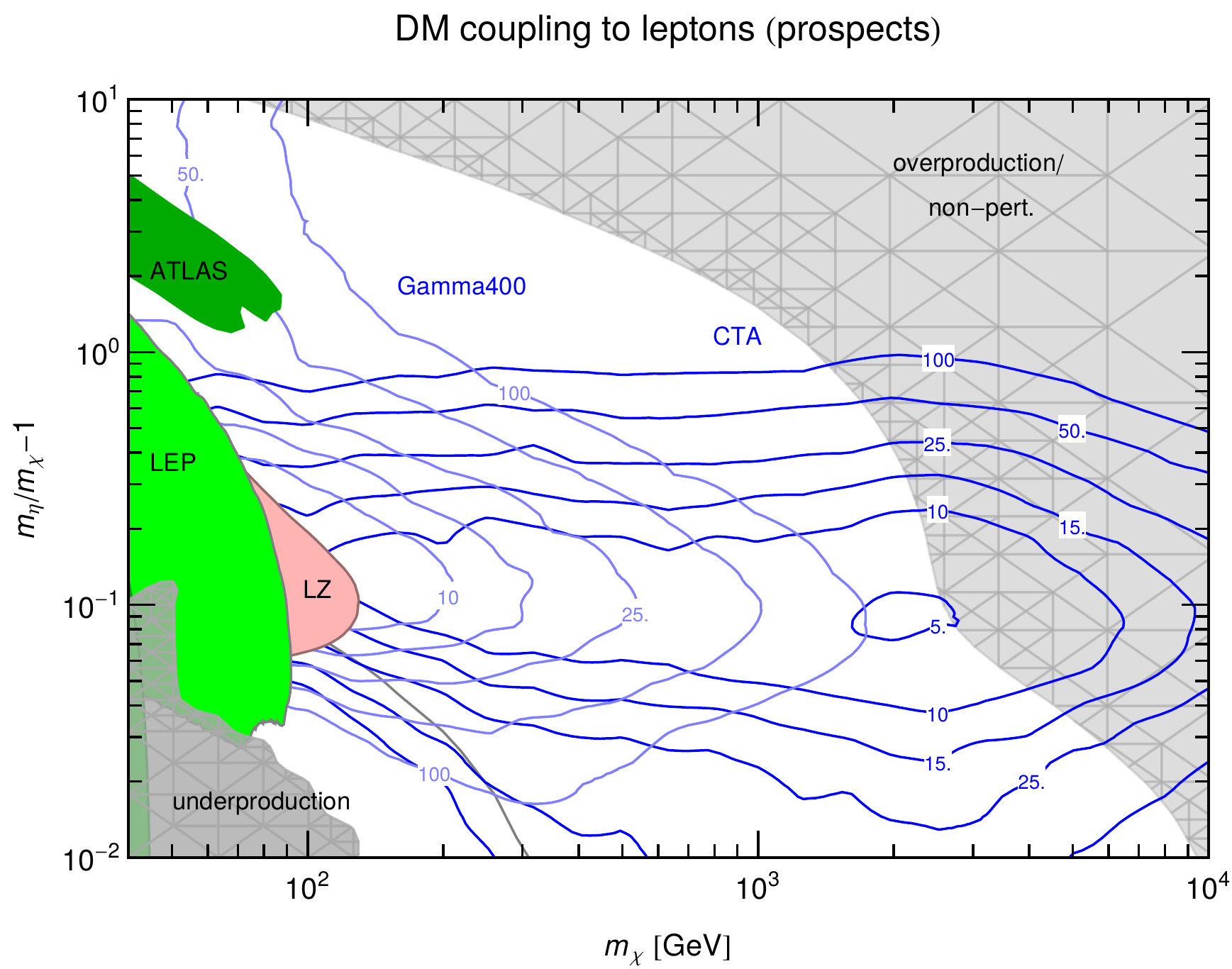}
  \end{center}
  \caption{\label{fig:leptons} Compilation of current constraints and prospects for thermally produced Majorana dark matter with uncolored mediator, coupling to right-handed muons.}
\end{figure}

For the case of an uncolored mediator, constraints and prospects are shown in Fig.\,\ref{fig:leptons}. Collider constraints from LEP and LHC partly exclude the region $m_\chi<100\,$GeV. The blue contour lines correspond to gamma ray constraints similarly as before. The reason why the sensitivity peaks at a mass splitting of ${\cal O}(10\%)$ is due to an interplay of two effects: first, the strength of the internal bremsstrahlung feature increases with smaller splitting. Second, the thermal coupling $y_{th}(m_\chi,m_\eta)$ decreases for $m_\eta\to m_\chi$, due to coannihilations. Although dark matter couples only to leptons at tree-level in this scenario, it may be possible to probe a small region of the parameter space with the future direct detection experiment LUX-ZEPLIN, due to the loop-induced anapole moment \cite{Kopp:2014tsa}. Nevertheless, indirect detection will remain the most sensitive probe for large parts of the parameter space.

\section{Conclusions}\label{sec:conclusions} 

We have reviewed the phenomenology of a model where the dark matter particle is a Majorana fermion that couples to a light Standard Model fermion via a Yukawa coupling with a scalar mediator. The simplest scenario contains only three free parameters, the dark matter mass, the mediator mass and the Yukawa coupling,  and leads, for appropriate choices of the parameters, to the observed dark matter abundance via thermal freeze-out. The small number of free parameters of the model allows us to systematically scan the parameter space, explore its rich phenomenology, and investigate the complementarity among the various search strategies. The most characteristic aspects of the model, and which make its phenomenology rather unique, are:
\begin{itemize}
\item  Requiring dark matter thermal production and perturbativity of the couplings leads to a finite parameter space, that could potentially be closed by experiments.
\item The annihilation process $\chi\chi \rightarrow f\bar f$ controls the dark matter abundance, however, it has a negligible cross section at present times. Instead, the dominant annihilation channels today are the higher order two-to-three process $\chi\chi\rightarrow f\bar f V$ and the one loop process $\chi\chi\rightarrow V V'$, with $V,V'$ gauge bosons.
\item When the dark matter particle and the scalar mediator are very degenerate in mass, the annihilation process $\chi\chi\rightarrow f\bar f \gamma$ leads to a sharp gamma-ray spectral feature which resembles a distorted gamma-ray line, and which could be detected in gamma-ray telescopes. The detection of such sharp feature in the gamma-ray sky would unambiguously point to a dark matter origin.
\item The tree level annihilation processes $\chi\chi\rightarrow f\bar f Z$ and the one loop processes $\chi\chi\rightarrow Z Z$ and $\chi\chi\rightarrow \gamma Z$ necessarily have a non-vanishing cross section if the dark matter particle couples to a Standard Model fermion. These processes lead to an exotic contribution to the antiproton flux, even for leptophilic dark matter particles.
\item For dark matter particles coupling to light quarks, the scattering rate with nucleons is enhanced in the mass degenerate limit, due to the resonant $s$-channel exchange of the scalar mediator. Away from the resonance, the spin-independent rate is highly suppressed since the coefficients of the dimension six operators vanish. Therefore, in this regime, the spin dependent scattering ought not be neglected. In fact, current experiments probing the spin-dependent interaction yield the dominant constraint over a wide region of the parameter space.
\item For dark matter particles coupling to quarks, the colored mediator plays a pivotal role in colliders, since it could be light enough to be directly produced in  parton collisions. For a Majorana dark matter particle which couples to a light quark, the total production cross section can be largely enhanced due to the contribution from quark-quark collisions. After being produced, the scalar mediator decays into a dark matter particle and a quark, which produces a jet. When the mass difference between the dark matter particle and the scalar mediator is large, the final state consists of two or more jets plus missing energy. On the other hand, when the mass difference is small,  the produced jets are too soft to be triggered and then the final state is invisible. However the process could be detected from the emission of a gluon off any of the colored states involved, producing a monojet signal. In either case, a correct description of the phenomenology of the model at colliders requires the inclusion of the colored mediator.
\end{itemize}

Within the scenario of a colored mediator, the theoretically viable parameter space for a thermally produced dark matter particle will be probed to a large extent by next-generation experiments, while the scenario of an uncolored mediator is much more challenging. Indirect, direct and collider constraints are covering largely complementary regions of parameter space, depending both on the dark matter mass and the mediator mass. Consequently, a null result at the next LHC run  would not preclude the possibility to observe a signal at XENON1T. Similarly, a negative result by XENON1T would leave open the possibility to observe a signal from internal bremsstrahlung at CTA. Nevertheless, in an optimistic case, a signal could be observed in more than one channel if the mass splitting between the mediator and the dark matter is of order one, and a combination of the characteristic signatures mentioned above would allow to pinpoint this dark matter model.

\section{Acknowledgments}
The work of AI was partially supported by the DFG cluster of excellence ``Origin and Structure of the Universe.'' We are grateful to Torsten Bringmann, Xiaoyuan Huang, Miguel Pato, Aaron Pierce, Sara Rydbeck, Nausheen Shah, Maximilian Totzauer, Andreas Weiler, Christoph Weniger and Sebastian Wild for discussions and collaborations.

\appendix

\section{Annihilation cross sections}
\label{ap:crosssections}
We include in this appendix the expressions for the relevant annihilation and coannihilation cross sections in the toy model discussed in Section \ref{sec:toy-model} in the limit $m_f\rightarrow 0$ and keeping the lowest order in the expansion in the relative dark matter velocity $v$ (unless otherwise specified). In the following formulas, $y$ is the Yukawa coupling between the Majorana dark matter particle $\chi$, the scalar $\eta$, and the right-handed Standard Model fermion $f_R$, which has electric charge $q_f$ and color $N_C$. The cross-sections for annihilations into $f_R\bar f_R$ were calculated in  Ref.~\citen{Cao:2009yy}, for $\gamma \gamma$ and $g g$ via a one-loop diagram in Refs.~\citen{Bergstrom:1997fh,Bern:1997ng}, for $\gamma Z$ in Ref.~\citen{Ullio:1997ke}, for $Z Z$ in Ref.~\citen{Ibarra:2014vya}, for  $f_R\bar f_R \gamma$ in Refs.~\citen{Bergstrom:1989jr,Flores:1989ru}, for $f_R\bar f_R Z$ in Refs.~\citen{Ciafaloni:2011sa,Bell:2011if,Garny:2011ii} and for $f_R\bar f_R g$ in Ref.~\citen{Flores:1989ru}. The cross sections for coannihilations were for example derived in Ref.~\citen{Garny:2012eb}.

\subsubsection*{Two-to-two annihilation into fermions}
\label{app:TwoToTwoAnnihilation}

 \begin{align}
\sigma v_{f\bar f} &= \frac{N_C \, y^4 \sqrt{1-4m_f^2/s}}
{32\, \pi \,  
   \left(m_{\eta}^2-m_f^2+m_{\chi }^2\right)^2} \Bigg( m_f^2 + \frac{ v^2 }{12 \left(m_{\eta}^2-m_f^2+m_{\chi }^2\right)^2}\Big( 8 m_\chi^6 -13 m_\chi^4m_f^2 \nonumber\\
& {} -5m_f^2(m_\eta^2-m_f^2)^2 +2m_\chi^2(4m_\eta^4-11m_\eta^2m_f^2+5m_f^4)\Big) + {\cal O}(v^4) \Bigg)
 \end{align}
where $s=4E_\chi^2=4\frac{m_\chi^2}{1-\frac{v^2}{4}}$. Note that we did not expand the square root in the velocity to avoid a singular behaviour that occurs for the particular case $1-m_f^2/m_\chi^2<v^2$. If this is not the case, the square root can be expanded in the velocity as well.

\subsubsection*{Two-to-two annihilations into gauge bosons via loops}
\label{app:TwoToTwoLoopsAnnihilation}

\begin{align}
(\sigma v)_{\gamma \gamma} &= \frac{N_C^2 \, q_f^4 \, \alpha_{\text{em}}^2 \, y^4}{256 \pi^3 \, m_{\chi}^2} \left[ \text{Li}_2 \left( -\frac{m_{\chi}^2}{m_{\eta}^2} \right) - \text{Li}_2 \left( \frac{m_{\chi}^2}{m_{\eta}^2} \right)\right]^2 \;,\\
(\sigma v)_{gg} &= \frac{2 \, \alpha_{\text{s}}^2 \, y^4}{256 \pi^3 \, m_{\chi}^2} \left[ \text{Li}_2 \left( -\frac{m_{\chi}^2}{m_{\eta}^2} \right) - \text{Li}_2 \left( \frac{m_{\chi}^2}{m_{\eta}^2} \right)\right]^2 \;,\\
(\sigma v)_{\gamma \text{Z}} &= \frac{\left| \mathcal{A}_{\gamma \text{Z}} \right|^2}{512 \, \pi^3 \, m_{\chi}^6 \, m_{\eta}^4 \left(1-\frac{m_{\text{Z}}^2}{4 \, m_{\chi}^2} \right) \left( 1-\frac{m_{\text{Z}}^4}{16 \, m_{\eta}^4} \right)^2}\;, \\
(\sigma v)_{\text{Z} \text{Z}} &= \frac{\left| \mathcal{A}_{\text{Z} \text{Z}} \right|^2}{1024 \, \pi^3 \, m_{\chi}^6 \, m_{\eta}^4 \, \sqrt{1-\frac{m_{\text{Z}}^2}{m_{\chi}^2}}}\;,
\end{align}
with ${\rm Li}_2(x)=\sum_{k=1}^\infty z^k/ k^2$ the dilogarithm function, while $\mathcal{A}_{\gamma \text{Z}}$ and  $\mathcal{A}_{\text{Z} \text{Z}}$ are defined by
\begin{align}
\mathcal{A}_{\gamma \text{Z}} &= N_C \, q_f^2 \, \alpha_{\text{em}} \, y^2 \, \tan \left( \theta_{\text{W}} \right) \left(1-\frac{m_{\text{Z}}^2}{4 \,m_{\eta}^2}\right) \Bigg\{ \nonumber \\
& m_{\text{Z}}^2 \left(\frac{m_{\chi}^4+m_{\eta}^4}{2}+\frac{m_{\text{Z}}^2 \left(m_{\eta}^2-m_{\chi}^2\right)}{4}+\frac{m_{\text{Z}}^4}{16}\right) \, \, \text{C}_0 \left(m_{\chi}^2,m_{\text{Z}}^2,\frac{m_{\text{Z}}^2}{2}-m_{\chi}^2,m_{\eta}^2,0,0\right) \nonumber \\
& + 2 \, m_{\eta}^2 \left(m_{\chi}^2-\frac{m_{\text{Z}}^2}{4}\right)^2 \text{C}_0 \left(m_{\chi}^2,0,\frac{m_{\text{Z}}^2}{2}-m_{\chi}^2,0,m_{\eta}^2,m_{\eta}^2\right) \nonumber\\
& +\left(m_{\eta}^2+\frac{m_{\text{Z}}^2}{4}\right) \left(2 m_{\chi}^4-m_{\chi}^2 m_{\text{Z}}^2+\frac{m_{\eta}^2 m_{\text{Z}}^2}{2}\right) \text{C}_0 \left(m_{\chi}^2,m_{\text{Z}}^2,\frac{m_{\text{Z}}^2}{2}-m_{\chi}^2,0,m_{\eta}^2,m_{\eta}^2\right) \nonumber\\
& + \, \frac{m_{\text{Z}}^2}{2} \left(m_{\eta}^2+\frac{m_{\text{Z}}^2}{4}\right) \, \, \left[2 \, \sqrt{\frac{4 m_{\eta}^2}{m_{\text{Z}}^2}-1} \, \, \text{arccot} \left(\sqrt{\frac{4 m_{\eta}^2}{m_{\text{Z}}^2}-1}\right)-\log \frac{m_{\text{Z}}^2}{m_{\eta}^2}+i \pi \right] \Bigg\} \, ,
\end{align}
\begin{align}
\mathcal{A}_{\text{ZZ}} &=  N_C \, q_f^2 \, \alpha_{\text{em}} \, y^2 \, \tan^2 \left( \theta_{\text{W}} \right) \Bigg\{ m_{\text{Z}}^2 \left(m_{\chi}^4+m_{\eta}^4-m_{\chi}^2 m_{\text{Z}}^2\right) \text{C}_0 \left(m_{\chi}^2,m_{\text{Z}}^2,-m_{\chi}^2+m_{\text{Z}}^2,m_{\eta}^2,0,0\right) \nonumber \\
& +\left[4 m_{\chi}^4 m_{\eta}^2+\left(-m_{\chi}^4-4 m_{\chi}^2 m_{\eta}^2+m_{\eta}^4\right) m_{\text{Z}}^2+m_{\chi}^2 m_{\text{Z}}^4\right] \text{C}_0 \left(m_{\chi}^2,m_{\text{Z}}^2,-m_{\chi}^2+m_{\text{Z}}^2,0,m_{\eta}^2,m_{\eta}^2\right) \nonumber \\
& + m_{\eta}^2 \, m_{\text{Z}}^2 \left[2 \sqrt{\frac{4 m_{\eta}^2}{m_{\text{Z}}^2}-1} \, \, \text{arccot} \left(\sqrt{\frac{4 m_{\eta}^2}{m_{\text{Z}}^2}-1}\right)-\text{log} \, \frac{m_{\text{Z}}^2}{m_{\eta}^2} +i \pi \right] \Bigg\} \, ,
\end{align}
$C_0$ being a Passarino-Veltman function. These expressions satisfy 
\begin{align}
\left( \sigma v \right)_{\text{Z} \text{Z}} \big|_{\tan \left( \theta_{\text{W}} \right) \equiv 1} \stackrel{m_{\text{Z}} \rightarrow 0}{\longrightarrow} \left( \sigma v \right)_{\gamma \gamma},~~~~~
\left( \sigma v \right)_{\gamma \text{Z}} \big|_{\tan \left( \theta_{\text{W}} \right) \equiv 1} \stackrel{m_{\text{Z}} \rightarrow 0}{\longrightarrow} 2 \, \left( \sigma v \right)_{\gamma \gamma} \,.
\end{align}

\subsubsection*{Two-to-three annihilations}
\label{app:TwoToThreeAnnihilations}

\begin{align}
\frac{d(\sigma v)_{f_R\bar f_R \gamma}}{dE_\gamma dE_f} & = \frac{q_f^2 N_C\alpha_{em}y^4 \left(1-\frac{E_\gamma}{m_\chi}\right) \left[\left(\frac{E_\gamma}{m_\chi}\right)^2-2\frac{E_\gamma}{m_\chi} \left(1-\frac{E_f}{m_\chi}\right)+2 \left(1-\frac{E_f}{m_\chi}\right)^2 \right]}{8\pi^2 m_\chi^4 \left(1-2\frac{E_f}{m_\chi}-\frac{m_\eta^2}{m_\chi^2}\right)^2 \left(3-2\frac{E_\gamma}{m_\chi}-2\frac{E_f}{m_\chi}+\frac{m_\eta^2}{m_\chi^2}\right)^2} \;, \\
\frac{d(\sigma v)_{f_R\bar f_R Z}}{dE_Z dE_f} & =  \frac{q_f^2N_c\tan^2(\theta_W)\alpha_{em}y^4 }{8\pi^2 m_\chi^4 \left(1-2\frac{E_f}{m_\chi}-\frac{m_\eta^2}{m_\chi^2}\right)^2 \left(3-2x-2\frac{E_f}{m_\chi}+\frac{m_\eta^2}{m_\chi^2}\right)^2}  \nonumber \\
& {} \times \Big\{ \left(1-\frac{E_Z}{m_\chi}\right)\left[\left(\frac{E_Z}{m_\chi}\right)^2-2\frac{E_Z}{m_\chi} \left(1-\frac{E_f}{m_\chi}\right)+2 \left(1-\frac{E_f}{m_\chi}\right)^2 \right] \nonumber \\
& {} + \left(\frac{m_Z}{m_\chi}\right)^2 \left[\left(\frac{E_Z}{m_\chi}\right)^2+2\left(\frac{E_f}{m_\chi}\right)^2+2\frac{E_Z}{m_\chi}\frac{E_f}{m_\chi} -4\frac{E_f}{m_\chi}\right]/4 \nonumber \\
&{} -\left(\frac{m_Z}{m_\chi}\right)^4/8 \Big \} \;, \\
\frac{d(\sigma v)_{f_R\bar f_R g}}{dE_\gamma dE_f} & =  \frac{\left(N_c^2-1\right)\alpha_{s}(m_\chi)y^4}{16\pi^2 m_\chi^4 \left(1-2\frac{E_f}{m_\chi}-\frac{m_\eta^2}{m_\chi^2}\right)^2\left(3-2\frac{E_g}{m_\chi}-2\frac{E_f}{m_\chi}+\frac{m_\eta^2}{m_\chi^2}\right)^2} \nonumber \\
&{} \times  \left(1-\frac{E_g}{m_\chi}\right) \left[\left(\frac{E_g}{m_\chi}\right)^2-2\frac{E_g}{m_\chi}(1-\frac{E_f}{m_\chi})+2\left(1-\frac{E_f}{m_\chi}\right)^2 \right]\;.
\end{align}
The spectra of gauge bosons are obtained by integrating the differential cross-section over the fermion energy, with integration limits given by $E_f^{\rm min/max}=m_\chi-(E_V\pm\sqrt{E_V^2-M_V^2})/2$. The total cross-section can be obtained by integrating over the remaining energy with limits $E_V^{\rm min}=M_V$ and $E_V^{\rm max}=m_\chi+M_V^2/(4m_\chi)$.  The corresponding expressions for annihilation into  left-handed fermions can be found in Ref.~\citen{Garny:2011ii}.

 For the case of annihilations into a fermion-antifermion pair and a photon, the total cross section is given by:
\begin{align}
  (\sigma v)_\text{3-body}
  \simeq  \frac{\alpha_\text{em} y^4 N_c q_f^2}{64\pi^2m_\chi^2}
  & \left\{ 
  (\frac{m_\eta^2}{m_\chi^2}+1) \left[ \frac{\pi^2}{6}-\ln^2\left( \frac{\frac{m_\eta^2}{m_\chi^2}+1}{2\frac{m_\eta^2}{m_\chi^2}} \right)
  -2\text{Li}_2\left( \frac{\frac{m_\eta^2}{m_\chi^2}+1}{2\frac{m_\eta^2}{m_\chi^2}} \right)\right]  \right. \nonumber \\
   &+ \left. \frac{4\frac{m_\eta^2}{m_\chi^2}+3}{\frac{m_\eta^2}{m_\chi^2}+1}+\frac{4\frac{m_\eta^2}{m_\chi^2}^2-3\frac{m_\eta^2}{m_\chi^2}-1}{2\frac{m_\eta^2}{m_\chi^2}}\ln\left(
  \frac{\frac{m_\eta^2}{m_\chi^2}-1}{\frac{m_\eta^2}{m_\chi^2}+1} \right)
  \right\}\;.
  \label{eqn:sv3}
\end{align}

\subsubsection*{Coannihilations}

\begin{align}
\sigma v(\chi\eta\to q g) & =  \frac{y^2g_s^2}{24\pi}\frac{1}{m_\eta(m_\eta+m_\chi)}\left(1-\frac{m_q^2}{(m_\eta+m_\chi)^2}\right)\,, \nonumber\\
\sigma v(\eta\bar\eta\to gg) & =  \frac{7g_s^4}{216 \pi m_\eta^2}\,, \nonumber\\
\sigma v(\eta\eta\to qq) & =  \frac{ y^4}{6\pi}\frac{m_\chi^2}{(m_\chi^2+m_\eta^2)^2}\,, \\
\sigma v(\eta\bar\eta\to q\bar q) & =  \frac{ y^4 m_q^2}{16\pi(m_\chi^2+m_\eta^2)^2}\nonumber\\
& + \frac{v^2}{432\pi} \left(\frac{g_s^4}{m_\eta^2}+ \frac{9y^4m_\eta^2}{(m_\chi^2+m_\eta^2)^2}-\frac{4y^2g_s^2}{m_\chi^2+m_\eta^2} \right) \,. \nonumber
\end{align}

\section{Effective operators for direct detection}
\label{app:DD}

We want to obtain the effective Lagrangian for the scattering with a parton, $\chi q \to \chi q$, and eventually for the scattering with a nucleon, by integrating out the scalar. Its equation of motion reads
\begin{align}
 (D_\mu D^\mu+m_\eta^2)\eta = -y\bar q_R\chi,\quad (D_\mu D^\mu+m_\eta^2)\eta^\dag = -y\bar\chi q_R \,.
\end{align}
By integrating out the scalar at tree level, one obtains the effective interaction Lagrangian
\begin{align}
{\cal L}_{eff} = y^2 \bar\chi q_R \frac{1}{D_\mu D^\mu + m_\eta^2} \bar q_R \chi + \mbox{h.c.}
\end{align}
The momentum of the scalar in the scattering is given by $p_\eta^2=(p_\chi+p_q)^2\approx (m_\chi+m_q)^2$.
Therefore, we would like to expand the denominator around
\begin{align}
\Delta m^2 \equiv m_\eta^2 - (m_\chi+m_q)^2\;.
\end{align}
Formally,
\begin{align}
  \frac{1}{D_\mu D^\mu + m_\eta^2} \bar q_R \chi = \sum_{n=0}^\infty \frac{(-D_\mu D^\mu-(m_\chi+m_q)^2)^n}{(\Delta m^2)^{n+1}}\bar q_R \chi 
\end{align}
Using the equations of motion $i\slashed{D}q=m_q q$ and $i\slashed{D}\chi=m_\chi\chi$,
\begin{align}
D_\mu D^\mu\bar q_R \chi = -(m_\chi^2+m_q^2)q_R \chi+2(D_\mu\bar q_R)( D^\mu\chi)\;.
\end{align}
By iterating this relation, we can rewrite the expansion as
\begin{align}
  \frac{1}{D_\mu D^\mu + m_\eta^2} \bar q_R \chi = \sum_{n=0}^\infty \frac{(2D_\mu^{\bar q} D^{\chi,\mu}+2m_\chi m_q)^n}{(\Delta m^2)^{n+1}}\bar q_R \chi\,,
\end{align}
where $D_\mu^{\bar q}$ acts on $\bar q_R$ and $D^{\chi,\mu}$ on $\chi$.
Finally we also use the Fierz identity in the form
\begin{align}
(\bar\chi q_R)(\bar q_R)\chi = \frac14 \bar\chi\gamma^\mu\chi \bar q_R \gamma_\mu q_R - \frac14 \bar\chi\gamma^\mu\gamma_5\chi \bar q_R \gamma_\mu\gamma_5 q_R\,,
\end{align}
where we have used that $q_R=P_R q_R$ is chiral. Putting everything together, and including the terms coming from the hermitean conjugate terms, one can formally write the effective Lagrangian as a sum of spin-independent (SI) and spin-dependent (SD) contributions,
\begin{align}
 {\cal L}_{eff}^{SI} & =  \frac{y^2}{4\Delta m^2}\sum_{n=1}^\infty  \left[ \left(\frac{2D_\mu^{\bar q} D^{\chi,\mu}+2m_\chi m_q}{\Delta m^2}\right)^n+\left(\frac{2D_\mu^{q} D^{\bar\chi,\mu}+2m_\chi m_q}{\Delta m^2}\right)^n\right] \nonumber\\
&\times \bar\chi\gamma^\mu\chi\, \bar q_R \gamma_\mu q_R  \\
 {\cal L}_{eff}^{SD} & = -\frac{y^2}{4\Delta m^2}\sum_{n=0}^\infty  \left[ \left(\frac{2D_\mu^{\bar q} D^{\chi,\mu}+2m_\chi m_q}{\Delta m^2}\right)^n+\left(\frac{2D_\mu^{q} D^{\bar\chi,\mu}+2m_\chi m_q}{\Delta m^2}\right)^n\right] \nonumber \\
& \times \bar\chi\gamma^\mu\gamma_5\chi\, \bar q_R \gamma_\mu\gamma_5 q_R 
\end{align}
The lowest orders correspond to the effective Lagrangian of Drees and Nojiri \cite{Drees:1993bu}. For the spin-dependent one,
the lowest order is 
\begin{align}
{\cal L}_{eff}^{SD} = -\frac{y^2}{2\Delta m^2}\bar\chi\gamma^\mu\gamma_5\chi \bar q_R \gamma_\mu\gamma_5 q_R\,,
\end{align}
which corresponds to Eq (1) of \citen{Drees:1993bu} for a chiral interaction with $a=-b=y$. The $n=0$ order of the spin-independent interaction vanishes because $\bar\chi\gamma^\mu\chi=0$ for Majorana spinors (however, {\it e.g.} $\bar\chi\gamma^\mu\partial_\nu\chi$ in non-zero). Using this property, the lowest non-zero contribution can be written as
\begin{align}
{\cal L}_{eff}^{SI} = -\frac{y^2}{2(\Delta m^2)^2}(\bar\chi\gamma^\mu D_\nu\chi) (\bar q_R \gamma_\mu D_\nu q_R-(D_\nu\bar q_R) \gamma_\mu q_R)
\end{align}
This corresponds to Eq (10) of \citen{Drees:1993bu}, again for a chiral interaction, and includes the contribution from the twist-2 operator. The expansion parameter is schematically given by
\begin{align}
\frac{2D_\mu^{q} D^{\bar\chi,\mu}+2m_\chi m_q}{\Delta m^2} \sim \frac{4m_\chi(E_q+m_q)}{\Delta m^2}\,,
\end{align}
where $E_q$ is a typical energy scale for the parton inside the nucleon (more precisely, the higher orders contain higher moments of the parton distribution function in the nucleon). When the scalar and the dark matter particle are nearly degenerate, the expansion parameter is thus of the order of $m_\chi E_q/(m_\eta^2-(m_\chi+m_q)^2)\sim E_q/(m_\eta-m_\chi-m_q)$.

\bibliographystyle{ws-ijmpd}
\bibliography{GIV}


\end{document}